\documentclass[reprint,superscriptaddress,nofootinbib,amsmath,amssymb,prd,floatfix]{revtex4-1}
\pdfoutput=1
\usepackage{graphicx}
\usepackage{bbold}
\usepackage{mathtools}
\usepackage[dvipsnames]{xcolor}
\usepackage{amsmath}
\allowdisplaybreaks
\usepackage{amssymb}

\usepackage{slashed}
\usepackage[utf8]{inputenc}
\usepackage[colorlinks=true,linkcolor=blue,bookmarksopen,bookmarksnumbered]{hyperref}
\usepackage{color}
\usepackage[normalem]{ulem}
\usepackage{soul}
\usepackage{units}
\usepackage{rotating}
\usepackage{hhline,multirow,tabularx}
\usepackage{bm}
\usepackage{hyperref}
\usepackage[export]{adjustbox}
\usepackage{mdframed}
\usepackage{comment}
\usepackage{natbib}

\def\bea#1\eea{\begin{align}#1\end{align}} 
\newcommand{\nnu}{\nonumber\\}
\newcommand{\bef}{\begin{figure}[htb]\centering}
\newcommand{\eef}{\end{figure}}

\newcommand{\bra}[1]{\left< #1 \right |}
\newcommand{\ket}[1]{\left | #1 \right >}

\usepackage{lipsum}
\begin{document}

\title{Studying chirality imbalance with quantum algorithms}

\author{Alexander M. Czajka}
\email{aczajka74@physics.ucla.edu}
\affiliation{Department of Physics and Astronomy, University of California, Los Angeles, CA 90095, USA}
\affiliation{Mani L. Bhaumik Institute for Theoretical Physics, University of California, Los Angeles, CA 90095, USA}

\author{Zhong-Bo Kang}
\email{zkang@ucla.edu}
\affiliation{Department of Physics and Astronomy, University of California, Los Angeles, CA 90095, USA}
\affiliation{Mani L. Bhaumik Institute for Theoretical Physics, University of California, Los Angeles, CA 90095, USA}
\affiliation{Center for Quantum Science and Engineering, University of California, Los Angeles, CA 90095, USA}
\affiliation{Center for Frontiers in Nuclear Science, Stony Brook University, Stony Brook, NY 11794, USA}

\author{Yuxuan Tee}
\email{yxtee0824@gmail.com}
\affiliation{Department of Physics and Astronomy, University of California, Los Angeles, CA 90095, USA}

\author{Fanyi Zhao}
\email{fanyizhao@physics.ucla.edu}
\affiliation{Department of Physics and Astronomy, University of California, Los Angeles, CA 90095, USA}
\affiliation{Mani L. Bhaumik Institute for Theoretical Physics, University of California, Los Angeles, CA 90095, USA}
\affiliation{Center for Quantum Science and Engineering, University of California, Los Angeles, CA 90095, USA}


\begin{abstract}
  To describe the chiral magnetic effect, the chiral chemical potential $\mu_5$ is introduced to imitate the impact of topological charge changing transitions in the quark-gluon plasma under the influence of an external magnetic field. We employ the $(1+1)$ dimensional Nambu-Jona-Lasinio (NJL) model to study the chiral phase structure and chirality charge density of strongly interacting matter with finite chiral chemical potential $\mu_5$ in a quantum simulator. By performing the Quantum imaginary time evolution (\texttt{QITE}) algorithm, we simulate the $(1+1)$ dimensional NJL model on the lattice at various temperature $T$ and chemical potentials $\mu,\ \mu_5$ and find that the quantum simulations are in good agreement with analytical calculations as well as exact diagonalization of the lattice Hamiltonian. 
\end{abstract}

\maketitle

\section{Introduction}
In quantum chromodynamics (QCD), several major challenges have gained considerable attention, including how the vacuum structures of QCD are affected in extreme environments~\cite{Fukushima:2010fe}. QCD research in hot and dense conditions is of great importance, not only from a purely theoretical perspective, but also for its numerous applications to the studies of the quark matter in the ultradense compact stars~\cite{Buballa:2003qv,McLerran:2007qj,McLerran:2008ua,PhysRevC.79.035807,Orsaria:2013hna,Ruggieri:2016lrn,Jimenez:2021ngg}, and the Quark-Gluon Plasma (QGP) which is abundantly produced in relativistic collisions of heavy ions~\cite{Gyulassy:2004zy,Soloveva:2020hpr}. Studying how non-perturbative features of QCD are affected by thermal excitations at high temperatures $T$ and by baryon-rich matter at finite chemical potentials $\mu$~\cite{Fukushima:2010bq} is highly interesting.

Besides the effects of finite $T$ and $\mu$, the influence of a strong magnetic field $B$ is an exciting topic relevant to phenomenology in relativistic heavy-ion collisions, where strong magnetic fields are generated in non-central collisions~\cite{Choudhury:2021jwd,Feng:2021pgf,STAR:2021mii,Milton:2021wku,Kharzeev:2022hqz,Kharzeev:2022ydx}. Many studies have been conducted on the effect of magnetic fields on the QCD vacuum~\cite{Adhikari:2021jff,Tawfik:2021eeb,Cao:2021rwx,Moreira:2021ety,Ding:2020hxw,Zhao:2019hta,Tawfik:2019rdd}, and it has been determined that magnetic fields $B$ act as a catalyst of dynamical chiral symmetry breaking~\cite{Gusynin:1999pq,Klimenko:1990rh,Klevansky:1989vi}. In the presence of a magnetic field, a finite current is induced along the direction of the field lines due to the anomalous production of an imbalance between right- and left-handed quarks, namely that the number of right-handed quarks $N_R$\footnote{More precisely, $N_R$ the number of right-handed quarks \textit{minus the number of left-handed antiquarks}, with $N_L$ defined analogously.} is not equal to the number of left-handed quarks $N_L$. This effect is known as the Chiral Magnetic Effect (CME)~\cite{Kharzeev:2007tn,Kharzeev:2007jp,Fukushima:2008xe}. 

The axial anomaly and topological objects in QCD are the fundamental physics of the CME. At low or zero temperatures, the change of non-trivial topological structure is related to instanton~\cite{Diakonov:2002fq,Schafer:1995pz} with the quantum tunneling effect. However, at finite temperatures, the transition is caused by sphalarons~\cite{Arnold:1987zg,Fukugita:1990gb,McLerran:1990de} and the chiral asymmetry shows up. Unbalanced left- and right-handed quarks can produce observable effects that can be used to investigate topological $\mathcal{P}$- and $\mathcal{CP}$-odd excitations~\cite{Witten:1979vv,Veneziano:1979ec,Schafer:1996wv,Vicari:2008jw,Kharzeev:2015znc,Bzdak:2019pkr,Chen:2019qoe}. Thus the CME is a phenomenologically and experimentally interesting effect of the strong magnetic field in heavy-ion collisions. In~\cite{Voloshin:2004vk}, an observable sensitive to local $\mathcal{P}$- and $\mathcal{CP}$-violation has been proposed for experiments. Measurements of charge correlations were made by STAR at RHIC~\cite{STAR:2009tro,STAR:2009wot,Hu:2021drc,STAR:2021mii,STAR:2021pwb,Zhao:2020utk}, where conclusive evidence of charge azimuthal correlations was observed, which could be a possible result from CME with local $\mathcal{P}$- and $\mathcal{CP}$-odd effects. Furthermore, consistent experimental data was provided by ALICE~\cite{ALICE:2012nhw,ALICE:2015cjr,Parmar:2017deh,ALICE:2017sss,ALICE:2020siw} and CMS~\cite{CMS:2017lrw,CMS:2016wfo} at the LHC, where the azimuthal correlator was measured to search for the CME in heavy-ion collisions. 

By introducing a finite chiral chemical potential $\mu_5$ that imitates the effects of the topological charge changing transitions, one can study the QCD phase diagram~\cite{Kharzeev:2020kgc} as well as the thermal behavior of the total chirality charge $N_5=N_R-N_L$ under the influence of an external magnetic field at finite temperature $T$ and baryon chemical potential $\mu$. At sufficiently high temperatures/densities, the strongly-interacting matter goes through a deconfinement phase transition from hadronic matter to quark-gluon plasma, and it is possible that a chirality charge is produced in the phase transition as a result of the flip of fermion helicity in the interaction with the gauge field. Moreover, it has been demonstrated that immediately after a heavy-ion collision, the chirality charge comes to and stays at an equilibrium value~\cite{Ruggieri:2016asg,Ruggieri:2016lrn,Ruggieri:2020qtq}. In light of these considerations, it is evident that exploring the chiral imbalance in the QCD phase diagrams is crucial for the description of heavy-ion collisions.

To study the chiral magnetic effect and the QCD chiral phase transition, the Nambu-Jona-Lasinio (NJL) model \cite{Nambu:1961tp,Nambu:1961fr} has been playing an important role for many years~\cite{PhysRevD.77.114028,COSTA2007431,Lu2015,doi:10.1142/S0217751X15501997,CUI2015172,PhysRevD.88.114019,PhysRevD.91.036006,PhysRevC.80.065805,PhysRevC.79.035807,PhysRevC.75.015805,PhysRevD.86.071502}. As an effective model for QCD, the NJL model is amenable to analytical calculations at finite temperature $T$ and chemical potentials $\mu$ and $\mu_5$.

In recent years, lattice QCD simulations have significantly improved our understanding of the QCD phase diagram at zero or small chemical potentials $\mu$~\cite{Yamamoto:2011gk,Braguta:2015owi,Braguta:2015zta,Alexandru:2015xva,Scorzato:2015qts,Braguta:2016aov}. However, as a result of the sign problem~\cite{Kapusta}, at finite $\mu$, Monte-Carlo simulation is unable to be directly applied because the fermion determinant becomes complex, and its phase fluctuations prohibit its interpretation as a probability density~\cite{Yamamoto:2011gk}. As a result, the sign problem is a fundamental impediment to comprehending the phase structure of nuclear matter. This is, however, not a flaw in QCD theory, but rather in the attempt to mimic quantum statistics via the functional integral using a classical Monte Carlo approach. 

Fortunately, as indicated in~\cite{Kharzeev:2020kgc,Feynman}, the statistical properties of a quantum computer can be used to obviate the necessity for a Monte Carlo study by modeling the lattice system on a quantum computer. And there have been abundant developments in applying quantum computing to solving physics problems. In recent years, it has been shown that using the current generation of Noisy Intermediate-Scale Quantum (NISQ) technology that quantum computers can solve complex problems like simulating thermal properties~\cite{Bauer,PhysRevA.61.022301,PhysRevLett.103.220502,PhysRevLett.108.080402,Temme2011,Yung754,Li:2021kcs,Zhang:2020uqo,Tomiya:2022chr,Xie:2022jgj,Davoudi:2022uzo}, evaluating ground states and real-time dynamics \cite{Arute:2020uxm,Ma2020,Kandala2019,PhysRevX.6.031007,Kandala2017,Peruzzo2014,PhysRevX.8.011021,Chiesa2019,Smith2019,Zhang2017,doi:10.1126/science.1232296,PhysRevB.101.014411,Feynman,doi:10.1126/science.273.5278.1073,DeJong:2020riy,deJong:2021wsd}, modeling many-body systems and relativistic effects ~\cite{wallraff_strong_2004,majer_coupling_2007,jordan_quantum_2012,zohar_simulating_2012,zohar_cold-atom_2013,banerjee_atomic_2013,banerjee_atomic_2012,wiese_ultracold_2013,wiese_towards_2014,jordan_quantum_2014,garcia-alvarez_fermion-fermion_2015,marcos_two-dimensional_2014,bazavov_gauge-invariant_2015,zohar_quantum_2015,mezzacapo_non-abelian_2015,dalmonte_lattice_2016,zohar_digital_2017,martinez_real-time_2016,bermudez_quantum_2017,gambetta_building_2017,krinner_spontaneous_2018,macridin_electron-phonon_2018,zache_quantum_2018,zhang_quantum_2018,klco_quantum-classical_2018,klco_digitization_2019,gustafson_quantum_2019,nuqs_collaboration_ensuremathsigma_2019,magnifico_real_2020,jordan_quantum_2019,lu_simulations_2019,klco_minimally-entangled_2020,lamm_simulation_2018,klco_su2_2020,alexandru_gluon_2019,mueller_deeply_2020,lamm_parton_2020,chakraborty_digital_2020,Bermudez:2018eyh,Ziegler:2020zkq,Ziegler:2021yua}, etc. Though digital quantum simulations on thermal physical systems were researched earlier on, finite-temperature physics is less well-known and still has to be improved on quantum computers~\cite{PRXQuantum.2.010317}. Several algorithms for imaginary time evolution on quantum computers, both with and without ansatz dependency, have been introduced in recent years. In particular, the Quantum Imaginary Time Evolution (\texttt{QITE}) algorithm applies a unitary operation to simulate imaginary time evolution and has been performed to simulate energy and magnetism in the Transverse Field Ising Model (TFIM)~\cite{Ville:2021hrl}, the chiral condensate in NJL model~\cite{Czajka:2021yll} and so on. This study, along with previous studies, demonstrates that NISQ quantum computers can provide consistent and correct answers to physical problems that cannot be solved efficiently or effectively using classical computing algorithms, indicating promising future applications of quantum computing in non-perturbative QCD and beyond.

The remainder of this paper is organized as follows: In Sec.~\ref{sec:background}, we provide a brief description of the $(1+1)$ dimensional NJL model and the \texttt{QITE} algorithm used for the quantum simulation. In Sec.~\ref{sec:analytic}, we show the analytic calculations of chiral condensate and chirality charge density at finite temperature, baryon and chiral chemical potentials. We then present and discuss our numerical results from the quantum simulation in comparison with analytical computations and exact diagonalization results in Sec.~\ref{sec:results}. Finally, our conclusions are summarized in Sec.~\ref{sec:conclusion}.

\section{Background}\label{sec:background}
In this section, we first briefly introduce the $(1+1)$-dimensional Nambu-Jona-Lasinio (NJL) model, and present the lattice discretization of the NJL Hamiltonian. Next, we provide a brief introduction of the \texttt{QITE} algorithm used for the quantum simulation.

\subsection{The NJL model in $(1+1)$ dimensions}\label{sec:njl}
The NJL model was defined in \cite{Nambu:1961tp,Nambu:1961fr} with the Lagrangian density
\bea
\label{eq:Lagrangian1}
\mathcal{L}_{\rm NJL}&=\bar{\psi}(i\slashed{\partial}-m)\psi+g\left[(\bar{\psi}\psi)^2+(\bar{\psi}i\gamma_5\psi)^2\right]\,,
\eea
where $m$ and $g$ represent the bare quark mass and coupling constant, respectively, and $\slashed{\partial} \equiv \gamma^\mu\partial_\mu$. The explicit representation of the $(1 + 1)$-dimensional Clifford algebra $\{\gamma^{\mu}, \gamma^\nu\} = 2\eta^{\mu\nu}$ used in this work is
\bea
\gamma_0=Z,\quad\gamma_1=-iY,\quad\gamma_5=\gamma_0\gamma_1=-X
\eea
where the Pauli gates are
\bea
X = \begin{pmatrix}
  0 & 1\\
  1 & 0
\end{pmatrix}, \ 
Y = \begin{pmatrix}
  0 & -i \\
  i & 0
\end{pmatrix}, \ 
Z = \begin{pmatrix}
  1 & 0 \\
  0 & -1
\end{pmatrix}
\eea
A simplified version of the NJL Model, the Gross-Neveu (GN) model \cite{Gross:1974jv}, is given by 
\bea
\label{eq:Lagrangian0}
\mathcal{L}&=\bar{\psi}(i\slashed{\partial}-m)\psi+g(\bar{\psi}\psi)^2\,.
\eea
To study the chiral phase transition and chirality imbalance in the GN model, we introduce additional terms related to non-zero chemical potential $\mu$ and chiral chemical potential $\mu_5$, which mimics the chiral imbalance between right- and left-chirality quarks coupled with the chirality charge density operator $n_5=\bar{\psi}\gamma_0\gamma_5\psi$. Therefore, the modified Lagrangian is
\bea\label{eq:Lagrangian}
\mathcal{L}=&\bar{\psi}(i\slashed{\partial}-m)\psi+g(\bar{\psi}\psi)^2+\mu\bar{\psi}\gamma_0\psi+\mu_5\bar{\psi}\gamma_0\gamma_5\psi\,.
\eea
In our previous work~\cite{Czajka:2021yll}, we have studied the behavior of the chiral condensate $\langle\bar\psi\psi\rangle$ at $\mu_5=0$ with finite and non-zero temperature $T$ and chemical potential $\mu$. The Hamiltonian $\mathcal H = i\psi^\dagger\partial_0\psi - \mathcal L$ corresponding to Eq.~\eqref{eq:Lagrangian} is given by
 \bea
\label{eq:Hamiltonian}
\mathcal{H}=&\bar{\psi}(i\gamma_1{\partial_1}+m)\psi-g(\bar{\psi}\psi)^2-\mu\bar{\psi}\gamma_0\psi\nnu
&-\mu_5\bar{\psi}\gamma_0\gamma_5\psi\,.
\eea
For clarification, when we mention ``NJL model'' in this work, we refer to the Hamiltonian given in Eq.~\eqref{eq:Hamiltonian}.

As in our previous work~\cite{Czajka:2021yll}, we first use a staggered fermion field $\chi_{2n},\ \chi_{2n+1}$ to discretize the Dirac fermion field $\psi(x)$. With the lattice spacing $a$ and $n=0,\cdots,\,N/2-1$ where $N$ is an even integer, one has~\cite{Borsanyi:2010cj,Staggered,Borsanyi:2013bia,Aoki:2005vt,HotQCD:2014kol,Aubin:2019usy}
\bea
\psi(x)=\frac{1}{\sqrt{a}}\begin{pmatrix}
   \chi_{2n}\ \\
   \chi_{2n+1}\
\end{pmatrix} .
\eea
Therefore, one obtains the following discrete approximations of the various operators appearing in the Hamiltonian $H = \int dx\, \mathcal H$ where periodic boundary conditions are considered,
\bea
\int dx\bar{\psi}i\gamma_1\partial_1\psi=&\ a\sum_{n=0}^{N/2-1}{\psi}_n^\dagger i\gamma_5\partial_1\psi_n\nnu
=&-\frac{i}{2a}\bigg[\sum_{n=0}^{N-2}\left(\chi_n^\dagger\chi_{n+1}-\chi_{n+1}^\dagger\chi_n\right)\nnu
&\hspace{1.cm}+\left(\chi_{N-1}^\dagger\chi_{0}-\chi_{0}^\dagger\chi_{N-1}\right)\bigg]\,,\\
\int dx\bar{\psi}\psi=&\ a\sum_{n=0}^{N/2-1}{\psi}_n^\dagger\gamma_0\psi_n=\sum_{n=0}^{N-1}(-1)^n\chi_n^\dagger\chi_n\,,\\
\int dx(\bar{\psi}\psi)^2=&\ a\sum_{n=0}^{N/2-1}({\psi}_n^\dagger\gamma_0\psi_n)^2\nnu
=&\ \frac{1}{a}\sum_{n=0}^{N/2-1}\left(\chi_{2n}^\dagger\chi_{2n}-\chi_{2n+1}^\dagger\chi_{2n+1}\right)^2\nnu
=&-\frac{2}{a}\sum_{n=0}^{N/2-1}\left(\chi_{2n}^\dagger\chi_{2n}\chi_{2n+1}^\dagger\chi_{2n+1}\right)\nnu
&+\frac{1}{a}\sum_{n=0}^{N-1}\left(\chi_{n}^\dagger\chi_{n}\right)^2\,,\\
\int dx\bar{\psi}\gamma_0\psi=&\ a\sum_{n=0}^{N/2-1}{\psi}_n^\dagger\psi_n=\sum_{n=0}^{N-1}\chi_n^\dagger\chi_n\,,\\
\int dx\bar{\psi}\gamma_0\gamma_5\psi=&\ a\sum_{n=0}^{N/2-1}\psi^\dagger_n\gamma_5\psi_n\nnu
=&-\sum_{n=0}^{N/2-1}\left(\chi_{2n}^\dagger\chi_{2n+1}+\chi_{2n+1}^\dagger\chi_{2n}\right)\,.
\eea
Subsequently, the Hamiltonian in Eq.~\eqref{eq:Hamiltonian} becomes
\bea
H=&\int dx \big[\bar{\psi}(m+i\gamma_1\partial_1-\mu\gamma_0-\mu_5\gamma_0\gamma_5)\psi-g(\bar{\psi}\psi)^2\big]\nnu
=&\,m\sum_{n=0}^{N-1}(-1)^n\chi_n^\dagger\chi_n-\frac{i}{2a}\bigg[\sum_{n=0}^{N-2}\left(\chi_n^\dagger\chi_{n+1}-\chi_{n+1}^\dagger\chi_n\right)\nnu
&+\left(\chi_{N-1}^\dagger\chi_{0}-\chi_{0}^\dagger\chi_{N-1}\right)\bigg]-\mu\sum_{n=0}^{N-1}\chi_n^\dagger\chi_n\nnu
&+{\mu_5}\sum_{n=0}^{N/2-1}\left(\chi_{2n}^\dagger\chi_{2n+1}+\chi_{2n+1}^\dagger\chi_{2n}\right)-\frac{g}{a}\sum_{n=0}^{N-1}\left(\chi_{n}^\dagger\chi_{n}\right)^2\nnu
&+\frac{2g}{a}\sum_{n=0}^{N/2-1}\left(\chi_{2n}^\dagger\chi_{2n}\chi_{2n+1}^\dagger\chi_{2n+1}\right)^2\label{eq:Hamiltonia_all}\,.
\eea
In order to implement the Hamiltonian to a quantum circuit, we write down the spin representation of the Hamiltonian using the Jordan-Wigner transformation \cite{Jordan1928},
\bea
\chi_{n}=\frac{X_{n}-iY_{n}}{2}\prod_{\mu=0}^{n-1}\left(-iZ_\mu\right)\,,\label{eq:jw-trans}
\eea
where $X_{n},\ Y_{n}$ and $Z_{n}$ are the Pauli-$X,\ Y$ and $Z$ matrices acting on the $n$-th lattice site. In such spin representation, the discrete approximations of the relevant operators are then given by
\bea
\int dx\bar{\psi}i\gamma_1\partial_1\psi=&\sum_{n=0}^{N-2}\frac{1}{4a}\left(X_{n}X_{n+1}+Y_{n}Y_{n+1}\right)\nnu
&\hspace{-1.2cm}+\frac{(-1)^{N/2}}{4a}\left(X_{N-1}X_{0}+Y_{N-1}Y_{0}\right)\prod_{i=1}^{N-2}Z_i\,,\label{eq:equi1}\\
\int dx\bar{\psi}\psi=&\sum_{n=0}^{N-1}(-1)^n\frac{Z_n}{2}\,,\\
\int dx(\bar{\psi}\psi)^2=&-\frac{1}{2a}\sum_{n=0}^{N/2-1}(\mathbb{1}+Z_{2n})(\mathbb{1}+Z_{2n+1})\nnu
&+\frac{1}{2a}\sum_{n=0}^{N-1}(\mathbb{1}+Z_n)\,,\label{eq:equi3}\\
\int dx\bar{\psi}\gamma_0\psi=&\sum_{n=0}^{N-1}\frac{Z_n}{2}\,,\\
\int dx\bar{\psi}\gamma_0\gamma_5\psi=&\frac{1}{2}\sum_{n=0}^{N/2-1}\left(X_{2n}Y_{2n+1}-Y_{2n}X_{2n+1}\right)\label{eq:equi8}\,.
\eea
In Eqs.~\eqref{eq:equi1} and ~\eqref{eq:equi3}, we have imposed periodic boundary conditions. With the relations in Eqs.~\eqref{eq:equi1}--\eqref{eq:equi8}, we decompose the total $(1+1)$-dimensional NJL Hamiltonian into 6 pieces, writing $\displaystyle H=\sum_{j=1}^6 H_j$ with
\bea
H_1=&\sum_{n=0}^{N/2-1}\frac{1}{4a}\left(X_{2n}X_{2n+1}+Y_{2n}Y_{2n+1}\right)\,,\label{Hamiltonia1}\\
H_2=&\sum_{n=1}^{N/2-1}\frac{1}{4a}\left(X_{2n-1}X_{2n}+Y_{2n-1}Y_{2n}\right)\label{Hamiltonia2}\\
  &+\frac{(-1)^{N/2}}{4a}\left(X_{N-1}X_{0}+Y_{N-1}Y_{0}\right)\prod_{i=1}^{N-2}Z_i\,,\nnu
H_3=&\frac{m}{2}\sum_{n=0}^{N-1}(-1)^nZ_n\,,\label{Hamiltonia3}\\
H_4=&\frac{g}{2a}\bigg(\sum_{n=0}^{N/2-1}(\mathbb{1}+Z_{2n})(\mathbb{1}+Z_{2n+1})-\sum_{n=0}^{N-1}(\mathbb{1}+Z_n)\bigg)\label{Hamiltonia4}\\
H_5=&-\frac{\mu}{2}\sum_{n=0}^{N-1}Z_n\,,\label{Hamiltonia5}\\
H_6=&-\frac{\mu_5}{2}\sum_{n=0}^{N/2-1}\left(X_{2n}Y_{2n+1}-Y_{2n}X_{2n+1}\right)\,.\label{Hamiltonia6}
\eea
Finally, with the decomposition of the Hamiltonian shown in Eqs.~\eqref{Hamiltonia1}--\eqref{Hamiltonia6}, we are able to perform the Suzuki-Trotter decomposition~\cite{trotter1959product,Suzuki} to study the effects of the chiral chemical potential $\mu_5$ on the finite temperature properties of the chiral condensate $\langle\bar{\psi}\psi\rangle$ and chirality charge density $n_5$ of the $(1+1)$-dimensional NJL model on a quantum simulator. 

\subsection{Quantum imaginary time evolution algorithm}\label{sec:qite}
In this section, we introduce the quantum imaginary time evolution (\texttt{QITE}) algorithm~\cite{Motta}, which is use for evaluating the temperature dependence of the NJL model for various values of the baryochemical potential $\mu$ and chiral chemical potential $\mu_5$. As pointed out in \cite{Motta}, compared with other techniques for quantum thermal averaging procedures~\cite{Terhal:1998yh,Temme:2009wa,https://doi.org/10.48550/arxiv.1603.02940,Brandao:2016mfe}, the \texttt{QITE} algorithm is advantageous in generating thermal averages of observables without any ancillae or deep circuits. Moreover, the \texttt{QITE} algorithm is more resource-efficient and requires exponentially less space and time in each iteration than its classical equivalents. 

Generally, for a given Hamiltonian $H$, one can approximate the (Euclidean) evolution operator $e^{-\beta H}$ by applying the Suzuki-Trotter decomposition~\cite{trotter1959product,Suzuki}
\bea
e^{- \beta H} = \left(e^{-\Delta\beta H}\right)^{N} + O(\Delta\beta^2),
\eea
where $\Delta\beta$ is a chosen imaginary time step and $N = \beta / \Delta\beta$ is the number of iterations needed to reach imaginary time $\beta=1/T$ with temperature $T$. However, since the evolution operator $e^{- \Delta\beta H}$ is not unitary, it cannot be implemented as a sequence of unitary quantum gates. In order to compute the Euclidean time evolution of a state $\ket \Psi$ on a quantum computer, one needs to approximate the action of the operator $e^{- \Delta\beta H}$ by some unitary operator. Fortunately, the \texttt{QITE} algorithm provides a procedure for doing this.

In the \texttt{QITE} algorithm, to approximate the Euclidean time evolution of $\ket \Psi$, a Hermitian operator $A$ is introduced such that the effect of the non-unitary operator $e^{- \Delta\beta H}$ on a quantum state $\ket\Psi$ is replicated by the unitary operator $e^{-i \Delta\beta A}$, namely\footnote{Recall that quantum states are represented by \textit{rays} $\{\alpha\ket\Psi : \alpha\in\mathbb C\}$ in a Hilbert state, not by the vectors themselves, since the normalization/phase of the state vectors are nonphysical.}
\bea
\frac{1}{\sqrt{c(\Delta\beta)}} e^{- \Delta\beta H} \ket{\Psi} \approx e^{-i \Delta\beta A} \ket{\Psi}\,,\label{eq:nonu_to_u}
\eea
where the normalization $c(\Delta\beta) = \langle\Psi|e^{-2\Delta\beta H}\ket{\Psi}$. 

When $\Delta\beta$ is very small, one is able to expand Eq.~\eqref{eq:nonu_to_u} up to $\mathcal{O}(\Delta\beta)$, truncating after the first nontrivial term. Then at imaginary time $\beta$, the change of the quantum states under the operators $e^{-\Delta\beta H}$ and $e^{-i\Delta\beta A}$ per small imaginary time interval $\Delta\beta$ can be represented by
\bea
\ket{\Delta\Psi_H(\beta)}&=\frac{1}{\Delta\beta}\left(\frac{1}{\sqrt{c(\Delta\beta)}} e^{- \Delta\beta H} \ket{\Psi(\beta)}-\ket{\Psi(\beta)}\right)\,,\\
\ket{\Delta\Psi_A(\beta)}&=\frac{1}{\Delta\beta}\bigg(e^{-i \Delta\beta A} \ket{\Psi(\beta)}-\ket{\Psi(\beta)}\bigg)\,,
\eea
As proposed in~\cite{Motta}, to determine the Hermitian operator $A$, we first parameterize it in terms of Pauli matrices as below
\bea
A(\bm a)=\sum_{\mu}a_{\mu}\hat{\sigma}_{\mu}\,.\label{eq:amu}
\eea
Here $\hat{\sigma}_{\mu}=\prod_{l}{\sigma}_{\mu_l,l}$ is a Pauli string and the subscript $\mu$ of $a_{\mu}$ labels the various Pauli strings. To evaluate the Hermitian operator $A$, we need to minimize the objective function $F(a)$ defined by
\bea
F(a)=&||\big(\ket{\Delta\Psi_H(\beta)}-\ket{\Delta_A\Psi(\beta)}\big)||^2\label{eq:fa}\\
=&||\,\ket{\Delta\Psi_H(\beta)}\,||^2+\sum_{\mu,\nu}a_\nu a_\mu\bra{\Psi(\beta)}\hat{\sigma}_\nu^\dagger\hat{\sigma}_\mu\ket{\Psi(\beta)}\nnu
&+i\sum_\mu \frac{a_\mu}{\sqrt{c(\Delta\beta)}}\bra{\Psi(\beta)} \big(H\hat{\sigma}_\mu-\hat{\sigma}_\mu^\dagger H\big)\ket{\Psi(\beta)}\,.\nonumber
\eea
The first term $||\,\ket{\Delta\Psi_H(\beta)}\,||^2$ is irrelevant to $a_\mu$. Thus, we take the derivative with respect to $a_\mu$ and set it equal to zero, yielding the linear equation $({\bm S}+{\bm S}^T)\,{\bm a}={\bm b}$, where the matrix ${\bm S}$ and vector ${\bm b}$ are defined by
\bea
S_{\mu\nu}&=\bra{\Psi(\beta)}\hat{\sigma}_\nu^\dagger\hat{\sigma}_\mu\ket{\Psi(\beta)}\,,
\\
b_\mu&=-\frac{i}{\sqrt{c(\Delta\beta)}}\bra{\Psi(\beta)} \big(H\hat{\sigma}_\mu+\hat{\sigma}_\mu^\dagger H\big)\ket{\Psi(\beta)}\,.
\eea
From this equation, we are able to solve for $a_\mu$ and evolve an initial quantum state under the unitary operator $e^{-i\Delta\beta A}$ to any imaginary time $\beta$ by Trotterization,
\bea
\ket{\Psi(\beta)}&=\left(e^{- i\Delta\beta A}\right)^N \ket{\Psi(0)}+\mathcal{O}(\Delta\beta)\,.
\eea

\begin{figure}
\centering
\includegraphics[width=0.4\textwidth,trim={0cm 1cm 2cm 2cm},clip]{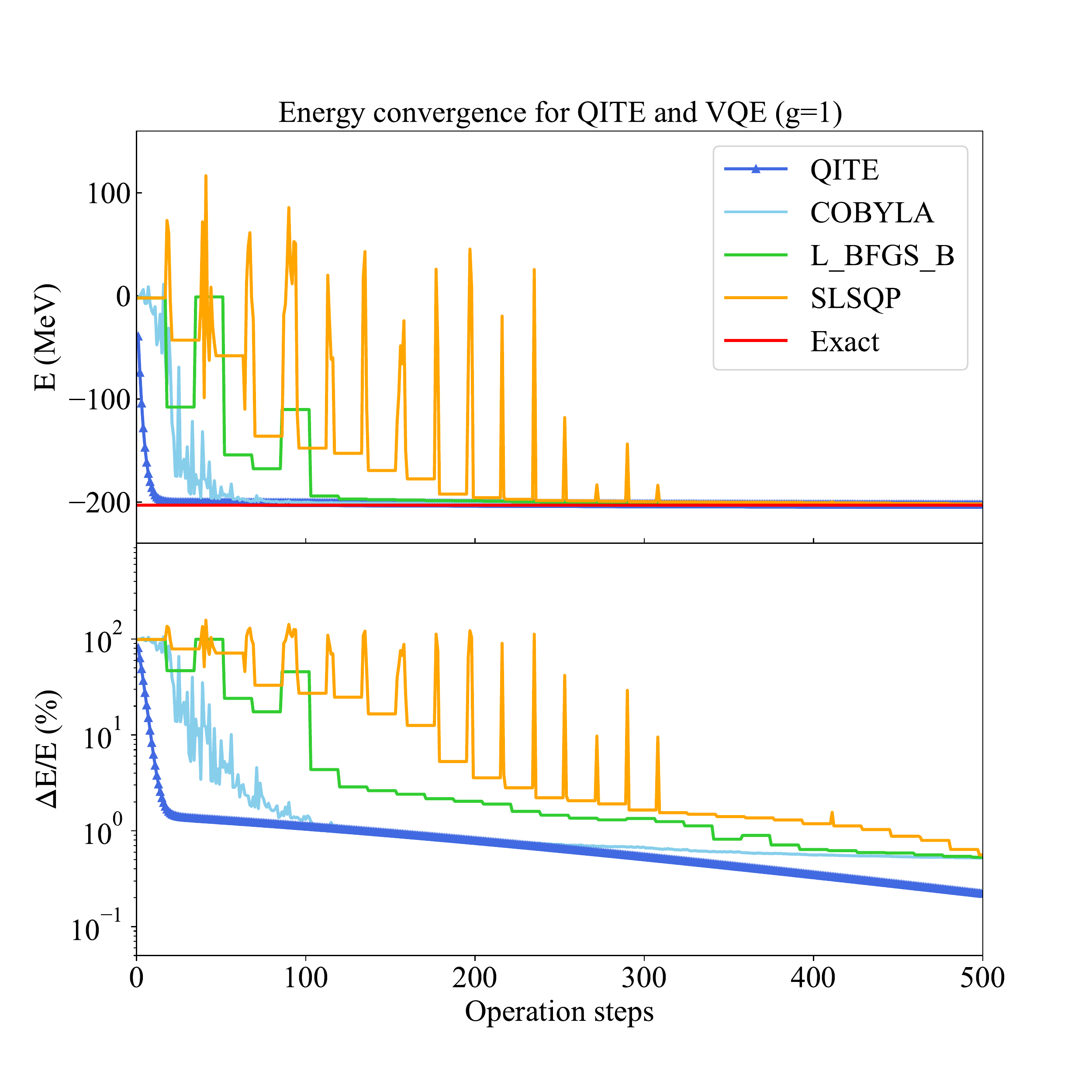}
\includegraphics[width=0.4\textwidth,trim={0cm 1cm 2cm 1cm},clip]{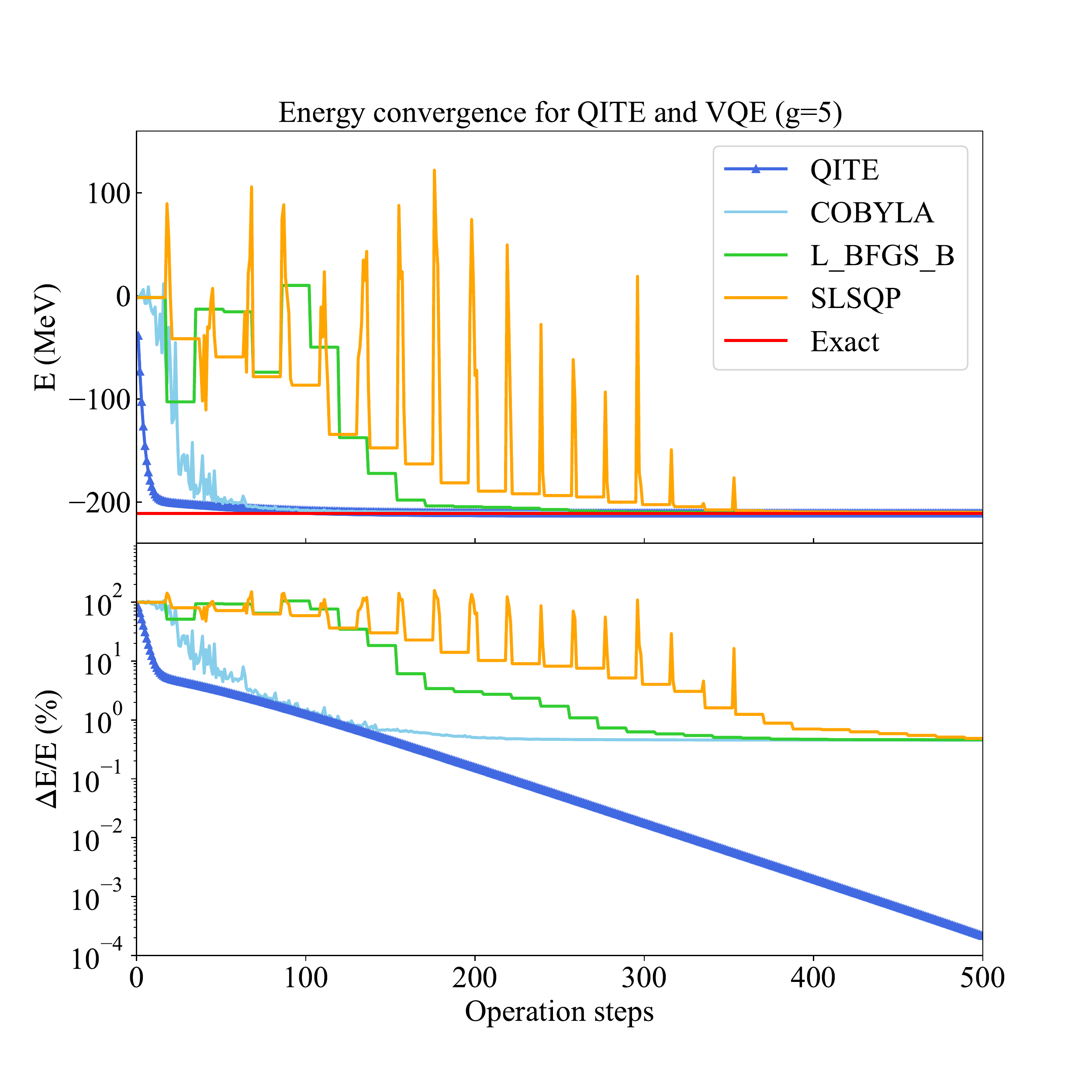}
\caption{Comparison between the \texttt{QITE} algorithms and \texttt{VQE} of various optimizers (\texttt{QOBYLA}, \texttt{L\_BFGS\_B} and \texttt{SLSQP}) for $g=1$ (top panel) and $g=5$ (bottom panel) to reach the ground state energy of the NJL Hamiltonian. We use 4 qubits for the NJL model with $m=100$ MeV, $g=1$, $\mu=100$ MeV and $\mu_5=10$ MeV. For the \texttt{VQE} algorithm, the $x$-axis represents the number of optimization steps while for the \texttt{QITE} algorithms, the $x$-axis is the number of thermal evolution steps $x=\beta/\Delta\beta$ ($\Delta\beta=0.001$). We chose three optimizers provided by \texttt{QISKIT} and set the maximum optimization steps as 500 for all optimizers.}\label{fig:vqe1}
\end{figure}

In the remainder of this section, we compare the performance of the \texttt{QITE} algorithm to the Variational Quantum Eigensolver (\texttt{VQE}) algorithms in obtaining the ground-state energy of the $(1+1)$-dimensional NJL model defined above. As pointed out in \cite{Motta}, the \texttt{QITE} algorithm is efficient for calculating ground-state energies. In Fig.~\ref{fig:vqe1}, we plot the ground-state energy with bare mass $m=100$ MeV, chemical potentials $\mu=100$ MeV and $\mu_5=10$ MeV at $g=1$ and $g=5$, respectively, as a function of the number of operation steps performed by the various agorithms. The \texttt{VQE} algorithms have been used to obtain many stellar results on NISQ hardware~\cite{Barison:2022drt,Johnson:2022lpl,Cao:2021uls,Omiya:2021vol,Johnson:2022lpl}, sparking a lot of attention recently. However, the reliance on an ansatz limits the effectiveness of the algorithm, as the part of the Hilbert space that the \texttt{VQE} can scan is influenced by the specific variational ansatz used, and the classical component of the algorithm requires optimization as well. The \texttt{QITE} algorithm, on the other hand, does not need an ansatz and evolves the prepared state closer to the ground state after each time-step in a controlled manner. The state should converge to the ground state provided the initial state has some overlap with it, with a \textit{quantifiable} error. 

As presented in Fig.~\ref{fig:vqe1}, we plot $500$ operation steps for both \texttt{QITE} (blue points with curve) and \texttt{VQE} algorithms (light-blue, green and orange curves for various optimizers). For the \texttt{QITE} algorithm, we chose imaginary time step $\Delta\beta=0.001$ and the quantum circuit is implemented in \texttt{QFORTE} \cite{stair2021qforte}, a quantum algorithms library based on \texttt{PYTHON}. For the \texttt{VQE} algorithm, operation steps are the optimizer steps and the results are given by \texttt{QISKIT}~\cite{Qiskit} of \texttt{IBMQ}~\cite{ibm}, where various optimizers are applied for comparison with \texttt{QITE} simulations. One can find that the \texttt{QITE} algorithm reaches the ground-state energy at a higher accuracy in less operation steps compared to the optimizers of \texttt{VQE} shown in the plot. In fact, one can see that the error of the \texttt{VQE} implementations ``levels off'' at around $1\%$, a consequence of the variational ansatz scanning a set that is a finite distance from the true vacuum.

Besides calculating the ground state energy, another powerful application of the \texttt{QITE} algorithm is to simulating the temperature dependence of a thermal process. In this work, we will use the \texttt{QITE} algorithm introduced above to generate the thermal state $|\Psi(\beta/2)\rangle$, then apply the following relation to obtain the thermal average of an observable $\hat{O}$,
\bea
\langle \hat{O}\rangle_\beta=\frac{{\rm Tr}(e^{-\beta \hat{H}}\hat{O})}{{\rm Tr}(e^{-\beta \hat{H}})}=\frac{\sum_{i\in\mathcal{S}}\bra{i}e^{-\beta \hat{H}/2}\hat{O}e^{-\beta \hat{H}/2}\ket{i}}{\sum_{i\in\mathcal{S}}\bra{i}e^{-\beta \hat{H}}\ket{i}}\,.
\eea
Here $\mathcal{S}$ is a complete set as the basis of the ground state~\cite{Motta}. Specifically, we will choose $\hat{O}=\bar{\psi}\psi$ and $\hat{O}=\bar{\psi}\gamma_0\gamma_5\psi$ for calculating the thermal average of chiral condensate $\langle\bar{\psi}\psi\rangle$ and chirality charge density $n_5=\langle\bar{\psi}\gamma_0\gamma_5\psi\rangle$.

\section{Analytical calculation}\label{sec:analytic}\label{sec:theory}
In this section, before reaching out to quantum simulations, we first provide the analytical calculations for the chiral condensate and chirality charge density at finite temperature $T$, baryochemical potential $\mu$ and chiral chemical potential $\mu_5$ using the Lagrangian provided in Eq.~\eqref{eq:Lagrangian} by minimizing the thermodynamic (Landau) potential.

\subsection{The Landau potential and chiral condensate}\label{sec:gap_eqn}
\begin{figure}
\centering
\includegraphics[width=0.35\textwidth,trim={2cm 5cm 0cm 2.7cm},clip]{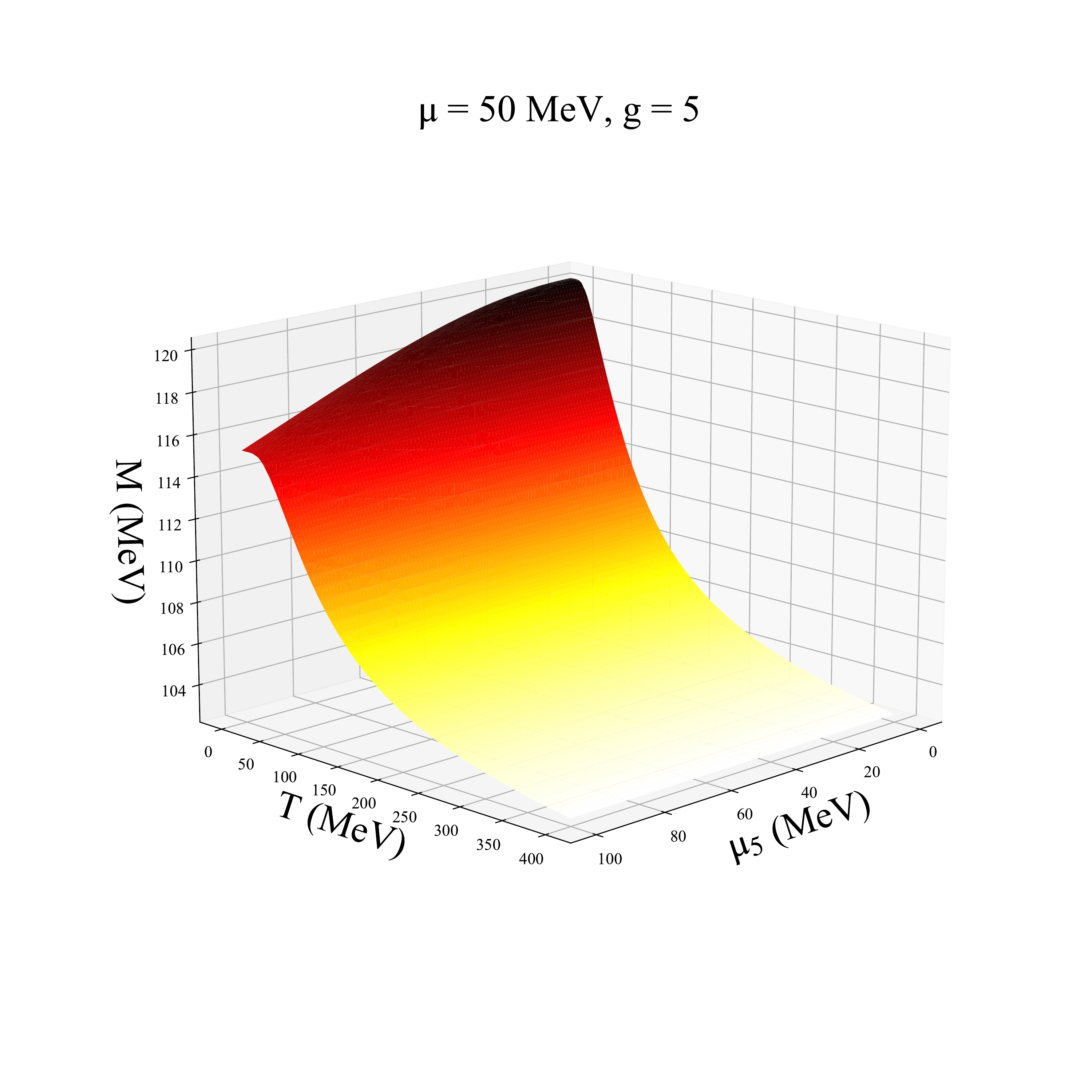}
\includegraphics[width=0.335\textwidth,trim={2cm 0cm 2cm 1cm},clip]{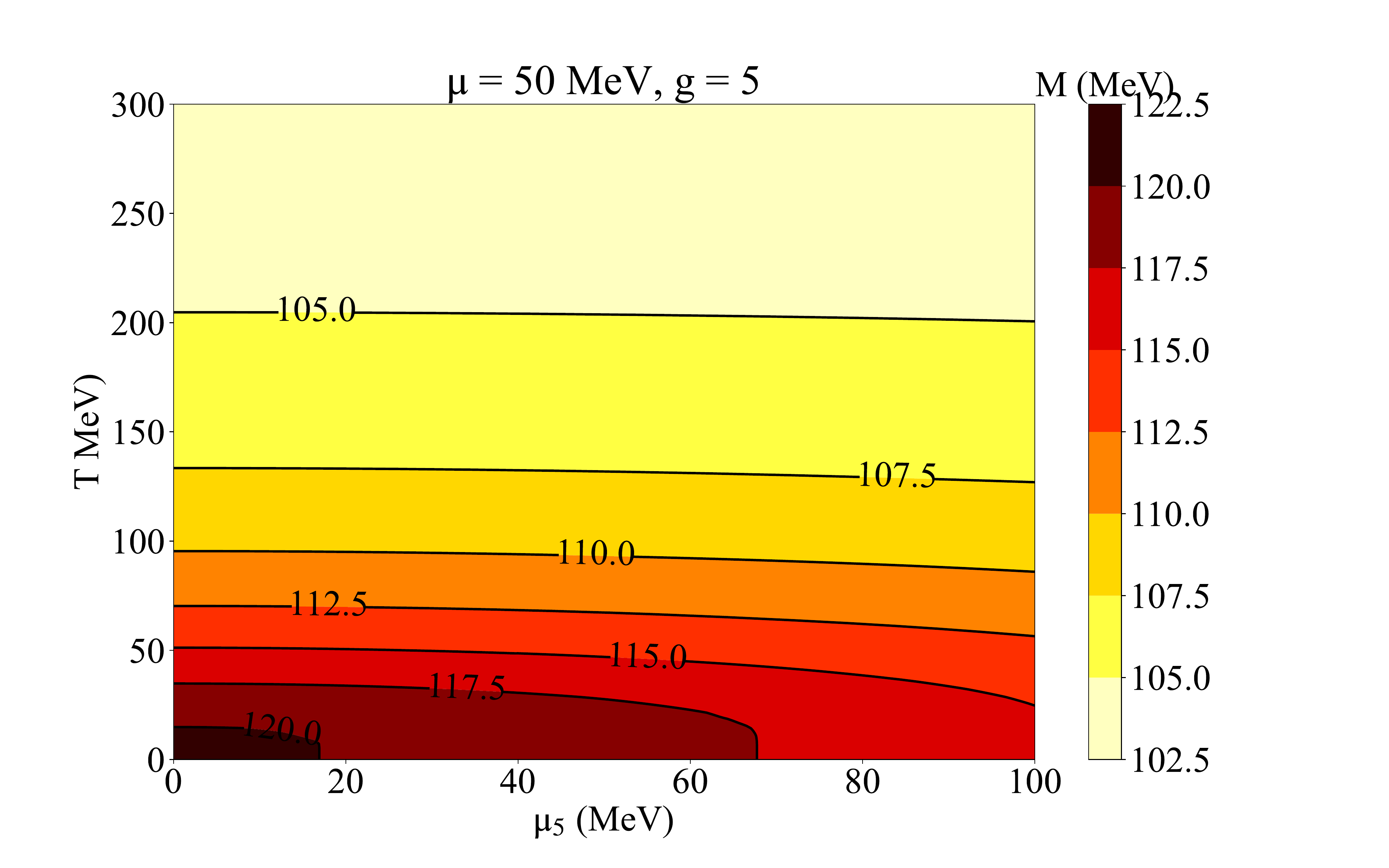}
\caption{The effective mass $M$ as a function of the chiral chemical potential $\mu_5$ and temperature $T$ at $g=5$, chemical potential $\mu=50$ MeV. Upper: 3D Surface Plot. Lower: Contour Plot.}\label{fig:3dplot1}
\end{figure}
We first provide theoretical calculations for the vacuum chiral condensate $\langle\bar\psi\psi\rangle$ at various temperatures $T$, chemical potentials $\mu$ and $\mu_5$ under the Hamiltonian $\mathcal{H}$ as defined in Eq.~\eqref{eq:Hamiltonian}. The chiral condensate $\langle\bar\psi\psi\rangle$, known to be an order parameter~\cite{RevModPhys.53.43,PhysRev.122.345,PhysRevD.24.450,POLYAKOV1978477,Fang:2018vkp} for the chiral phase transition in the chiral limit ($m\sim 0$), has been studied in the \textit{mean field} approximation. That is to say, as introduced in~\cite{Gross:1974jv,WALECKA1974491}, one writes $\bar\psi\psi = \langle\bar\psi\psi\rangle + \sigma$ with a constant\footnote{Here ``constant'' means unchanged with the coordinates. $\langle\bar\psi\psi\rangle$, known as the {\it global} chiral condensate~\cite{Ohata:2020myj,Carabba:2021xmc}, is coordinate-independent and distinguished from the {\it local} chiral condensate $\langle\bar\psi(x)\psi(x)\rangle$ which depend on the coordinates. De facto, $\langle\bar\psi\psi\rangle$ is a function of temperature $T$ and chemical potentials $\mu$ and $\mu_5$.} term $\langle\bar\psi\psi\rangle$ and a small real scalar field $\sigma$, corresponding to fluctuations about the vacuum value, and then drop terms that are $\mathcal O (\sigma^2)$. Then the four-fermion contact interaction $g(\bar\psi\psi)^2$ can be written as
\bea
g(\bar\psi\psi)^2=&g\left(\langle\bar\psi\psi\rangle + \sigma\right)^2\\
= &2g\bar\psi\psi\langle\bar\psi\psi\rangle-g\langle\bar\psi\psi\rangle^2+ \mathcal O(\sigma^2).\nonumber
\eea
Furthermore, by defining the effective mass $M(\mu,\mu_5,T)$ by
\bea
M(\mu,\mu_5,T) = m - 2g\langle\bar\psi\psi\rangle(\mu,\mu_5,T),
\label{eq:relation}
\eea
the Lagrangian is given by $\mathcal L = \mathcal L_\mathrm{eff} + \mathcal O(\sigma^2)$, where
\bea
\label{Leff}
\mathcal L_\mathrm{eff} =& \bar\psi(i\slashed\partial - M + \mu\gamma_0+\mu_5\gamma_0\gamma_5)\psi - \frac{(M - m)^2}{4g}\nnu 
=& \mathcal L_\mathrm{Dirac} - \mathcal V\,,
\eea
where the potential $\mathcal V$ is related to the chiral condensate as well as the effective mass
\bea
\mathcal V = g\langle\bar\psi\psi\rangle^2= (M - m)^2/4g.
\eea
Following~\cite{Kapusta,Buballa:2003qv}, the Grand Canonical Potential $\Omega_\mathrm{Dirac}$ of $\mathcal L_\mathrm{Dirac}$ with mass $M$ is given in the following form
\bea
&\Omega_\mathrm{Dirac}(\mu, \mu_5, T; M)= -\frac{2}{\pi}\sum_{s=\pm 1}\int_0^\infty \bigg[T\ln(1 + e^{-\beta(\omega_{k,s} + \mu)}) \nnu
&+ T\ln(1 + e^{-\beta(\omega_{k,s} - \mu)})+\omega_{k,s}\bigg]dk\,,
\eea
where the energy spectrum of the free fermions $\omega_{k,s} = \sqrt{(k+s\mu_5)^2 + M^2}$ with $s=\pm 1$. Then, by adding the potential $\mathcal V = (M - m)^2/4g$, we obtain the grand canonical potential for the NJL model as follows
\bea
\label{GC}
&\Omega(\mu, \mu_5, T; M)= \mathcal{V}-\frac{2}{\pi}\sum_s\int_0^\infty \bigg[T\ln(1 + e^{-\beta(\omega_{k,s} + \mu)})\nnu
& + T\ln(1 + e^{-\beta(\omega_{k,s} - \mu)})+\omega_{k,s}\bigg]dk\,.
\eea

Due to the divergent behavior of this quantity, one has to regularize it. In this work, for comparison with numerical results with the lattice spacing $a$, the natural momentum cutoff is $\Lambda = \pi/a$. With this hard momentum cutoff imposed for the integral shown in Eq.~\eqref{GC}, one is able to determine the effective mass $M$ at fixed values of $\mu,~\mu_5$ and $T$ numerically by minimizing $\Omega(\mu, \mu_5, T; M)$ in regard to $M$, namely solving the gap equation,
\bea
\frac{\partial \Omega(\mu, \mu_5, T; M)}{\partial M} = 0\,.
\eea
Then the chiral condensate is given by $\langle\bar\psi\psi\rangle = (m - M)/2g$ following Eq.~\eqref{eq:relation}. 

As an example, in Fig.~\ref{fig:3dplot1}, we present the effective mass $M$ plot as a function of temperature $T$ and chiral chemical potential $\mu_5$ with bare mass $m=100$ MeV, coupling constant $g=5$, lattice spacing $a=1$ MeV$^{-1}$ and chemical potential $\mu=50$ MeV. As expected, at high $\mu_5$ or $T$, one has $M \rightarrow m$, corresponding to a restoration of chiral symmetry. Conversely, at low $(\mu_5^2 + T^2)$, one finds a dynamically generated mass of around $\Delta m \equiv M-m \sim 20$ MeV. Therefore, a free field theory is expected at asymptotically high temperatures/chemical potentials~\cite{Gross:1974jv}.

\subsection{The chirality charge density}\label{sec:n5theory}
In a magnetic field, under the imbalance of right/left-handed chirality, a finite induced current is produced along the magnetic field. Specifically, when the number of right-handed quarks, $N_R$ is unequal to that of the left-handed quarks $N_L$, positive charge is separated from negative charge along the magnetic field, which is the so-called ``Chiral Magnetic Effect''~\cite{Kharzeev:2007tn,Kharzeev:2007jp,Fukushima:2008xe}. The axial anomaly and topological objects in QCD are the fundamental physics of the CME. Unbalanced left- and right-handed quarks can produce observable effects that can be used to investigate topological $\mathcal{P}$- and $\mathcal{CP}$-odd excitations~\cite{Witten:1979vv,Veneziano:1979ec,Schafer:1996wv,Vicari:2008jw}.

In Eq.~\eqref{eq:Lagrangian}, we have introduced an additional term with the chiral chemical potential $\mu_5$ coupled with the chirality charge density operator $n_5=\bar{\psi}\gamma_0\gamma_5\psi$, which has been known as a distinctive characteristic in hot and dense QCD matter and is not conserved as a consequence of the chiral anomaly.  In the presence of a chiral chemical potential $\mu_5$, a manifestation of the chiral imbalance, the non-vanishing finite chirality charge density $n_5$ is required for the CME effect and recognized as a unique property of hot and dense QCD matter. Despite the fact that the chiral chemical potential $\mu_5$ is introduced for studying topological charge fluctuations, it is considered as a time-independent quantity that represents the chiral imbalance. And the chirality charge density $n_5=\langle\bar{\psi}\gamma_0\gamma_5\psi\rangle$ is a constant in the coordinates like the chiral condensate $\langle\bar{\psi}\psi\rangle$~\footnote{Here the coordinate-independent chirality charge density $n_5=\langle\bar{\psi}\gamma_0\gamma_5\psi\rangle$ is also a {\it global} quantity, like the {\it global} chiral condensate defined in~\cite{Ohata:2020myj,Carabba:2021xmc} and it will depend on the temperature $T$ and the chemical potentials $\mu$ and $\mu_5$.}.

When describing the induced electric current density as a function of the chirality density, the relationship between $n_5$ and $\mu_5$ is important and the chirality charge density $n_5$ can be calculated by~\cite{Fukushima:2010fe}
\bea
n_5=&-\frac{\partial\Omega(\mu, \mu_5, T; M)}{\partial\mu_5}\,,\label{eq:mu5_calc}
\eea
where the grand potential $\Omega(\mu, \mu_5, T; M)$ is given in Eq.~\eqref{GC}. In the next section, we will present the analytical calculations of $n_5$ in comparison with \texttt{QITE} and exact diagonalization results.

\section{Results}\label{sec:results}
In this subsection, we study the finite temperature behaviors of the chiral condensate $\langle\bar{\psi}\psi\rangle$ and chirality charge density $n_5=\langle\bar{\psi}\gamma_0\gamma_5\psi\rangle$ in the $(1+1)$-dimensional NJL model given in Eq.~\eqref{eq:Hamiltonia_all}. As emphasized in previous sections, we apply a quantum algorithm to simulate the thermal behaviours of physical observables. To demonstrate the reliability of our simulations, we will provide all the result from three different approaches:
\begin{enumerate}
  \item \texttt{QITE} simulations of the thermal observable;
  \item Exact diagonalizations from the discretization of the NJL Hamiltonian;
  \item Analytical calculations given by solving the gap equation numerically.
\end{enumerate}
For the consistency among the three procedures, we have chosen the same bare mass $m=100$ MeV and lattice spacing $a=1$ MeV$^{-1}$. The coupling constant $g$ at $g=1$ and $g=5$ are applied for testing the effects of the four-fermion interaction term in the Lagrangian.

To apply the quantum circuits, many quantum simulation packages have been developed and give similar results for quantum simulations. These quantum simulators, such as \texttt{PYQUILL}~\cite{https://doi.org/10.48550/arxiv.1608.03355} (Rigetti), \texttt{TEQUILA}~\cite{Kottmann_2021}, \texttt{Q\#}~\cite{qsharp} (Microsoft), \texttt{QISKIT}~\cite{gadi_aleksandrowicz_2019_2562111} (IBM), \texttt{QFORTE}~\cite{stair2021qforte}, \texttt{XACC}~\cite{McCaskey_2020}, \texttt{FQE}~\cite{Rubin:2021znj} \texttt{CIRQ}~\cite{cirq_developers_2021_5182845}, are \texttt{PYTHON} software libraries and the outputs are the expected outputs of an ideal quantum computer. One can find a list of general quantum simulation packages in~\cite{Bharti:2021zez} and some implementations of the quantum algorithms are listed in~\cite{Anand:2021xbq}. To execute the \texttt{QITE} algorithm, we construct a quantum circuit using the open-source software package \texttt{QFORTE}~\cite{stair2021qforte}, where many useful quantum algorithms have been implemented.
\begin{figure}[htp]
\centering
\includegraphics[width=0.45\textwidth,trim={2cm 2.5cm 2cm 0},clip]{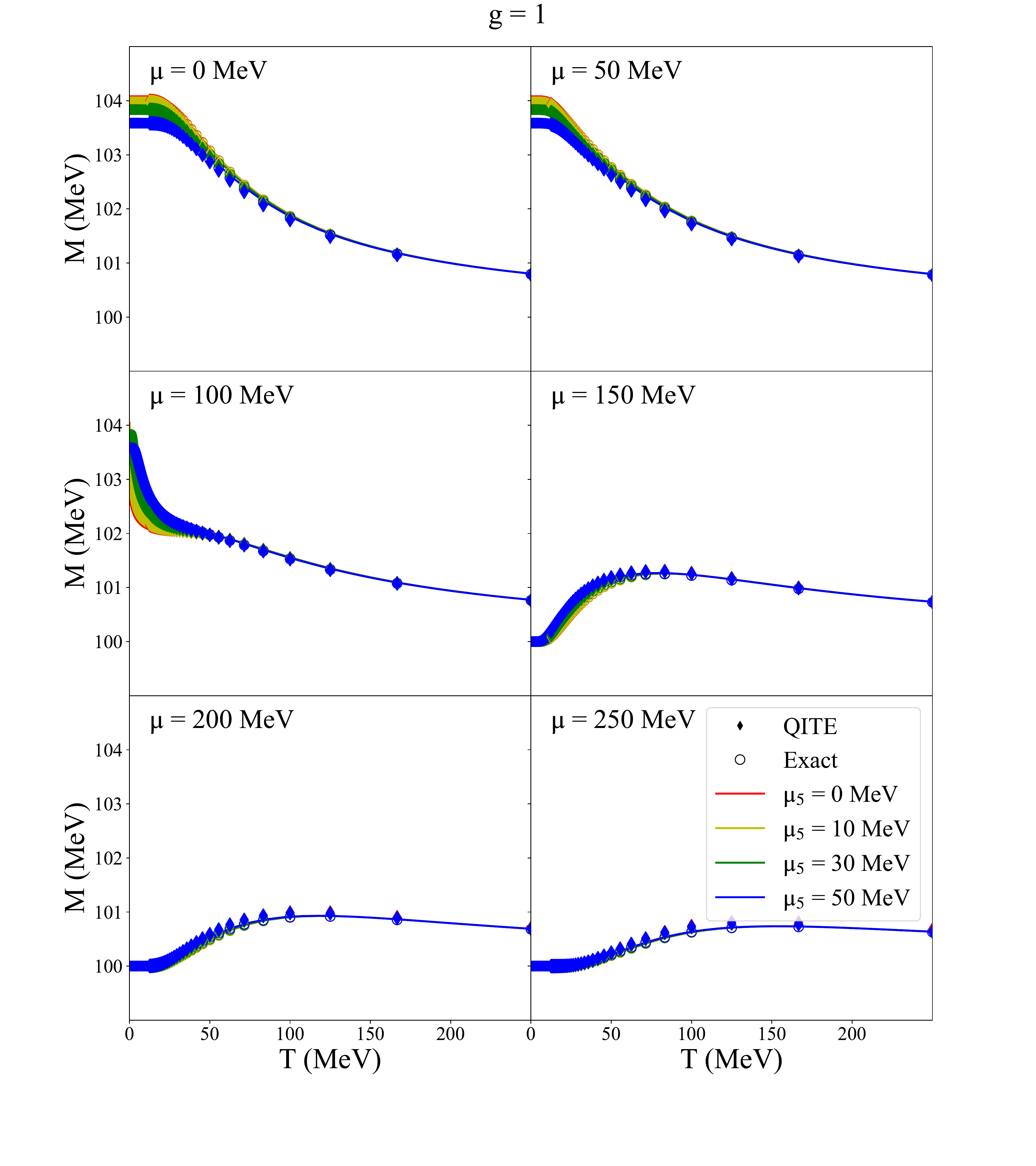}
\includegraphics[width=0.45\textwidth,trim={2cm 3cm 2cm 0},clip]{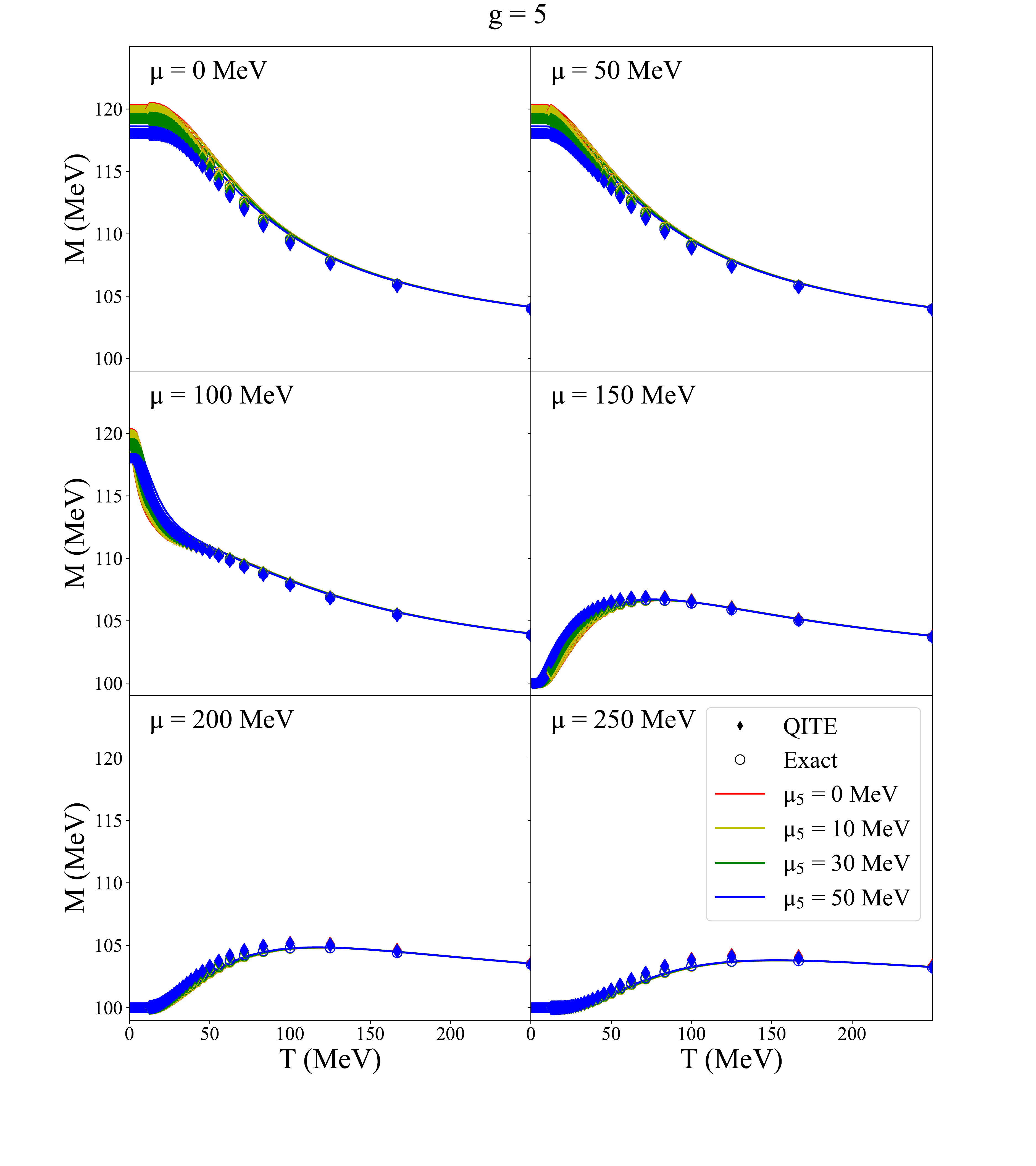}
\caption{Chiral Condensate $\langle
\bar\psi\psi\rangle$ as a function of temperature $T$ MeV at coupling constant $g=1$ (top panel) and $g=5$ (bottom panel) with a fixed chemical potential $\mu$ in each panel. Filled diamond points are given by \texttt{QITE} algorithm, hollow circle points are from exact diagonalization and solid curves are calculated analytically.}\label{fig:mu_g1}
\end{figure}

\begin{figure}
\centering
\includegraphics[width=0.45\textwidth,trim={2cm 1.5cm 2cm 0},clip]{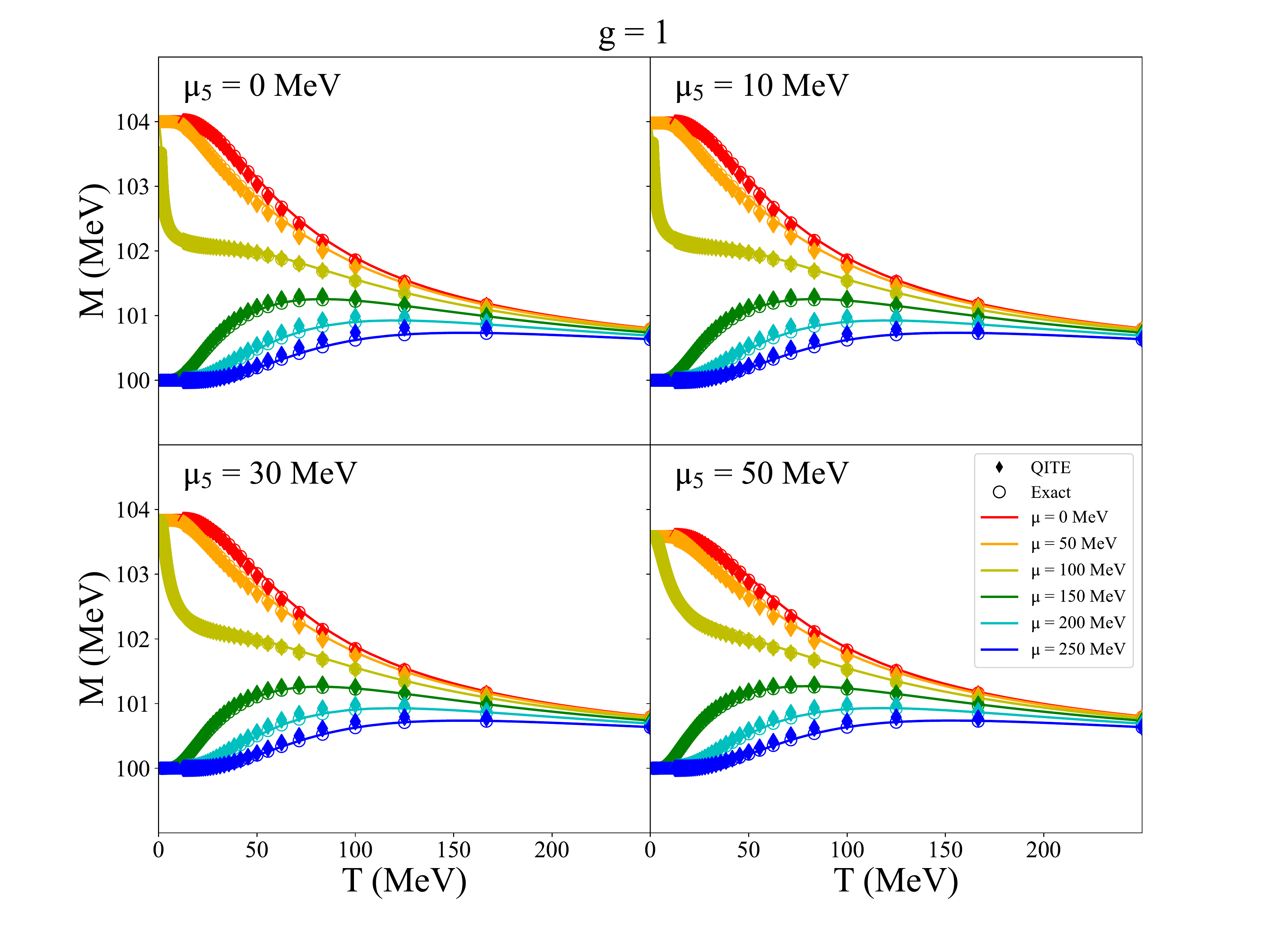}
\includegraphics[width=0.45\textwidth,trim={2cm 1.5cm 2cm 0},clip]{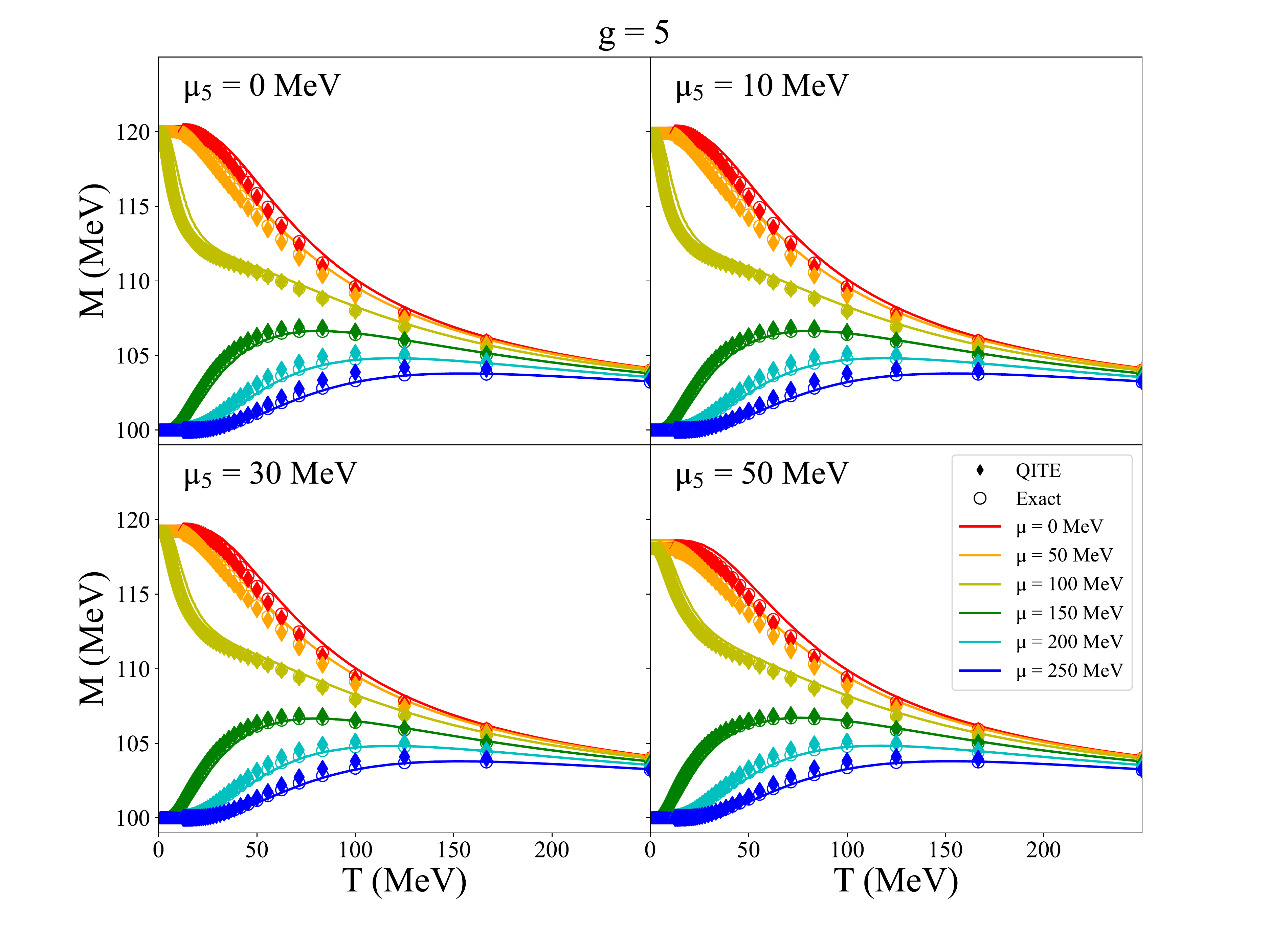}
\caption{Chiral Condensate $\langle
\bar\psi\psi\rangle$ as a function of temperature $T$ MeV at coupling constant $g=1$ (top panel) and $g=5$ (bottom panel) at a fixed chiral chemical potential $\mu_5$ in each panel.}\label{fig:mu5_g1}
\end{figure}

\begin{figure}
\centering
\includegraphics[width=0.45\textwidth,trim={2cm 2.5cm 2cm 0},clip]{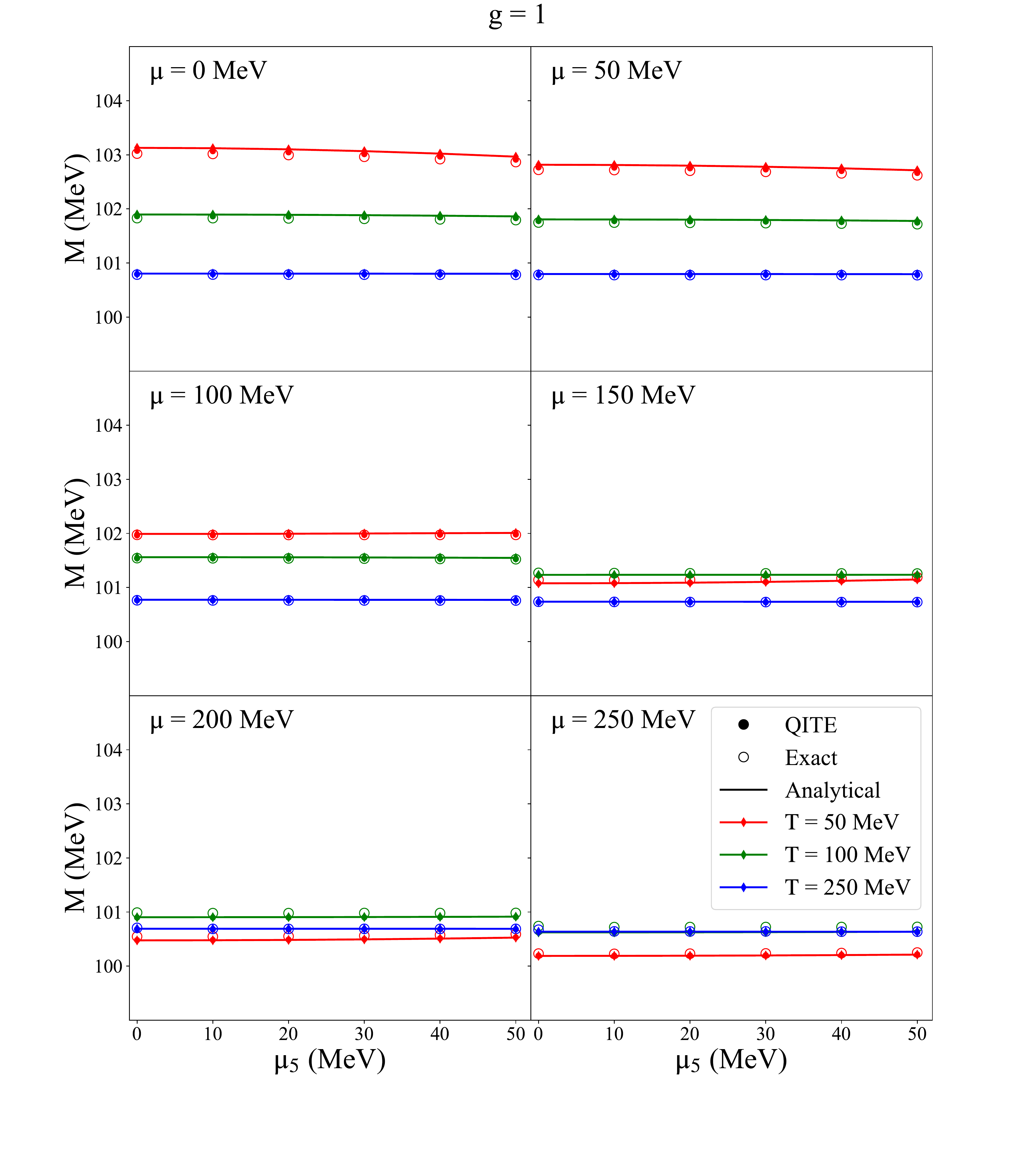}
\includegraphics[width=0.45\textwidth,trim={2cm 3cm 2cm 0},clip]{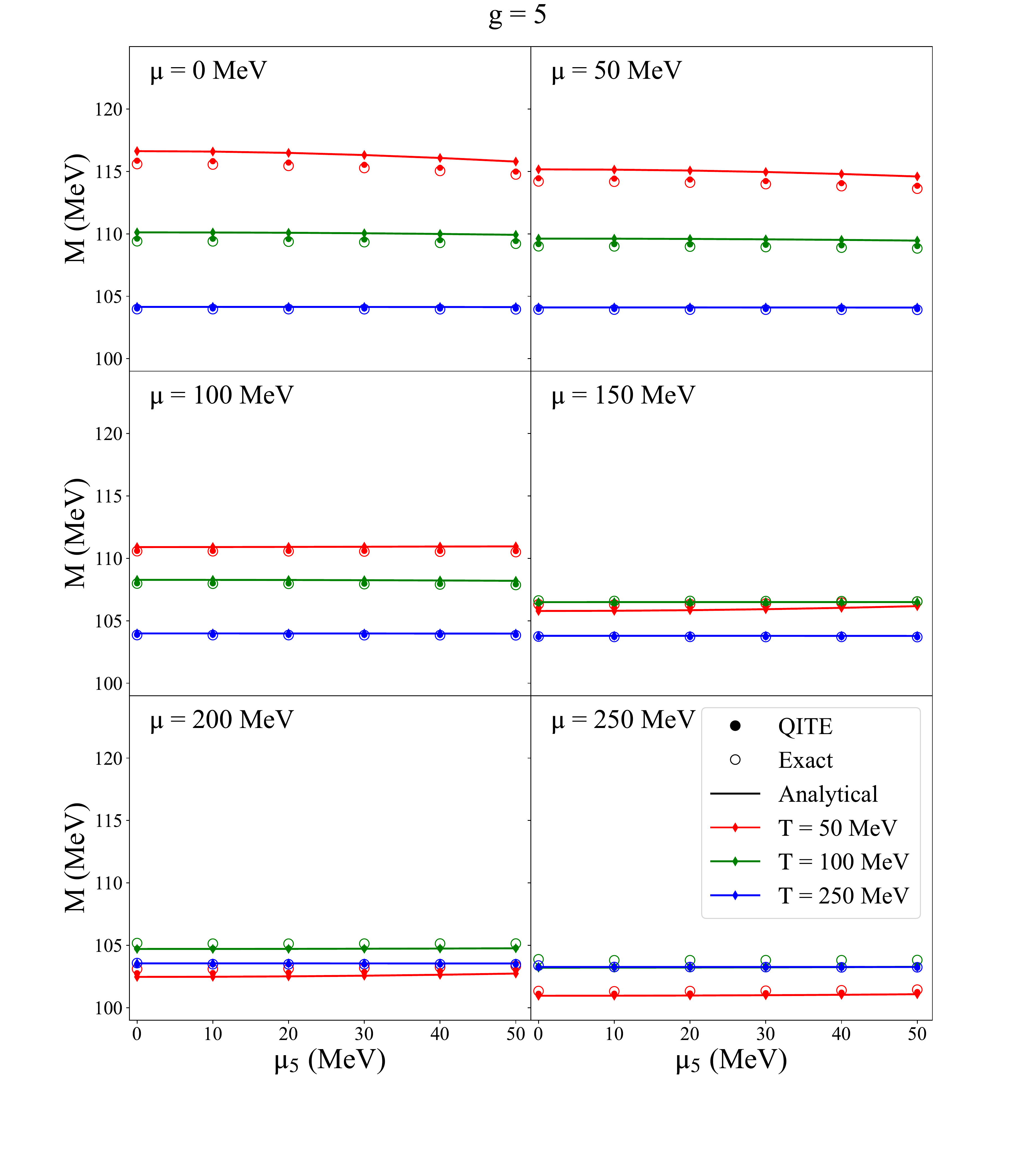}
\caption{Chiral Condensate $\langle
\bar\psi\psi\rangle$ as a function of $\mu_5$ at coupling constant $g=1$ (top panel) and $g=5$ (bottom panel).}\label{fig:Tmu5_g1}
\end{figure}

\begin{figure}[htp]
\centering
\includegraphics[width=0.43\textwidth,trim={2cm 2.5cm 2cm 0},clip]{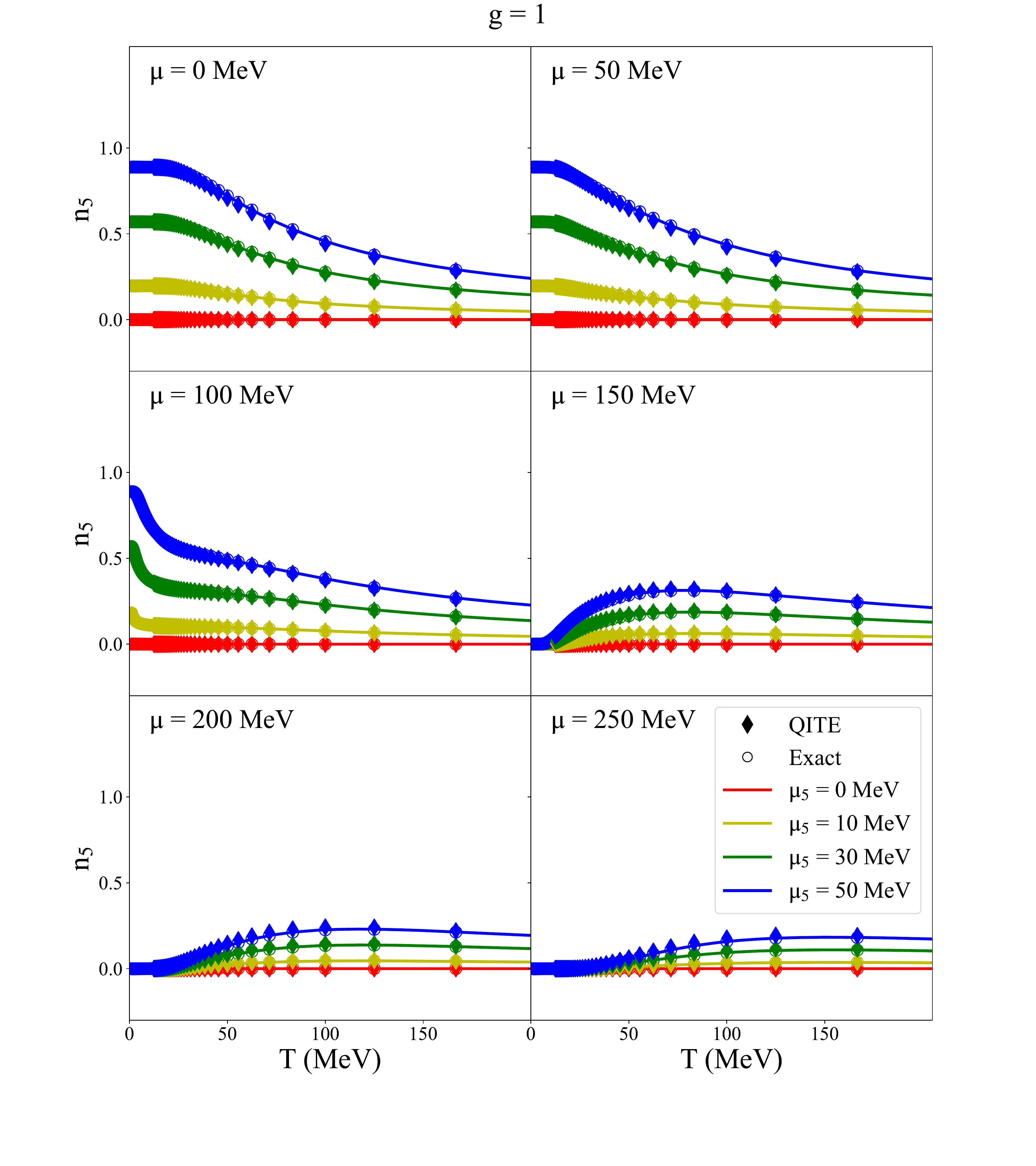}
\includegraphics[width=0.43\textwidth,trim={2cm 3cm 2cm 0},clip]{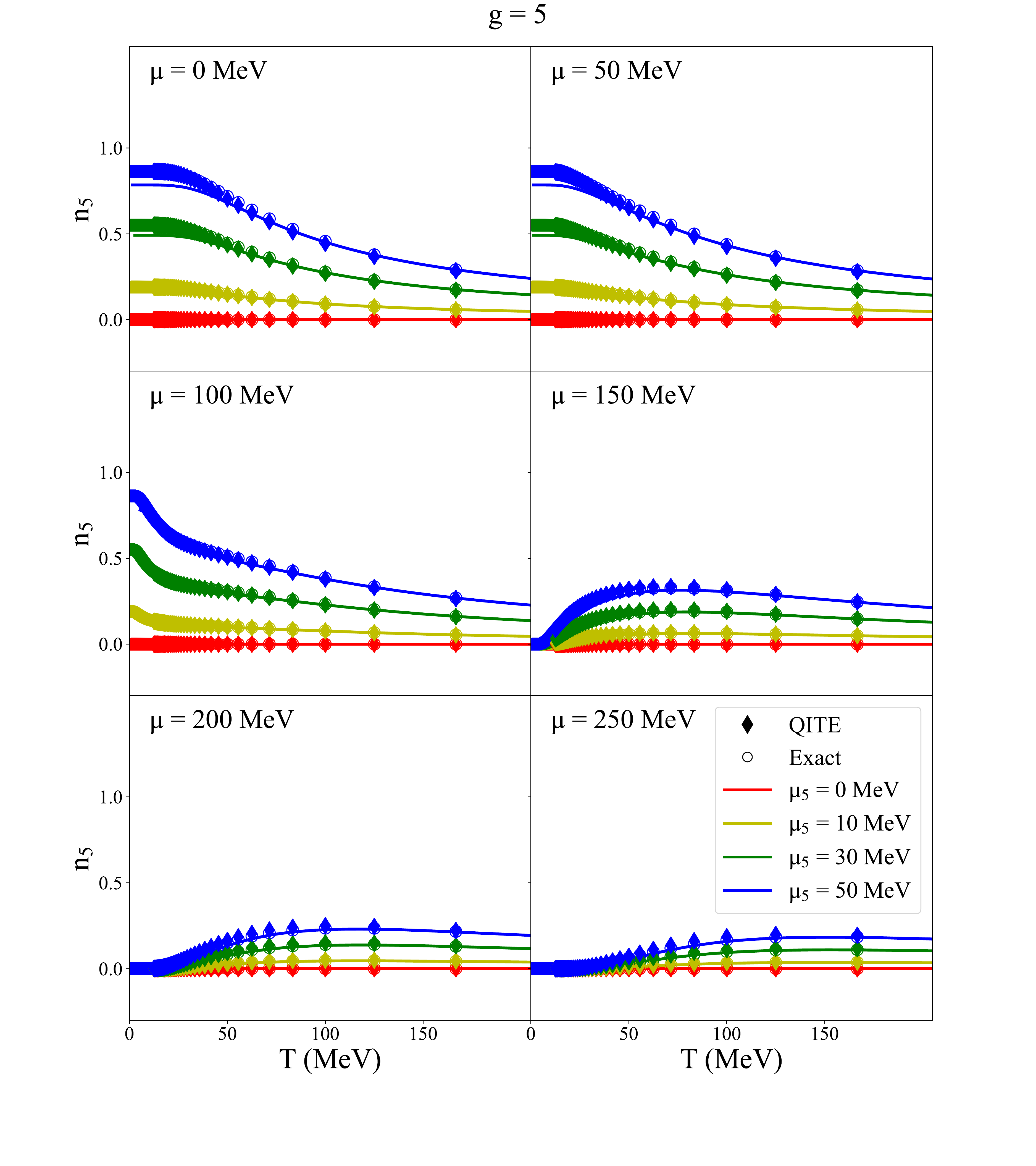}
\caption{Chirality charge density $n_5$ as a function of temperature $T$ at $g=1$ (top panel) and $g=5$ (bottom panel) at a fixed chemical potential $\mu = 0, 50, 100, \cdots, 250$ MeV in each panel.}\label{fig:n5_1}
\end{figure}

\begin{figure}
\centering
\includegraphics[width=0.43\textwidth,trim={2cm 1.5cm 2cm 0cm},clip]{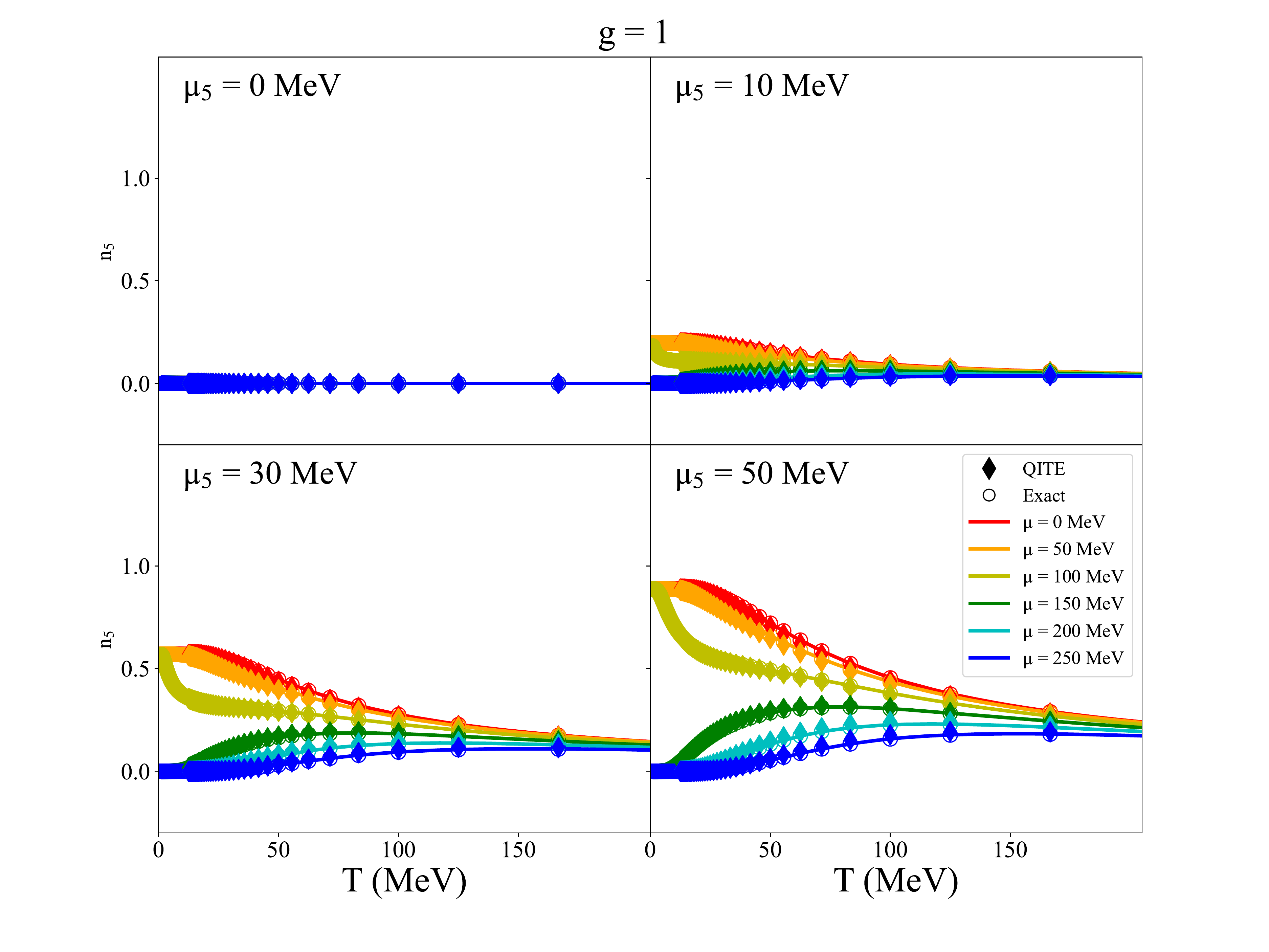}
\includegraphics[width=0.43\textwidth,trim={2cm 1.5cm 2cm 0cm},clip]{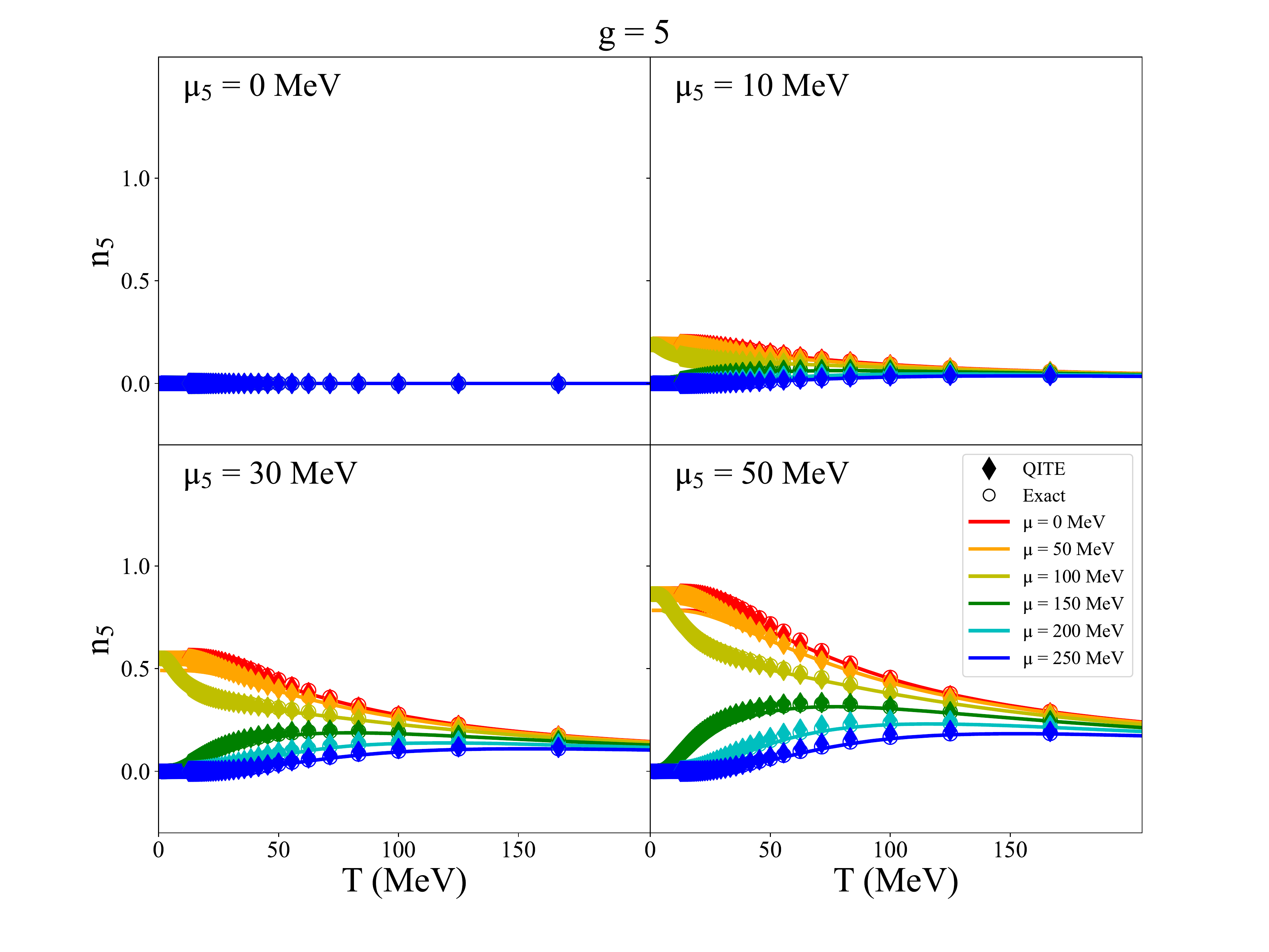}
\caption{Chirality charge density $n_5$ as a function of temperature at $g=1$ (top panel) and $g=5$ (bottom panel) with a fixed chiral chemical potential $\mu_5$ in each panel.}\label{fig:n5_3}
\end{figure}

\subsection{Chiral condensate}\label{sec:pheno1}
To study the effects of the chiral imbalance on the the chiral condensate at finite temperatures and chemical potentials, we present plots of the effective mass $M$ at various chemical potentials $\mu$, $\mu_5$ and temperatures $T$ in Figs.~\ref{fig:mu_g1},~\ref{fig:mu5_g1} and~\ref{fig:Tmu5_g1}, respectively. In Fig.~\ref{fig:mu_g1}, we compare the temperature dependence of the effective mass $M$ at $g=1$ and $g=5$ as a function of temperature $T$ at chiral chemical potential $\mu_5=0,\ 10,\ 30$ and $50$ MeV with a fixed baryochemical potentials $\mu\in\{0,\ 50, \cdots, 250\}$ MeV in each panel, where filled diamond points are given by the \texttt{QITE} algorithm, hollow circle points are from exact diagonalization and solid curve results are calculated analytically. Between $\mu=100$ MeV and $\mu=150$ MeV panels, the pattern of the effective mass changed from decreasing ($\mu\leq 100$ MeV) to increasing with temperature ($\mu\geq 150$ MeV). Also for $\mu\leq 100$ MeV panels, the effective mass $M$ at smaller $\mu_5$ is larger, while for $\mu\geq 150$ MeV panels, the effective mass $M$ at smaller $\mu_5$ is smaller.

In Fig.~\ref{fig:mu5_g1}, we fix the value of chiral chemical potentials $\mu_5$ in each panel and plot the $M\mathrm{-}T$ curves at various baryochemical potentials $\mu=0,\ 50, \cdots,250$ MeV at $g=1$ and $g=5$. A non-trivial phase transition at $100<\mu<150$ MeV is observed in all the panels and this has also been pointed out when interpreting Fig.~\ref{fig:mu5_g1}. The patterns look similar at various $\mu_5$ and we find that at smaller $\mu_5$, the effective mass $M$ changes more rapidly as a function of $T$. 

To better illustrate the effects of $\mu_5$, we present the effective mass $M$ as a function of chiral chemical potentials $\mu_5$ and show the results at several temperatures and baryochemical potentials in Fig.~\ref{fig:Tmu5_g1} for $g=1$ and $g=5$. The effective mass changes slightly as the chiral chemical potential increases, indicating that the chiral chemical potential plays a lesser role in chiral symmetry breaking than the temperature and chemical potential, at least in this model. At $\mu\leq 100$ MeV, $M$ is lower for higher temperature, which is expected by the asymptotic freedom. While as $\mu$ increases, the effective mass $M$ becomes smaller at lower temperatures.

\subsection{Chirality charge density}\label{sec:pheno2}
In this subsection, we present the results for the chirality charge density $n_5$ using the \texttt{QITE} algorithm in comparison with analytical calculations as introduced in Sec.~\ref{sec:n5theory} and exact diagonalization.

In Fig.~\ref{fig:n5_1}, we plot the chirality charge density as a function of temperature $T$ at coupling constant $g=1$ and $g=5$. We use different colors to distinguish various chiral chemical potentials $\mu_5=0,\ 10,\ 30$ and $50$ MeV and in each panel, the chemical potential $\mu$ is fixed at $0,\ 50,\ 100, \cdots $ or $250$ MeV. At $\mu_5=0$ MeV, one obtains $n_5=0$ since there is no other mechanism for generating a nonzero $\mathcal{N}_5=\bar{\psi}\gamma_0\gamma_5\psi=\psi^\dagger_R\psi_R-\psi^\dagger_L\psi_L$. As a result, only when $\mu_5\ne 0$, there exists non-zero $n_5=\langle\mathcal{N}_5\rangle$. 

Similar to what was observed in the effective mass plots in Fig.~\ref{fig:mu_g1}, a non-trivial phase transition is observed between $\mu=100$ MeV and $\mu=150$ MeV. For $\mu\leq 100$ MeV, the chirality charge density decreases as temperature rises, while for $\mu\geq 150$ MeV, the chirality charge density first increases then drops and converges to $0$ with the increasing of temperature. 

In Fig.~\ref{fig:n5_3}, we plot the chirality charge density $n_5$ as a function of temperature $T$ at coupling constants $g=1$ and $g=5$. We use different colors to distinguish various chemical potentials $\mu\in\{0,\ 50, \cdots, 250\}$ MeV and in each panel, the chemical potential $\mu_5$ is fixed at $0,\ 10,\ 30$ or $50$ MeV. At $\mu_5 = 0$ MeV, the chirality charge density is $0$ at all temperatures and chemical potentials, as expected from the previous figure. At non-zero $\mu_5$ values, the phase transition between $\mu = 100$ MeV and $\mu = 150$ MeV can also be observed: for $\mu \leq 100$ MeV, the chirality charge density begins at some non-zero value at $T = 0$ MeV and decreases with temperature, while for curves $\mu \geq 150$ MeV, the chirality charge density begins at $0$ at $T=0$ MeV and first increases before decreasing and converging to 0 with increasing temperature. At greater values of $\mu_5$, curves with $\mu \leq 100$ MeV begin at higher chirality charge densities.

\begin{figure}
\centering
\includegraphics[width=0.43\textwidth,trim={2cm 2.5cm 2cm 0},clip]{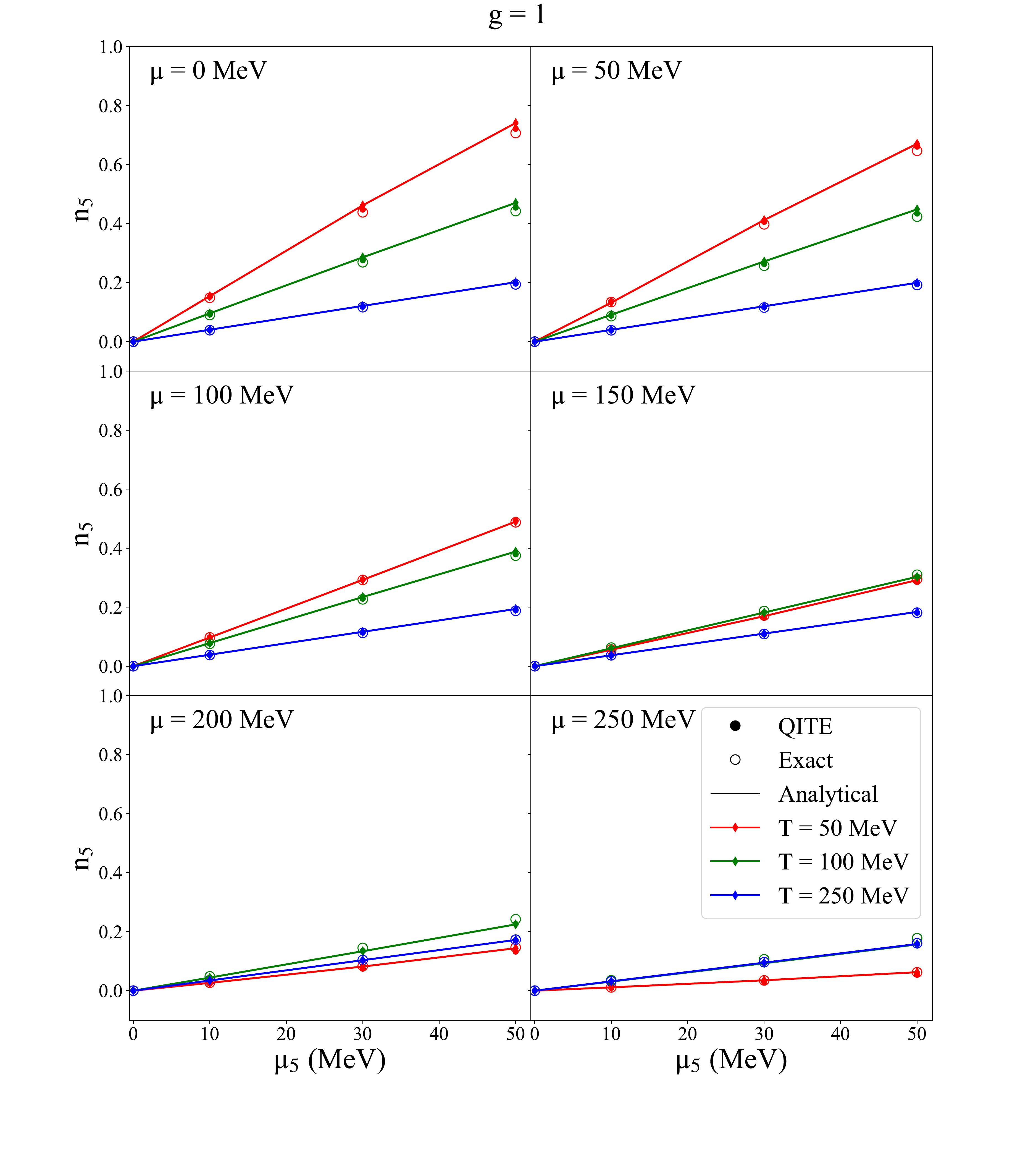}
\includegraphics[width=0.43\textwidth,trim={2cm 2.5cm 2cm 0},clip]{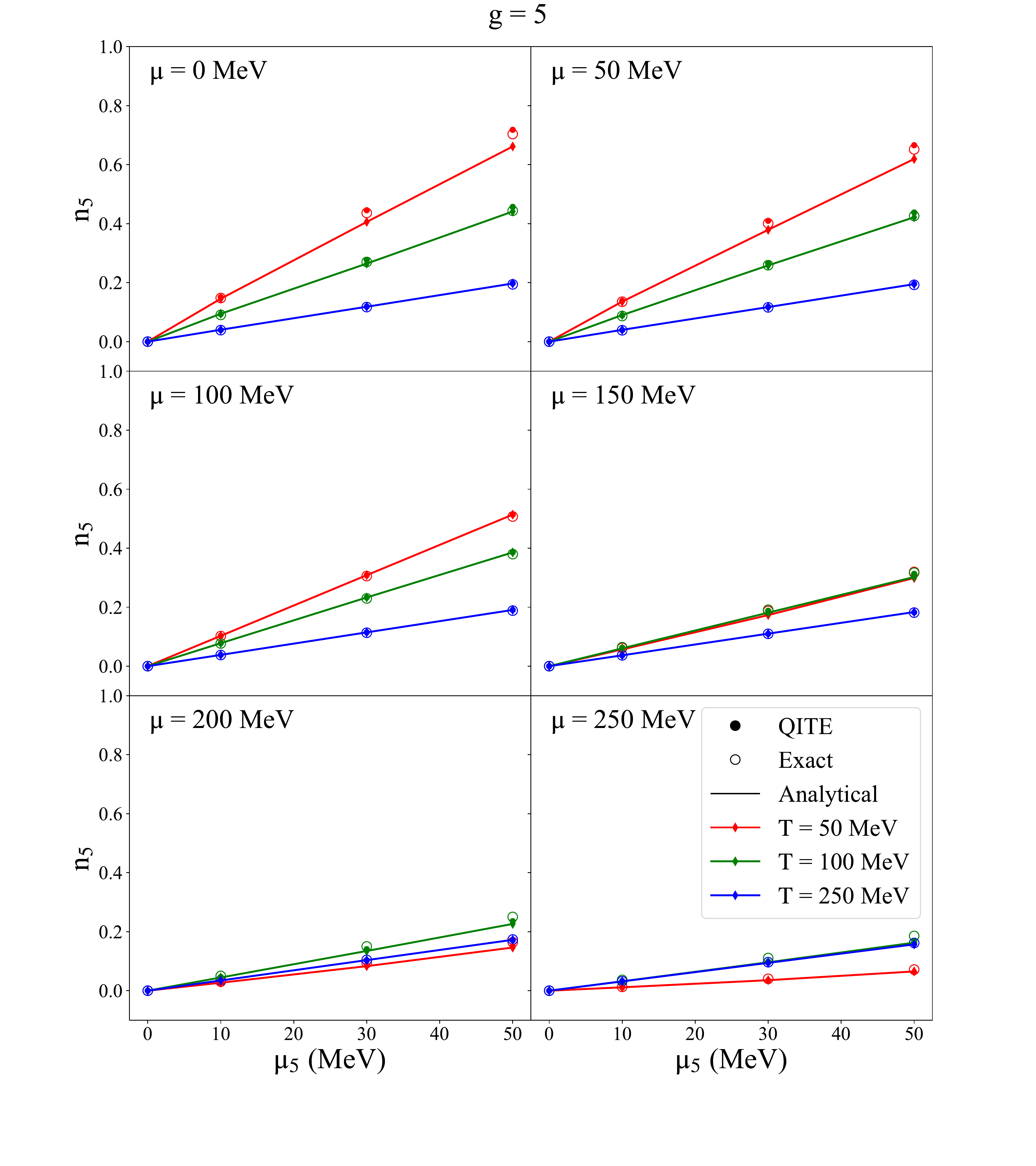}
\caption{Chirality charge density $n_5$ as a function of chiral chemical potential $\mu_5$ at $g=1$ (top panel) and $g=5$ (bottom panel) with a fixed temperature $T$ in each panel.}\label{fig:n5_5}
\end{figure}

In Fig.~\ref{fig:n5_5}, we plot the chirality charge density $n_5$ as a function of chiral chemical potential $\mu_5$ at coupling constants $g=1$ and $g=5$. Colors are used to distinguish between different various temperatures $T = 50,\ 100,\ $ or $200$ MeV, and the chemical potential is fixed at $\mu = 0,\ 50,\cdots, 250$ MeV in each panel. At all chemical potentials $\mu$ and temperatures $T$, we see that (approximately) $n_5 \propto \mu_5$: the chirality charge density $n_5$ begins at $0$ MeV at $\mu_5 = 0$ MeV, and increases linearly with $\mu_5$. The chiral chemical potential $\mu_5$, at least in this model, is thus a direct measure of chiral imbalance in the plasma. Higher chemical potentials $\mu$ result in a slower increase of $n_5$ with $\mu_5$. Once again, a phase transition can be observed between $\mu \leq 100$ MeV and $\mu \geq 150$ MeV. For $\mu \leq 100$ MeV, at each $\mu_5$, $n_5$ decreases with increasing temperature. For $\mu \geq 150$ MeV, $n_5$ first increases with temperature from $T = 50$ MeV to $T = 100$ MeV, but then decreases from $T = 100$ MeV to $T = 200$ MeV.

\section{Conclusion}\label{sec:conclusion}
In conclusion, using the \texttt{QITE} algorithm, we performed a quantum simulation for the chiral phase transition of the 1+1 dimensional NJL model at finite temperature, baryochemical potential and chiral chemical potential. Specifically, we use a 4-qubit quantum circuit to simulate the NJL Hamiltonian and find that the results among the digital quantum simulation, exact diagonalization, and analytical analysis are all consistent, implying that quantum computing will be a promising tool in simulating finite-temperature behaviors in the future for QCD.

The quantum simulations' efficacy has been shown to be insensitive to the chemical potentials $\mu$ and $\mu_5$ (and possibly other external parameters), which opens up new possibilities for studying finite density effects in QCD and other field theories. Owing to technological constraints (nonperturbative dynamics, Monte Carlo failure due to the sign problem, etc.) restricting the use of typical computing methods, this element of quark physics remains substantially unexplored in comparison to other fields. With the development of scalable quantum computing technologies on the horizon, Lattice QCD calculations on quantum computers are becoming not only conceivable, but also practical. 

This research, together with prior research, shows that NISQ quantum computers may produce consistent and correct answers to physical issues that cannot be solved efficiently or effectively using classical computing algorithms, reflecting bright prospects for future applications of quantum computing to non-perturbative QCD in the NISQ era and beyond.

\section*{Acknowledgements}
We thank Henry Ma for discussions and collaborations at early stages of this work. This work is supported by the National Science Foundation under grant No.~PHY-1945471.

\bibliography{refs}

\begin{thebibliography}{187}%
\makeatletter
\providecommand \@ifxundefined [1]{%
 \@ifx{#1\undefined}
}%
\providecommand \@ifnum [1]{%
 \ifnum #1\expandafter \@firstoftwo
 \else \expandafter \@secondoftwo
 \fi
}%
\providecommand \@ifx [1]{%
 \ifx #1\expandafter \@firstoftwo
 \else \expandafter \@secondoftwo
 \fi
}%
\providecommand \natexlab [1]{#1}%
\providecommand \enquote  [1]{``#1''}%
\providecommand \bibnamefont  [1]{#1}%
\providecommand \bibfnamefont [1]{#1}%
\providecommand \citenamefont [1]{#1}%
\providecommand \href@noop [0]{\@secondoftwo}%
\providecommand \href [0]{\begingroup \@sanitize@url \@href}%
\providecommand \@href[1]{\@@startlink{#1}\@@href}%
\providecommand \@@href[1]{\endgroup#1\@@endlink}%
\providecommand \@sanitize@url [0]{\catcode `\\12\catcode `\$12\catcode
  `\&12\catcode `\#12\catcode `\^12\catcode `\_12\catcode `\%12\relax}%
\providecommand \@@startlink[1]{}%
\providecommand \@@endlink[0]{}%
\providecommand \url  [0]{\begingroup\@sanitize@url \@url }%
\providecommand \@url [1]{\endgroup\@href {#1}{\urlprefix }}%
\providecommand \urlprefix  [0]{URL }%
\providecommand \Eprint [0]{\href }%
\providecommand \doibase [0]{http://dx.doi.org/}%
\providecommand \selectlanguage [0]{\@gobble}%
\providecommand \bibinfo  [0]{\@secondoftwo}%
\providecommand \bibfield  [0]{\@secondoftwo}%
\providecommand \translation [1]{[#1]}%
\providecommand \BibitemOpen [0]{}%
\providecommand \bibitemStop [0]{}%
\providecommand \bibitemNoStop [0]{.\EOS\space}%
\providecommand \EOS [0]{\spacefactor3000\relax}%
\providecommand \BibitemShut  [1]{\csname bibitem#1\endcsname}%
\let\auto@bib@innerbib\@empty
\bibitem [{\citenamefont {Fukushima}\ \emph {et~al.}(2010)\citenamefont
  {Fukushima}, \citenamefont {Ruggieri},\ and\ \citenamefont
  {Gatto}}]{Fukushima:2010fe}%
  \BibitemOpen
  \bibfield  {author} {\bibinfo {author} {\bibfnamefont {K.}~\bibnamefont
  {Fukushima}}, \bibinfo {author} {\bibfnamefont {M.}~\bibnamefont {Ruggieri}},
  \ and\ \bibinfo {author} {\bibfnamefont {R.}~\bibnamefont {Gatto}},\ }\href
  {\doibase 10.1103/PhysRevD.81.114031} {\bibfield  {journal} {\bibinfo
  {journal} {Phys. Rev. D}\ }\textbf {\bibinfo {volume} {81}},\ \bibinfo
  {pages} {114031} (\bibinfo {year} {2010})},\ \Eprint
  {http://arxiv.org/abs/1003.0047} {arXiv:1003.0047 [hep-ph]} \BibitemShut
  {NoStop}%
\bibitem [{\citenamefont {Buballa}(2005)}]{Buballa:2003qv}%
  \BibitemOpen
  \bibfield  {author} {\bibinfo {author} {\bibfnamefont {M.}~\bibnamefont
  {Buballa}},\ }\href {\doibase 10.1016/j.physrep.2004.11.004} {\bibfield
  {journal} {\bibinfo  {journal} {Phys. Rept.}\ }\textbf {\bibinfo {volume}
  {407}},\ \bibinfo {pages} {205} (\bibinfo {year} {2005})},\ \Eprint
  {http://arxiv.org/abs/hep-ph/0402234} {arXiv:hep-ph/0402234} \BibitemShut
  {NoStop}%
\bibitem [{\citenamefont {McLerran}\ and\ \citenamefont
  {Pisarski}(2007)}]{McLerran:2007qj}%
  \BibitemOpen
  \bibfield  {author} {\bibinfo {author} {\bibfnamefont {L.}~\bibnamefont
  {McLerran}}\ and\ \bibinfo {author} {\bibfnamefont {R.~D.}\ \bibnamefont
  {Pisarski}},\ }\href {\doibase 10.1016/j.nuclphysa.2007.08.013} {\bibfield
  {journal} {\bibinfo  {journal} {Nucl. Phys. A}\ }\textbf {\bibinfo {volume}
  {796}},\ \bibinfo {pages} {83} (\bibinfo {year} {2007})},\ \Eprint
  {http://arxiv.org/abs/0706.2191} {arXiv:0706.2191 [hep-ph]} \BibitemShut
  {NoStop}%
\bibitem [{\citenamefont {McLerran}\ \emph {et~al.}(2009)\citenamefont
  {McLerran}, \citenamefont {Redlich},\ and\ \citenamefont
  {Sasaki}}]{McLerran:2008ua}%
  \BibitemOpen
  \bibfield  {author} {\bibinfo {author} {\bibfnamefont {L.}~\bibnamefont
  {McLerran}}, \bibinfo {author} {\bibfnamefont {K.}~\bibnamefont {Redlich}}, \
  and\ \bibinfo {author} {\bibfnamefont {C.}~\bibnamefont {Sasaki}},\ }\href
  {\doibase 10.1016/j.nuclphysa.2009.04.001} {\bibfield  {journal} {\bibinfo
  {journal} {Nucl. Phys. A}\ }\textbf {\bibinfo {volume} {824}},\ \bibinfo
  {pages} {86} (\bibinfo {year} {2009})},\ \Eprint
  {http://arxiv.org/abs/0812.3585} {arXiv:0812.3585 [hep-ph]} \BibitemShut
  {NoStop}%
\bibitem [{\citenamefont {Menezes}\ \emph
  {et~al.}(2009{\natexlab{a}})\citenamefont {Menezes}, \citenamefont {Pinto},
  \citenamefont {Avancini}, \citenamefont {Mart\'{\i}nez},\ and\ \citenamefont
  {Provid\^encia}}]{PhysRevC.79.035807}%
  \BibitemOpen
  \bibfield  {author} {\bibinfo {author} {\bibfnamefont {D.~P.}\ \bibnamefont
  {Menezes}}, \bibinfo {author} {\bibfnamefont {M.~B.}\ \bibnamefont {Pinto}},
  \bibinfo {author} {\bibfnamefont {S.~S.}\ \bibnamefont {Avancini}}, \bibinfo
  {author} {\bibfnamefont {A.~P.}\ \bibnamefont {Mart\'{\i}nez}}, \ and\
  \bibinfo {author} {\bibfnamefont {C.}~\bibnamefont {Provid\^encia}},\ }\href
  {\doibase 10.1103/PhysRevC.79.035807} {\bibfield  {journal} {\bibinfo
  {journal} {Phys. Rev. C}\ }\textbf {\bibinfo {volume} {79}},\ \bibinfo
  {pages} {035807} (\bibinfo {year} {2009}{\natexlab{a}})}\BibitemShut
  {NoStop}%
\bibitem [{\citenamefont {Orsaria}\ \emph {et~al.}(2014)\citenamefont
  {Orsaria}, \citenamefont {Rodrigues}, \citenamefont {Weber},\ and\
  \citenamefont {Contrera}}]{Orsaria:2013hna}%
  \BibitemOpen
  \bibfield  {author} {\bibinfo {author} {\bibfnamefont {M.}~\bibnamefont
  {Orsaria}}, \bibinfo {author} {\bibfnamefont {H.}~\bibnamefont {Rodrigues}},
  \bibinfo {author} {\bibfnamefont {F.}~\bibnamefont {Weber}}, \ and\ \bibinfo
  {author} {\bibfnamefont {G.~A.}\ \bibnamefont {Contrera}},\ }\href {\doibase
  10.1103/PhysRevC.89.015806} {\bibfield  {journal} {\bibinfo  {journal} {Phys.
  Rev. C}\ }\textbf {\bibinfo {volume} {89}},\ \bibinfo {pages} {015806}
  (\bibinfo {year} {2014})},\ \Eprint {http://arxiv.org/abs/1308.1657}
  {arXiv:1308.1657 [nucl-th]} \BibitemShut {NoStop}%
\bibitem [{\citenamefont {Ruggieri}\ and\ \citenamefont
  {Peng}(2016)}]{Ruggieri:2016lrn}%
  \BibitemOpen
  \bibfield  {author} {\bibinfo {author} {\bibfnamefont {M.}~\bibnamefont
  {Ruggieri}}\ and\ \bibinfo {author} {\bibfnamefont {G.~X.}\ \bibnamefont
  {Peng}},\ }\href {\doibase 10.1103/PhysRevD.93.094021} {\bibfield  {journal}
  {\bibinfo  {journal} {Phys. Rev. D}\ }\textbf {\bibinfo {volume} {93}},\
  \bibinfo {pages} {094021} (\bibinfo {year} {2016})},\ \Eprint
  {http://arxiv.org/abs/1602.08994} {arXiv:1602.08994 [hep-ph]} \BibitemShut
  {NoStop}%
\bibitem [{\citenamefont {Jim\'enez}(2021)}]{Jimenez:2021ngg}%
  \BibitemOpen
  \bibfield  {author} {\bibinfo {author} {\bibfnamefont {J.~C.}\ \bibnamefont
  {Jim\'enez}},\ }\emph {\bibinfo {title} {{Interacting quark matter effects on
  the structure of compact stars}}},\ \href@noop {} {\bibinfo {type} {Other
  thesis}} (\bibinfo {year} {2021}),\ \Eprint {http://arxiv.org/abs/2104.03551}
  {arXiv:2104.03551 [hep-ph]} \BibitemShut {NoStop}%
\bibitem [{\citenamefont {Gyulassy}\ and\ \citenamefont
  {McLerran}(2005)}]{Gyulassy:2004zy}%
  \BibitemOpen
  \bibfield  {author} {\bibinfo {author} {\bibfnamefont {M.}~\bibnamefont
  {Gyulassy}}\ and\ \bibinfo {author} {\bibfnamefont {L.}~\bibnamefont
  {McLerran}},\ }\href {\doibase 10.1016/j.nuclphysa.2004.10.034} {\bibfield
  {journal} {\bibinfo  {journal} {Nucl. Phys. A}\ }\textbf {\bibinfo {volume}
  {750}},\ \bibinfo {pages} {30} (\bibinfo {year} {2005})},\ \Eprint
  {http://arxiv.org/abs/nucl-th/0405013} {arXiv:nucl-th/0405013} \BibitemShut
  {NoStop}%
\bibitem [{\citenamefont {Soloveva}\ \emph {et~al.}(2021)\citenamefont
  {Soloveva}, \citenamefont {Fuseau}, \citenamefont {Aichelin},\ and\
  \citenamefont {Bratkovskaya}}]{Soloveva:2020hpr}%
  \BibitemOpen
  \bibfield  {author} {\bibinfo {author} {\bibfnamefont {O.}~\bibnamefont
  {Soloveva}}, \bibinfo {author} {\bibfnamefont {D.}~\bibnamefont {Fuseau}},
  \bibinfo {author} {\bibfnamefont {J.}~\bibnamefont {Aichelin}}, \ and\
  \bibinfo {author} {\bibfnamefont {E.}~\bibnamefont {Bratkovskaya}},\ }\href
  {\doibase 10.1103/PhysRevC.103.054901} {\bibfield  {journal} {\bibinfo
  {journal} {Phys. Rev. C}\ }\textbf {\bibinfo {volume} {103}},\ \bibinfo
  {pages} {054901} (\bibinfo {year} {2021})},\ \Eprint
  {http://arxiv.org/abs/2011.03505} {arXiv:2011.03505 [nucl-th]} \BibitemShut
  {NoStop}%
\bibitem [{\citenamefont {Fukushima}\ and\ \citenamefont
  {Hatsuda}(2011)}]{Fukushima:2010bq}%
  \BibitemOpen
  \bibfield  {author} {\bibinfo {author} {\bibfnamefont {K.}~\bibnamefont
  {Fukushima}}\ and\ \bibinfo {author} {\bibfnamefont {T.}~\bibnamefont
  {Hatsuda}},\ }\href {\doibase 10.1088/0034-4885/74/1/014001} {\bibfield
  {journal} {\bibinfo  {journal} {Rept. Prog. Phys.}\ }\textbf {\bibinfo
  {volume} {74}},\ \bibinfo {pages} {014001} (\bibinfo {year} {2011})},\
  \Eprint {http://arxiv.org/abs/1005.4814} {arXiv:1005.4814 [hep-ph]}
  \BibitemShut {NoStop}%
\bibitem [{\citenamefont {Choudhury}\ \emph {et~al.}(2022)\citenamefont
  {Choudhury} \emph {et~al.}}]{Choudhury:2021jwd}%
  \BibitemOpen
  \bibfield  {author} {\bibinfo {author} {\bibfnamefont {S.}~\bibnamefont
  {Choudhury}} \emph {et~al.},\ }\href {\doibase 10.1088/1674-1137/ac2a1f}
  {\bibfield  {journal} {\bibinfo  {journal} {Chin. Phys. C}\ }\textbf
  {\bibinfo {volume} {46}},\ \bibinfo {pages} {014101} (\bibinfo {year}
  {2022})},\ \Eprint {http://arxiv.org/abs/2105.06044} {arXiv:2105.06044
  [nucl-ex]} \BibitemShut {NoStop}%
\bibitem [{\citenamefont {Feng}\ \emph {et~al.}(2022)\citenamefont {Feng},
  \citenamefont {Zhao}, \citenamefont {Li}, \citenamefont {Xu},\ and\
  \citenamefont {Wang}}]{Feng:2021pgf}%
  \BibitemOpen
  \bibfield  {author} {\bibinfo {author} {\bibfnamefont {Y.}~\bibnamefont
  {Feng}}, \bibinfo {author} {\bibfnamefont {J.}~\bibnamefont {Zhao}}, \bibinfo
  {author} {\bibfnamefont {H.}~\bibnamefont {Li}}, \bibinfo {author}
  {\bibfnamefont {H.-j.}\ \bibnamefont {Xu}}, \ and\ \bibinfo {author}
  {\bibfnamefont {F.}~\bibnamefont {Wang}},\ }\href {\doibase
  10.1103/PhysRevC.105.024913} {\bibfield  {journal} {\bibinfo  {journal}
  {Phys. Rev. C}\ }\textbf {\bibinfo {volume} {105}},\ \bibinfo {pages}
  {024913} (\bibinfo {year} {2022})},\ \Eprint
  {http://arxiv.org/abs/2106.15595} {arXiv:2106.15595 [nucl-ex]} \BibitemShut
  {NoStop}%
\bibitem [{\citenamefont {Abdallah}\ \emph
  {et~al.}(2022{\natexlab{a}})\citenamefont {Abdallah} \emph
  {et~al.}}]{STAR:2021mii}%
  \BibitemOpen
  \bibfield  {author} {\bibinfo {author} {\bibfnamefont {M.}~\bibnamefont
  {Abdallah}} \emph {et~al.} (\bibinfo {collaboration} {STAR}),\ }\href
  {\doibase 10.1103/PhysRevC.105.014901} {\bibfield  {journal} {\bibinfo
  {journal} {Phys. Rev. C}\ }\textbf {\bibinfo {volume} {105}},\ \bibinfo
  {pages} {014901} (\bibinfo {year} {2022}{\natexlab{a}})},\ \Eprint
  {http://arxiv.org/abs/2109.00131} {arXiv:2109.00131 [nucl-ex]} \BibitemShut
  {NoStop}%
\bibitem [{\citenamefont {Milton}\ \emph {et~al.}(2021)\citenamefont {Milton},
  \citenamefont {Wang}, \citenamefont {Sergeeva}, \citenamefont {Shi},
  \citenamefont {Liao},\ and\ \citenamefont {Huang}}]{Milton:2021wku}%
  \BibitemOpen
  \bibfield  {author} {\bibinfo {author} {\bibfnamefont {R.}~\bibnamefont
  {Milton}}, \bibinfo {author} {\bibfnamefont {G.}~\bibnamefont {Wang}},
  \bibinfo {author} {\bibfnamefont {M.}~\bibnamefont {Sergeeva}}, \bibinfo
  {author} {\bibfnamefont {S.}~\bibnamefont {Shi}}, \bibinfo {author}
  {\bibfnamefont {J.}~\bibnamefont {Liao}}, \ and\ \bibinfo {author}
  {\bibfnamefont {H.~Z.}\ \bibnamefont {Huang}},\ }\href {\doibase
  10.1103/PhysRevC.104.064906} {\bibfield  {journal} {\bibinfo  {journal}
  {Phys. Rev. C}\ }\textbf {\bibinfo {volume} {104}},\ \bibinfo {pages}
  {064906} (\bibinfo {year} {2021})},\ \Eprint
  {http://arxiv.org/abs/2110.01435} {arXiv:2110.01435 [nucl-th]} \BibitemShut
  {NoStop}%
\bibitem [{\citenamefont {Kharzeev}\ \emph {et~al.}(2022)\citenamefont
  {Kharzeev}, \citenamefont {Liao},\ and\ \citenamefont
  {Shi}}]{Kharzeev:2022hqz}%
  \BibitemOpen
  \bibfield  {author} {\bibinfo {author} {\bibfnamefont {D.~E.}\ \bibnamefont
  {Kharzeev}}, \bibinfo {author} {\bibfnamefont {J.}~\bibnamefont {Liao}}, \
  and\ \bibinfo {author} {\bibfnamefont {S.}~\bibnamefont {Shi}},\ }\href@noop
  {} {\  (\bibinfo {year} {2022})},\ \Eprint {http://arxiv.org/abs/2205.00120}
  {arXiv:2205.00120 [nucl-th]} \BibitemShut {NoStop}%
\bibitem [{\citenamefont {Kharzeev}(2022)}]{Kharzeev:2022ydx}%
  \BibitemOpen
  \bibfield  {author} {\bibinfo {author} {\bibfnamefont {D.~E.}\ \bibnamefont
  {Kharzeev}}\ }(\bibinfo {year} {2022})\ \Eprint
  {http://arxiv.org/abs/2204.10903} {arXiv:2204.10903 [hep-ph]} \BibitemShut
  {NoStop}%
\bibitem [{\citenamefont {Adhikari}(2022)}]{Adhikari:2021jff}%
  \BibitemOpen
  \bibfield  {author} {\bibinfo {author} {\bibfnamefont {P.}~\bibnamefont
  {Adhikari}},\ }\href {\doibase 10.1016/j.nuclphysb.2021.115627} {\bibfield
  {journal} {\bibinfo  {journal} {Nucl. Phys. B}\ }\textbf {\bibinfo {volume}
  {974}},\ \bibinfo {pages} {115627} (\bibinfo {year} {2022})},\ \Eprint
  {http://arxiv.org/abs/2111.06196} {arXiv:2111.06196 [hep-ph]} \BibitemShut
  {NoStop}%
\bibitem [{\citenamefont {Tawfik}\ and\ \citenamefont
  {Diab}(2021)}]{Tawfik:2021eeb}%
  \BibitemOpen
  \bibfield  {author} {\bibinfo {author} {\bibfnamefont {A.~N.}\ \bibnamefont
  {Tawfik}}\ and\ \bibinfo {author} {\bibfnamefont {A.~M.}\ \bibnamefont
  {Diab}},\ }\href {\doibase 10.1140/epja/s10050-021-00501-z} {\bibfield
  {journal} {\bibinfo  {journal} {Eur. Phys. J. A}\ }\textbf {\bibinfo {volume}
  {57}},\ \bibinfo {pages} {200} (\bibinfo {year} {2021})},\ \Eprint
  {http://arxiv.org/abs/2106.04576} {arXiv:2106.04576 [hep-ph]} \BibitemShut
  {NoStop}%
\bibitem [{\citenamefont {Cao}(2021)}]{Cao:2021rwx}%
  \BibitemOpen
  \bibfield  {author} {\bibinfo {author} {\bibfnamefont {G.}~\bibnamefont
  {Cao}},\ }\href {\doibase 10.1140/epja/s10050-021-00570-0} {\bibfield
  {journal} {\bibinfo  {journal} {Eur. Phys. J. A}\ }\textbf {\bibinfo {volume}
  {57}},\ \bibinfo {pages} {264} (\bibinfo {year} {2021})},\ \Eprint
  {http://arxiv.org/abs/2103.00456} {arXiv:2103.00456 [hep-ph]} \BibitemShut
  {NoStop}%
\bibitem [{\citenamefont {Moreira}\ \emph {et~al.}(2021)\citenamefont
  {Moreira}, \citenamefont {Costa},\ and\ \citenamefont
  {Restrepo}}]{Moreira:2021ety}%
  \BibitemOpen
  \bibfield  {author} {\bibinfo {author} {\bibfnamefont {J.~a.}\ \bibnamefont
  {Moreira}}, \bibinfo {author} {\bibfnamefont {P.}~\bibnamefont {Costa}}, \
  and\ \bibinfo {author} {\bibfnamefont {T.~E.}\ \bibnamefont {Restrepo}},\
  }\href {\doibase 10.1140/epja/s10050-021-00440-9} {\bibfield  {journal}
  {\bibinfo  {journal} {Eur. Phys. J. A}\ }\textbf {\bibinfo {volume} {57}},\
  \bibinfo {pages} {123} (\bibinfo {year} {2021})},\ \Eprint
  {http://arxiv.org/abs/2101.12004} {arXiv:2101.12004 [hep-ph]} \BibitemShut
  {NoStop}%
\bibitem [{\citenamefont {Ding}\ \emph {et~al.}(2021)\citenamefont {Ding},
  \citenamefont {Li}, \citenamefont {Tomiya}, \citenamefont {Wang},\ and\
  \citenamefont {Zhang}}]{Ding:2020hxw}%
  \BibitemOpen
  \bibfield  {author} {\bibinfo {author} {\bibfnamefont {H.~T.}\ \bibnamefont
  {Ding}}, \bibinfo {author} {\bibfnamefont {S.~T.}\ \bibnamefont {Li}},
  \bibinfo {author} {\bibfnamefont {A.}~\bibnamefont {Tomiya}}, \bibinfo
  {author} {\bibfnamefont {X.~D.}\ \bibnamefont {Wang}}, \ and\ \bibinfo
  {author} {\bibfnamefont {Y.}~\bibnamefont {Zhang}},\ }\href {\doibase
  10.1103/PhysRevD.104.014505} {\bibfield  {journal} {\bibinfo  {journal}
  {Phys. Rev. D}\ }\textbf {\bibinfo {volume} {104}},\ \bibinfo {pages}
  {014505} (\bibinfo {year} {2021})},\ \Eprint
  {http://arxiv.org/abs/2008.00493} {arXiv:2008.00493 [hep-lat]} \BibitemShut
  {NoStop}%
\bibitem [{\citenamefont {Zhao}\ and\ \citenamefont
  {Wang}(2019)}]{Zhao:2019hta}%
  \BibitemOpen
  \bibfield  {author} {\bibinfo {author} {\bibfnamefont {J.}~\bibnamefont
  {Zhao}}\ and\ \bibinfo {author} {\bibfnamefont {F.}~\bibnamefont {Wang}},\
  }\href {\doibase 10.1016/j.ppnp.2019.05.001} {\bibfield  {journal} {\bibinfo
  {journal} {Prog. Part. Nucl. Phys.}\ }\textbf {\bibinfo {volume} {107}},\
  \bibinfo {pages} {200} (\bibinfo {year} {2019})},\ \Eprint
  {http://arxiv.org/abs/1906.11413} {arXiv:1906.11413 [nucl-ex]} \BibitemShut
  {NoStop}%
\bibitem [{\citenamefont {Tawfik}\ \emph {et~al.}(2019)\citenamefont {Tawfik},
  \citenamefont {Diab},\ and\ \citenamefont {Hussein}}]{Tawfik:2019rdd}%
  \BibitemOpen
  \bibfield  {author} {\bibinfo {author} {\bibfnamefont {A.~N.}\ \bibnamefont
  {Tawfik}}, \bibinfo {author} {\bibfnamefont {A.~M.}\ \bibnamefont {Diab}}, \
  and\ \bibinfo {author} {\bibfnamefont {M.~T.}\ \bibnamefont {Hussein}},\
  }\href {\doibase 10.1088/1674-1137/43/3/034103} {\bibfield  {journal}
  {\bibinfo  {journal} {Chin. Phys. C}\ }\textbf {\bibinfo {volume} {43}},\
  \bibinfo {pages} {034103} (\bibinfo {year} {2019})},\ \Eprint
  {http://arxiv.org/abs/1901.03293} {arXiv:1901.03293 [hep-ph]} \BibitemShut
  {NoStop}%
\bibitem [{\citenamefont {Gusynin}\ \emph {et~al.}(1999)\citenamefont
  {Gusynin}, \citenamefont {Miransky},\ and\ \citenamefont
  {Shovkovy}}]{Gusynin:1999pq}%
  \BibitemOpen
  \bibfield  {author} {\bibinfo {author} {\bibfnamefont {V.~P.}\ \bibnamefont
  {Gusynin}}, \bibinfo {author} {\bibfnamefont {V.~A.}\ \bibnamefont
  {Miransky}}, \ and\ \bibinfo {author} {\bibfnamefont {I.~A.}\ \bibnamefont
  {Shovkovy}},\ }\href {\doibase 10.1016/S0550-3213(99)00573-8} {\bibfield
  {journal} {\bibinfo  {journal} {Nucl. Phys. B}\ }\textbf {\bibinfo {volume}
  {563}},\ \bibinfo {pages} {361} (\bibinfo {year} {1999})},\ \Eprint
  {http://arxiv.org/abs/hep-ph/9908320} {arXiv:hep-ph/9908320} \BibitemShut
  {NoStop}%
\bibitem [{\citenamefont {Klimenko}(1991)}]{Klimenko:1990rh}%
  \BibitemOpen
  \bibfield  {author} {\bibinfo {author} {\bibfnamefont {K.~G.}\ \bibnamefont
  {Klimenko}},\ }\href {\doibase 10.1007/BF01015908} {\bibfield  {journal}
  {\bibinfo  {journal} {Teor. Mat. Fiz.}\ }\textbf {\bibinfo {volume} {89}},\
  \bibinfo {pages} {211} (\bibinfo {year} {1991})}\BibitemShut {NoStop}%
\bibitem [{\citenamefont {Klevansky}\ and\ \citenamefont
  {Lemmer}(1989)}]{Klevansky:1989vi}%
  \BibitemOpen
  \bibfield  {author} {\bibinfo {author} {\bibfnamefont {S.~P.}\ \bibnamefont
  {Klevansky}}\ and\ \bibinfo {author} {\bibfnamefont {R.~H.}\ \bibnamefont
  {Lemmer}},\ }\href {\doibase 10.1103/PhysRevD.39.3478} {\bibfield  {journal}
  {\bibinfo  {journal} {Phys. Rev. D}\ }\textbf {\bibinfo {volume} {39}},\
  \bibinfo {pages} {3478} (\bibinfo {year} {1989})}\BibitemShut {NoStop}%
\bibitem [{\citenamefont {Kharzeev}\ and\ \citenamefont
  {Zhitnitsky}(2007)}]{Kharzeev:2007tn}%
  \BibitemOpen
  \bibfield  {author} {\bibinfo {author} {\bibfnamefont {D.}~\bibnamefont
  {Kharzeev}}\ and\ \bibinfo {author} {\bibfnamefont {A.}~\bibnamefont
  {Zhitnitsky}},\ }\href {\doibase 10.1016/j.nuclphysa.2007.10.001} {\bibfield
  {journal} {\bibinfo  {journal} {Nucl. Phys. A}\ }\textbf {\bibinfo {volume}
  {797}},\ \bibinfo {pages} {67} (\bibinfo {year} {2007})},\ \Eprint
  {http://arxiv.org/abs/0706.1026} {arXiv:0706.1026 [hep-ph]} \BibitemShut
  {NoStop}%
\bibitem [{\citenamefont {Kharzeev}\ \emph {et~al.}(2008)\citenamefont
  {Kharzeev}, \citenamefont {McLerran},\ and\ \citenamefont
  {Warringa}}]{Kharzeev:2007jp}%
  \BibitemOpen
  \bibfield  {author} {\bibinfo {author} {\bibfnamefont {D.~E.}\ \bibnamefont
  {Kharzeev}}, \bibinfo {author} {\bibfnamefont {L.~D.}\ \bibnamefont
  {McLerran}}, \ and\ \bibinfo {author} {\bibfnamefont {H.~J.}\ \bibnamefont
  {Warringa}},\ }\href {\doibase 10.1016/j.nuclphysa.2008.02.298} {\bibfield
  {journal} {\bibinfo  {journal} {Nucl. Phys. A}\ }\textbf {\bibinfo {volume}
  {803}},\ \bibinfo {pages} {227} (\bibinfo {year} {2008})},\ \Eprint
  {http://arxiv.org/abs/0711.0950} {arXiv:0711.0950 [hep-ph]} \BibitemShut
  {NoStop}%
\bibitem [{\citenamefont {Fukushima}\ \emph {et~al.}(2008)\citenamefont
  {Fukushima}, \citenamefont {Kharzeev},\ and\ \citenamefont
  {Warringa}}]{Fukushima:2008xe}%
  \BibitemOpen
  \bibfield  {author} {\bibinfo {author} {\bibfnamefont {K.}~\bibnamefont
  {Fukushima}}, \bibinfo {author} {\bibfnamefont {D.~E.}\ \bibnamefont
  {Kharzeev}}, \ and\ \bibinfo {author} {\bibfnamefont {H.~J.}\ \bibnamefont
  {Warringa}},\ }\href {\doibase 10.1103/PhysRevD.78.074033} {\bibfield
  {journal} {\bibinfo  {journal} {Phys. Rev. D}\ }\textbf {\bibinfo {volume}
  {78}},\ \bibinfo {pages} {074033} (\bibinfo {year} {2008})},\ \Eprint
  {http://arxiv.org/abs/0808.3382} {arXiv:0808.3382 [hep-ph]} \BibitemShut
  {NoStop}%
\bibitem [{\citenamefont {Diakonov}(2003)}]{Diakonov:2002fq}%
  \BibitemOpen
  \bibfield  {author} {\bibinfo {author} {\bibfnamefont {D.}~\bibnamefont
  {Diakonov}},\ }\href {\doibase 10.1016/S0146-6410(03)90014-7} {\bibfield
  {journal} {\bibinfo  {journal} {Prog. Part. Nucl. Phys.}\ }\textbf {\bibinfo
  {volume} {51}},\ \bibinfo {pages} {173} (\bibinfo {year} {2003})},\ \Eprint
  {http://arxiv.org/abs/hep-ph/0212026} {arXiv:hep-ph/0212026} \BibitemShut
  {NoStop}%
\bibitem [{\citenamefont {Sch\"afer}\ and\ \citenamefont
  {Shuryak}(1996)}]{Schafer:1995pz}%
  \BibitemOpen
  \bibfield  {author} {\bibinfo {author} {\bibfnamefont {T.}~\bibnamefont
  {Sch\"afer}}\ and\ \bibinfo {author} {\bibfnamefont {E.~V.}\ \bibnamefont
  {Shuryak}},\ }\href {\doibase 10.1103/PhysRevD.53.6522} {\bibfield  {journal}
  {\bibinfo  {journal} {Phys. Rev. D}\ }\textbf {\bibinfo {volume} {53}},\
  \bibinfo {pages} {6522} (\bibinfo {year} {1996})},\ \Eprint
  {http://arxiv.org/abs/hep-ph/9509337} {arXiv:hep-ph/9509337} \BibitemShut
  {NoStop}%
\bibitem [{\citenamefont {Arnold}\ and\ \citenamefont
  {McLerran}(1988)}]{Arnold:1987zg}%
  \BibitemOpen
  \bibfield  {author} {\bibinfo {author} {\bibfnamefont {P.~B.}\ \bibnamefont
  {Arnold}}\ and\ \bibinfo {author} {\bibfnamefont {L.~D.}\ \bibnamefont
  {McLerran}},\ }\href {\doibase 10.1103/PhysRevD.37.1020} {\bibfield
  {journal} {\bibinfo  {journal} {Phys. Rev. D}\ }\textbf {\bibinfo {volume}
  {37}},\ \bibinfo {pages} {1020} (\bibinfo {year} {1988})}\BibitemShut
  {NoStop}%
\bibitem [{\citenamefont {Fukugita}\ and\ \citenamefont
  {Yanagida}(1990)}]{Fukugita:1990gb}%
  \BibitemOpen
  \bibfield  {author} {\bibinfo {author} {\bibfnamefont {M.}~\bibnamefont
  {Fukugita}}\ and\ \bibinfo {author} {\bibfnamefont {T.}~\bibnamefont
  {Yanagida}},\ }\href {\doibase 10.1103/PhysRevD.42.1285} {\bibfield
  {journal} {\bibinfo  {journal} {Phys. Rev. D}\ }\textbf {\bibinfo {volume}
  {42}},\ \bibinfo {pages} {1285} (\bibinfo {year} {1990})}\BibitemShut
  {NoStop}%
\bibitem [{\citenamefont {McLerran}\ \emph {et~al.}(1991)\citenamefont
  {McLerran}, \citenamefont {Mottola},\ and\ \citenamefont
  {Shaposhnikov}}]{McLerran:1990de}%
  \BibitemOpen
  \bibfield  {author} {\bibinfo {author} {\bibfnamefont {L.~D.}\ \bibnamefont
  {McLerran}}, \bibinfo {author} {\bibfnamefont {E.}~\bibnamefont {Mottola}}, \
  and\ \bibinfo {author} {\bibfnamefont {M.~E.}\ \bibnamefont {Shaposhnikov}},\
  }\href {\doibase 10.1103/PhysRevD.43.2027} {\bibfield  {journal} {\bibinfo
  {journal} {Phys. Rev. D}\ }\textbf {\bibinfo {volume} {43}},\ \bibinfo
  {pages} {2027} (\bibinfo {year} {1991})}\BibitemShut {NoStop}%
\bibitem [{\citenamefont {Witten}(1979)}]{Witten:1979vv}%
  \BibitemOpen
  \bibfield  {author} {\bibinfo {author} {\bibfnamefont {E.}~\bibnamefont
  {Witten}},\ }\href {\doibase 10.1016/0550-3213(79)90031-2} {\bibfield
  {journal} {\bibinfo  {journal} {Nucl. Phys. B}\ }\textbf {\bibinfo {volume}
  {156}},\ \bibinfo {pages} {269} (\bibinfo {year} {1979})}\BibitemShut
  {NoStop}%
\bibitem [{\citenamefont {Veneziano}(1979)}]{Veneziano:1979ec}%
  \BibitemOpen
  \bibfield  {author} {\bibinfo {author} {\bibfnamefont {G.}~\bibnamefont
  {Veneziano}},\ }\href {\doibase 10.1016/0550-3213(79)90332-8} {\bibfield
  {journal} {\bibinfo  {journal} {Nucl. Phys. B}\ }\textbf {\bibinfo {volume}
  {159}},\ \bibinfo {pages} {213} (\bibinfo {year} {1979})}\BibitemShut
  {NoStop}%
\bibitem [{\citenamefont {Sch\"afer}\ and\ \citenamefont
  {Shuryak}(1998)}]{Schafer:1996wv}%
  \BibitemOpen
  \bibfield  {author} {\bibinfo {author} {\bibfnamefont {T.}~\bibnamefont
  {Sch\"afer}}\ and\ \bibinfo {author} {\bibfnamefont {E.~V.}\ \bibnamefont
  {Shuryak}},\ }\href {\doibase 10.1103/RevModPhys.70.323} {\bibfield
  {journal} {\bibinfo  {journal} {Rev. Mod. Phys.}\ }\textbf {\bibinfo {volume}
  {70}},\ \bibinfo {pages} {323} (\bibinfo {year} {1998})},\ \Eprint
  {http://arxiv.org/abs/hep-ph/9610451} {arXiv:hep-ph/9610451} \BibitemShut
  {NoStop}%
\bibitem [{\citenamefont {Vicari}\ and\ \citenamefont
  {Panagopoulos}(2009)}]{Vicari:2008jw}%
  \BibitemOpen
  \bibfield  {author} {\bibinfo {author} {\bibfnamefont {E.}~\bibnamefont
  {Vicari}}\ and\ \bibinfo {author} {\bibfnamefont {H.}~\bibnamefont
  {Panagopoulos}},\ }\href {\doibase 10.1016/j.physrep.2008.10.001} {\bibfield
  {journal} {\bibinfo  {journal} {Phys. Rept.}\ }\textbf {\bibinfo {volume}
  {470}},\ \bibinfo {pages} {93} (\bibinfo {year} {2009})},\ \Eprint
  {http://arxiv.org/abs/0803.1593} {arXiv:0803.1593 [hep-th]} \BibitemShut
  {NoStop}%
\bibitem [{\citenamefont {Kharzeev}\ \emph {et~al.}(2016)\citenamefont
  {Kharzeev}, \citenamefont {Liao}, \citenamefont {Voloshin},\ and\
  \citenamefont {Wang}}]{Kharzeev:2015znc}%
  \BibitemOpen
  \bibfield  {author} {\bibinfo {author} {\bibfnamefont {D.~E.}\ \bibnamefont
  {Kharzeev}}, \bibinfo {author} {\bibfnamefont {J.}~\bibnamefont {Liao}},
  \bibinfo {author} {\bibfnamefont {S.~A.}\ \bibnamefont {Voloshin}}, \ and\
  \bibinfo {author} {\bibfnamefont {G.}~\bibnamefont {Wang}},\ }\href {\doibase
  10.1016/j.ppnp.2016.01.001} {\bibfield  {journal} {\bibinfo  {journal} {Prog.
  Part. Nucl. Phys.}\ }\textbf {\bibinfo {volume} {88}},\ \bibinfo {pages} {1}
  (\bibinfo {year} {2016})},\ \Eprint {http://arxiv.org/abs/1511.04050}
  {arXiv:1511.04050 [hep-ph]} \BibitemShut {NoStop}%
\bibitem [{\citenamefont {Bzdak}\ \emph {et~al.}(2020)\citenamefont {Bzdak},
  \citenamefont {Esumi}, \citenamefont {Koch}, \citenamefont {Liao},
  \citenamefont {Stephanov},\ and\ \citenamefont {Xu}}]{Bzdak:2019pkr}%
  \BibitemOpen
  \bibfield  {author} {\bibinfo {author} {\bibfnamefont {A.}~\bibnamefont
  {Bzdak}}, \bibinfo {author} {\bibfnamefont {S.}~\bibnamefont {Esumi}},
  \bibinfo {author} {\bibfnamefont {V.}~\bibnamefont {Koch}}, \bibinfo {author}
  {\bibfnamefont {J.}~\bibnamefont {Liao}}, \bibinfo {author} {\bibfnamefont
  {M.}~\bibnamefont {Stephanov}}, \ and\ \bibinfo {author} {\bibfnamefont
  {N.}~\bibnamefont {Xu}},\ }\href {\doibase 10.1016/j.physrep.2020.01.005}
  {\bibfield  {journal} {\bibinfo  {journal} {Phys. Rept.}\ }\textbf {\bibinfo
  {volume} {853}},\ \bibinfo {pages} {1} (\bibinfo {year} {2020})},\ \Eprint
  {http://arxiv.org/abs/1906.00936} {arXiv:1906.00936 [nucl-th]} \BibitemShut
  {NoStop}%
\bibitem [{\citenamefont {Chen}\ and\ \citenamefont
  {Feng}(2020)}]{Chen:2019qoe}%
  \BibitemOpen
  \bibfield  {author} {\bibinfo {author} {\bibfnamefont {B.-X.}\ \bibnamefont
  {Chen}}\ and\ \bibinfo {author} {\bibfnamefont {S.-Q.}\ \bibnamefont
  {Feng}},\ }\href {\doibase 10.1088/1674-1137/44/2/024104} {\bibfield
  {journal} {\bibinfo  {journal} {Chin. Phys. C}\ }\textbf {\bibinfo {volume}
  {44}},\ \bibinfo {pages} {024104} (\bibinfo {year} {2020})},\ \Eprint
  {http://arxiv.org/abs/1909.10836} {arXiv:1909.10836 [hep-ph]} \BibitemShut
  {NoStop}%
\bibitem [{\citenamefont {Voloshin}(2004)}]{Voloshin:2004vk}%
  \BibitemOpen
  \bibfield  {author} {\bibinfo {author} {\bibfnamefont {S.~A.}\ \bibnamefont
  {Voloshin}},\ }\href {\doibase 10.1103/PhysRevC.70.057901} {\bibfield
  {journal} {\bibinfo  {journal} {Phys. Rev. C}\ }\textbf {\bibinfo {volume}
  {70}},\ \bibinfo {pages} {057901} (\bibinfo {year} {2004})},\ \Eprint
  {http://arxiv.org/abs/hep-ph/0406311} {arXiv:hep-ph/0406311} \BibitemShut
  {NoStop}%
\bibitem [{\citenamefont {Abelev}\ \emph {et~al.}(2010)\citenamefont {Abelev}
  \emph {et~al.}}]{STAR:2009tro}%
  \BibitemOpen
  \bibfield  {author} {\bibinfo {author} {\bibfnamefont {B.~I.}\ \bibnamefont
  {Abelev}} \emph {et~al.} (\bibinfo {collaboration} {STAR}),\ }\href {\doibase
  10.1103/PhysRevC.81.054908} {\bibfield  {journal} {\bibinfo  {journal} {Phys.
  Rev. C}\ }\textbf {\bibinfo {volume} {81}},\ \bibinfo {pages} {054908}
  (\bibinfo {year} {2010})},\ \Eprint {http://arxiv.org/abs/0909.1717}
  {arXiv:0909.1717 [nucl-ex]} \BibitemShut {NoStop}%
\bibitem [{\citenamefont {Abelev}\ \emph {et~al.}(2009)\citenamefont {Abelev}
  \emph {et~al.}}]{STAR:2009wot}%
  \BibitemOpen
  \bibfield  {author} {\bibinfo {author} {\bibfnamefont {B.~I.}\ \bibnamefont
  {Abelev}} \emph {et~al.} (\bibinfo {collaboration} {STAR}),\ }\href {\doibase
  10.1103/PhysRevLett.103.251601} {\bibfield  {journal} {\bibinfo  {journal}
  {Phys. Rev. Lett.}\ }\textbf {\bibinfo {volume} {103}},\ \bibinfo {pages}
  {251601} (\bibinfo {year} {2009})},\ \Eprint {http://arxiv.org/abs/0909.1739}
  {arXiv:0909.1739 [nucl-ex]} \BibitemShut {NoStop}%
\bibitem [{\citenamefont {Hu}(2022)}]{Hu:2021drc}%
  \BibitemOpen
  \bibfield  {author} {\bibinfo {author} {\bibfnamefont {Y.}~\bibnamefont {Hu}}
  (\bibinfo {collaboration} {STAR}),\ }\href {\doibase
  10.1051/epjconf/202225913013} {\bibfield  {journal} {\bibinfo  {journal} {EPJ
  Web Conf.}\ }\textbf {\bibinfo {volume} {259}},\ \bibinfo {pages} {13013}
  (\bibinfo {year} {2022})},\ \Eprint {http://arxiv.org/abs/2110.15937}
  {arXiv:2110.15937 [nucl-ex]} \BibitemShut {NoStop}%
\bibitem [{\citenamefont {Abdallah}\ \emph
  {et~al.}(2022{\natexlab{b}})\citenamefont {Abdallah} \emph
  {et~al.}}]{STAR:2021pwb}%
  \BibitemOpen
  \bibfield  {author} {\bibinfo {author} {\bibfnamefont {M.}~\bibnamefont
  {Abdallah}} \emph {et~al.} (\bibinfo {collaboration} {STAR}),\ }\href
  {\doibase 10.1103/PhysRevLett.128.092301} {\bibfield  {journal} {\bibinfo
  {journal} {Phys. Rev. Lett.}\ }\textbf {\bibinfo {volume} {128}},\ \bibinfo
  {pages} {092301} (\bibinfo {year} {2022}{\natexlab{b}})},\ \Eprint
  {http://arxiv.org/abs/2106.09243} {arXiv:2106.09243 [nucl-ex]} \BibitemShut
  {NoStop}%
\bibitem [{\citenamefont {Zhao}(2021)}]{Zhao:2020utk}%
  \BibitemOpen
  \bibfield  {author} {\bibinfo {author} {\bibfnamefont {J.}~\bibnamefont
  {Zhao}} (\bibinfo {collaboration} {STAR}),\ }\href {\doibase
  10.1016/j.nuclphysa.2020.121766} {\bibfield  {journal} {\bibinfo  {journal}
  {Nucl. Phys. A}\ }\textbf {\bibinfo {volume} {1005}},\ \bibinfo {pages}
  {121766} (\bibinfo {year} {2021})},\ \Eprint
  {http://arxiv.org/abs/2002.09410} {arXiv:2002.09410 [nucl-ex]} \BibitemShut
  {NoStop}%
\bibitem [{\citenamefont {Abelev}\ \emph {et~al.}(2013)\citenamefont {Abelev}
  \emph {et~al.}}]{ALICE:2012nhw}%
  \BibitemOpen
  \bibfield  {author} {\bibinfo {author} {\bibfnamefont {B.}~\bibnamefont
  {Abelev}} \emph {et~al.} (\bibinfo {collaboration} {ALICE}),\ }\href
  {\doibase 10.1103/PhysRevLett.110.012301} {\bibfield  {journal} {\bibinfo
  {journal} {Phys. Rev. Lett.}\ }\textbf {\bibinfo {volume} {110}},\ \bibinfo
  {pages} {012301} (\bibinfo {year} {2013})},\ \Eprint
  {http://arxiv.org/abs/1207.0900} {arXiv:1207.0900 [nucl-ex]} \BibitemShut
  {NoStop}%
\bibitem [{\citenamefont {Adam}\ \emph {et~al.}(2016)\citenamefont {Adam} \emph
  {et~al.}}]{ALICE:2015cjr}%
  \BibitemOpen
  \bibfield  {author} {\bibinfo {author} {\bibfnamefont {J.}~\bibnamefont
  {Adam}} \emph {et~al.} (\bibinfo {collaboration} {ALICE}),\ }\href {\doibase
  10.1103/PhysRevC.93.044903} {\bibfield  {journal} {\bibinfo  {journal} {Phys.
  Rev. C}\ }\textbf {\bibinfo {volume} {93}},\ \bibinfo {pages} {044903}
  (\bibinfo {year} {2016})},\ \Eprint {http://arxiv.org/abs/1512.05739}
  {arXiv:1512.05739 [nucl-ex]} \BibitemShut {NoStop}%
\bibitem [{\citenamefont {Parmar}(2018)}]{Parmar:2017deh}%
  \BibitemOpen
  \bibfield  {author} {\bibinfo {author} {\bibfnamefont {S.}~\bibnamefont
  {Parmar}} (\bibinfo {collaboration} {ALICE}),\ }\href {\doibase
  10.1007/978-3-319-73171-1_189} {\bibfield  {journal} {\bibinfo  {journal}
  {Springer Proc. Phys.}\ }\textbf {\bibinfo {volume} {203}},\ \bibinfo {pages}
  {785} (\bibinfo {year} {2018})},\ \Eprint {http://arxiv.org/abs/1703.09496}
  {arXiv:1703.09496 [hep-ex]} \BibitemShut {NoStop}%
\bibitem [{\citenamefont {Acharya}\ \emph {et~al.}(2018)\citenamefont {Acharya}
  \emph {et~al.}}]{ALICE:2017sss}%
  \BibitemOpen
  \bibfield  {author} {\bibinfo {author} {\bibfnamefont {S.}~\bibnamefont
  {Acharya}} \emph {et~al.} (\bibinfo {collaboration} {ALICE}),\ }\href
  {\doibase 10.1016/j.physletb.2017.12.021} {\bibfield  {journal} {\bibinfo
  {journal} {Phys. Lett. B}\ }\textbf {\bibinfo {volume} {777}},\ \bibinfo
  {pages} {151} (\bibinfo {year} {2018})},\ \Eprint
  {http://arxiv.org/abs/1709.04723} {arXiv:1709.04723 [nucl-ex]} \BibitemShut
  {NoStop}%
\bibitem [{\citenamefont {Acharya}\ \emph {et~al.}(2020)\citenamefont {Acharya}
  \emph {et~al.}}]{ALICE:2020siw}%
  \BibitemOpen
  \bibfield  {author} {\bibinfo {author} {\bibfnamefont {S.}~\bibnamefont
  {Acharya}} \emph {et~al.} (\bibinfo {collaboration} {ALICE}),\ }\href
  {\doibase 10.1007/JHEP09(2020)160} {\bibfield  {journal} {\bibinfo  {journal}
  {JHEP}\ }\textbf {\bibinfo {volume} {09}},\ \bibinfo {pages} {160} (\bibinfo
  {year} {2020})},\ \Eprint {http://arxiv.org/abs/2005.14640} {arXiv:2005.14640
  [nucl-ex]} \BibitemShut {NoStop}%
\bibitem [{\citenamefont {Sirunyan}\ \emph {et~al.}(2018)\citenamefont
  {Sirunyan} \emph {et~al.}}]{CMS:2017lrw}%
  \BibitemOpen
  \bibfield  {author} {\bibinfo {author} {\bibfnamefont {A.~M.}\ \bibnamefont
  {Sirunyan}} \emph {et~al.} (\bibinfo {collaboration} {CMS}),\ }\href
  {\doibase 10.1103/PhysRevC.97.044912} {\bibfield  {journal} {\bibinfo
  {journal} {Phys. Rev. C}\ }\textbf {\bibinfo {volume} {97}},\ \bibinfo
  {pages} {044912} (\bibinfo {year} {2018})},\ \Eprint
  {http://arxiv.org/abs/1708.01602} {arXiv:1708.01602 [nucl-ex]} \BibitemShut
  {NoStop}%
\bibitem [{\citenamefont {Khachatryan}\ \emph {et~al.}(2017)\citenamefont
  {Khachatryan} \emph {et~al.}}]{CMS:2016wfo}%
  \BibitemOpen
  \bibfield  {author} {\bibinfo {author} {\bibfnamefont {V.}~\bibnamefont
  {Khachatryan}} \emph {et~al.} (\bibinfo {collaboration} {CMS}),\ }\href
  {\doibase 10.1103/PhysRevLett.118.122301} {\bibfield  {journal} {\bibinfo
  {journal} {Phys. Rev. Lett.}\ }\textbf {\bibinfo {volume} {118}},\ \bibinfo
  {pages} {122301} (\bibinfo {year} {2017})},\ \Eprint
  {http://arxiv.org/abs/1610.00263} {arXiv:1610.00263 [nucl-ex]} \BibitemShut
  {NoStop}%
\bibitem [{\citenamefont {Kharzeev}\ and\ \citenamefont
  {Kikuchi}(2020)}]{Kharzeev:2020kgc}%
  \BibitemOpen
  \bibfield  {author} {\bibinfo {author} {\bibfnamefont {D.~E.}\ \bibnamefont
  {Kharzeev}}\ and\ \bibinfo {author} {\bibfnamefont {Y.}~\bibnamefont
  {Kikuchi}},\ }\href {\doibase 10.1103/PhysRevResearch.2.023342} {\bibfield
  {journal} {\bibinfo  {journal} {Phys. Rev. Res.}\ }\textbf {\bibinfo {volume}
  {2}},\ \bibinfo {pages} {023342} (\bibinfo {year} {2020})},\ \Eprint
  {http://arxiv.org/abs/2001.00698} {arXiv:2001.00698 [hep-ph]} \BibitemShut
  {NoStop}%
\bibitem [{\citenamefont {Ruggieri}\ \emph {et~al.}(2016)\citenamefont
  {Ruggieri}, \citenamefont {Peng},\ and\ \citenamefont
  {Chernodub}}]{Ruggieri:2016asg}%
  \BibitemOpen
  \bibfield  {author} {\bibinfo {author} {\bibfnamefont {M.}~\bibnamefont
  {Ruggieri}}, \bibinfo {author} {\bibfnamefont {G.~X.}\ \bibnamefont {Peng}},
  \ and\ \bibinfo {author} {\bibfnamefont {M.}~\bibnamefont {Chernodub}},\
  }\href {\doibase 10.1103/PhysRevD.94.054011} {\bibfield  {journal} {\bibinfo
  {journal} {Phys. Rev. D}\ }\textbf {\bibinfo {volume} {94}},\ \bibinfo
  {pages} {054011} (\bibinfo {year} {2016})},\ \Eprint
  {http://arxiv.org/abs/1606.03287} {arXiv:1606.03287 [hep-ph]} \BibitemShut
  {NoStop}%
\bibitem [{\citenamefont {Ruggieri}\ \emph {et~al.}(2020)\citenamefont
  {Ruggieri}, \citenamefont {Chernodub},\ and\ \citenamefont
  {Lu}}]{Ruggieri:2020qtq}%
  \BibitemOpen
  \bibfield  {author} {\bibinfo {author} {\bibfnamefont {M.}~\bibnamefont
  {Ruggieri}}, \bibinfo {author} {\bibfnamefont {M.~N.}\ \bibnamefont
  {Chernodub}}, \ and\ \bibinfo {author} {\bibfnamefont {Z.-Y.}\ \bibnamefont
  {Lu}},\ }\href {\doibase 10.1103/PhysRevD.102.014031} {\bibfield  {journal}
  {\bibinfo  {journal} {Phys. Rev. D}\ }\textbf {\bibinfo {volume} {102}},\
  \bibinfo {pages} {014031} (\bibinfo {year} {2020})},\ \Eprint
  {http://arxiv.org/abs/2004.09393} {arXiv:2004.09393 [hep-ph]} \BibitemShut
  {NoStop}%
\bibitem [{\citenamefont {Nambu}\ and\ \citenamefont
  {Jona-Lasinio}(1961{\natexlab{a}})}]{Nambu:1961tp}%
  \BibitemOpen
  \bibfield  {author} {\bibinfo {author} {\bibfnamefont {Y.}~\bibnamefont
  {Nambu}}\ and\ \bibinfo {author} {\bibfnamefont {G.}~\bibnamefont
  {Jona-Lasinio}},\ }\href {\doibase 10.1103/PhysRev.122.345} {\bibfield
  {journal} {\bibinfo  {journal} {Phys. Rev.}\ }\textbf {\bibinfo {volume}
  {122}},\ \bibinfo {pages} {345} (\bibinfo {year}
  {1961}{\natexlab{a}})}\BibitemShut {NoStop}%
\bibitem [{\citenamefont {Nambu}\ and\ \citenamefont
  {Jona-Lasinio}(1961{\natexlab{b}})}]{Nambu:1961fr}%
  \BibitemOpen
  \bibfield  {author} {\bibinfo {author} {\bibfnamefont {Y.}~\bibnamefont
  {Nambu}}\ and\ \bibinfo {author} {\bibfnamefont {G.}~\bibnamefont
  {Jona-Lasinio}},\ }\href {\doibase 10.1103/PhysRev.124.246} {\bibfield
  {journal} {\bibinfo  {journal} {Phys. Rev.}\ }\textbf {\bibinfo {volume}
  {124}},\ \bibinfo {pages} {246} (\bibinfo {year}
  {1961}{\natexlab{b}})}\BibitemShut {NoStop}%
\bibitem [{\citenamefont {Fukushima}(2008)}]{PhysRevD.77.114028}%
  \BibitemOpen
  \bibfield  {author} {\bibinfo {author} {\bibfnamefont {K.}~\bibnamefont
  {Fukushima}},\ }\href {\doibase 10.1103/PhysRevD.77.114028} {\bibfield
  {journal} {\bibinfo  {journal} {Phys. Rev. D}\ }\textbf {\bibinfo {volume}
  {77}},\ \bibinfo {pages} {114028} (\bibinfo {year} {2008})}\BibitemShut
  {NoStop}%
\bibitem [{\citenamefont {Costa}\ \emph {et~al.}(2007)\citenamefont {Costa},
  \citenamefont {{de Sousa}}, \citenamefont {Ruivo},\ and\ \citenamefont
  {Kalinovsky}}]{COSTA2007431}%
  \BibitemOpen
  \bibfield  {author} {\bibinfo {author} {\bibfnamefont {P.}~\bibnamefont
  {Costa}}, \bibinfo {author} {\bibfnamefont {C.}~\bibnamefont {{de Sousa}}},
  \bibinfo {author} {\bibfnamefont {M.}~\bibnamefont {Ruivo}}, \ and\ \bibinfo
  {author} {\bibfnamefont {Y.}~\bibnamefont {Kalinovsky}},\ }\href {\doibase
  https://doi.org/10.1016/j.physletb.2007.02.045} {\bibfield  {journal}
  {\bibinfo  {journal} {Physics Letters B}\ }\textbf {\bibinfo {volume}
  {647}},\ \bibinfo {pages} {431} (\bibinfo {year} {2007})}\BibitemShut
  {NoStop}%
\bibitem [{\citenamefont {Lu}\ \emph {et~al.}(2015)\citenamefont {Lu},
  \citenamefont {Du}, \citenamefont {Cui},\ and\ \citenamefont
  {Zong}}]{Lu2015}%
  \BibitemOpen
  \bibfield  {author} {\bibinfo {author} {\bibfnamefont {Y.}~\bibnamefont
  {Lu}}, \bibinfo {author} {\bibfnamefont {Y.-L.}\ \bibnamefont {Du}}, \bibinfo
  {author} {\bibfnamefont {Z.-F.}\ \bibnamefont {Cui}}, \ and\ \bibinfo
  {author} {\bibfnamefont {H.-S.}\ \bibnamefont {Zong}},\ }\href {\doibase
  10.1140/epjc/s10052-015-3720-2} {\bibfield  {journal} {\bibinfo  {journal}
  {The European Physical Journal C}\ }\textbf {\bibinfo {volume} {75}},\
  \bibinfo {pages} {495} (\bibinfo {year} {2015})}\BibitemShut {NoStop}%
\bibitem [{\citenamefont {Du}\ \emph {et~al.}(2015)\citenamefont {Du},
  \citenamefont {Lu}, \citenamefont {Xu}, \citenamefont {Cui}, \citenamefont
  {Shi},\ and\ \citenamefont {Zong}}]{doi:10.1142/S0217751X15501997}%
  \BibitemOpen
  \bibfield  {author} {\bibinfo {author} {\bibfnamefont {Y.-L.}\ \bibnamefont
  {Du}}, \bibinfo {author} {\bibfnamefont {Y.}~\bibnamefont {Lu}}, \bibinfo
  {author} {\bibfnamefont {S.-S.}\ \bibnamefont {Xu}}, \bibinfo {author}
  {\bibfnamefont {Z.-F.}\ \bibnamefont {Cui}}, \bibinfo {author} {\bibfnamefont
  {C.}~\bibnamefont {Shi}}, \ and\ \bibinfo {author} {\bibfnamefont {H.-S.}\
  \bibnamefont {Zong}},\ }\href {\doibase 10.1142/S0217751X15501997} {\bibfield
   {journal} {\bibinfo  {journal} {International Journal of Modern Physics A}\
  }\textbf {\bibinfo {volume} {30}},\ \bibinfo {pages} {1550199} (\bibinfo
  {year} {2015})},\ \Eprint
  {http://arxiv.org/abs/https://doi.org/10.1142/S0217751X15501997}
  {https://doi.org/10.1142/S0217751X15501997} \BibitemShut {NoStop}%
\bibitem [{\citenamefont {Cui}\ \emph {et~al.}(2015)\citenamefont {Cui},
  \citenamefont {Hou}, \citenamefont {Shi}, \citenamefont {Wang},\ and\
  \citenamefont {Zong}}]{CUI2015172}%
  \BibitemOpen
  \bibfield  {author} {\bibinfo {author} {\bibfnamefont {Z.-F.}\ \bibnamefont
  {Cui}}, \bibinfo {author} {\bibfnamefont {F.-Y.}\ \bibnamefont {Hou}},
  \bibinfo {author} {\bibfnamefont {Y.-M.}\ \bibnamefont {Shi}}, \bibinfo
  {author} {\bibfnamefont {Y.-L.}\ \bibnamefont {Wang}}, \ and\ \bibinfo
  {author} {\bibfnamefont {H.-S.}\ \bibnamefont {Zong}},\ }\href {\doibase
  https://doi.org/10.1016/j.aop.2015.03.025} {\bibfield  {journal} {\bibinfo
  {journal} {Annals of Physics}\ }\textbf {\bibinfo {volume} {358}},\ \bibinfo
  {pages} {172} (\bibinfo {year} {2015})},\ \bibinfo {note} {school of Physics
  at Nanjing University}\BibitemShut {NoStop}%
\bibitem [{\citenamefont {Du}\ \emph {et~al.}(2013)\citenamefont {Du},
  \citenamefont {Cui}, \citenamefont {Xia},\ and\ \citenamefont
  {Zong}}]{PhysRevD.88.114019}%
  \BibitemOpen
  \bibfield  {author} {\bibinfo {author} {\bibfnamefont {Y.-l.}\ \bibnamefont
  {Du}}, \bibinfo {author} {\bibfnamefont {Z.-f.}\ \bibnamefont {Cui}},
  \bibinfo {author} {\bibfnamefont {Y.-h.}\ \bibnamefont {Xia}}, \ and\
  \bibinfo {author} {\bibfnamefont {H.-s.}\ \bibnamefont {Zong}},\ }\href
  {\doibase 10.1103/PhysRevD.88.114019} {\bibfield  {journal} {\bibinfo
  {journal} {Phys. Rev. D}\ }\textbf {\bibinfo {volume} {88}},\ \bibinfo
  {pages} {114019} (\bibinfo {year} {2013})}\BibitemShut {NoStop}%
\bibitem [{\citenamefont {Shi}\ \emph {et~al.}(2015)\citenamefont {Shi},
  \citenamefont {Yang}, \citenamefont {Xia}, \citenamefont {Cui}, \citenamefont
  {Liu},\ and\ \citenamefont {Zong}}]{PhysRevD.91.036006}%
  \BibitemOpen
  \bibfield  {author} {\bibinfo {author} {\bibfnamefont {S.}~\bibnamefont
  {Shi}}, \bibinfo {author} {\bibfnamefont {Y.-C.}\ \bibnamefont {Yang}},
  \bibinfo {author} {\bibfnamefont {Y.-H.}\ \bibnamefont {Xia}}, \bibinfo
  {author} {\bibfnamefont {Z.-F.}\ \bibnamefont {Cui}}, \bibinfo {author}
  {\bibfnamefont {X.-J.}\ \bibnamefont {Liu}}, \ and\ \bibinfo {author}
  {\bibfnamefont {H.-S.}\ \bibnamefont {Zong}},\ }\href {\doibase
  10.1103/PhysRevD.91.036006} {\bibfield  {journal} {\bibinfo  {journal} {Phys.
  Rev. D}\ }\textbf {\bibinfo {volume} {91}},\ \bibinfo {pages} {036006}
  (\bibinfo {year} {2015})}\BibitemShut {NoStop}%
\bibitem [{\citenamefont {Menezes}\ \emph
  {et~al.}(2009{\natexlab{b}})\citenamefont {Menezes}, \citenamefont {Pinto},
  \citenamefont {Avancini},\ and\ \citenamefont
  {Provid\^encia}}]{PhysRevC.80.065805}%
  \BibitemOpen
  \bibfield  {author} {\bibinfo {author} {\bibfnamefont {D.~P.}\ \bibnamefont
  {Menezes}}, \bibinfo {author} {\bibfnamefont {M.~B.}\ \bibnamefont {Pinto}},
  \bibinfo {author} {\bibfnamefont {S.~S.}\ \bibnamefont {Avancini}}, \ and\
  \bibinfo {author} {\bibfnamefont {C.}~\bibnamefont {Provid\^encia}},\ }\href
  {\doibase 10.1103/PhysRevC.80.065805} {\bibfield  {journal} {\bibinfo
  {journal} {Phys. Rev. C}\ }\textbf {\bibinfo {volume} {80}},\ \bibinfo
  {pages} {065805} (\bibinfo {year} {2009}{\natexlab{b}})}\BibitemShut
  {NoStop}%
\bibitem [{\citenamefont {Ghosh}\ \emph {et~al.}(2007)\citenamefont {Ghosh},
  \citenamefont {Mandal},\ and\ \citenamefont
  {Chakrabarty}}]{PhysRevC.75.015805}%
  \BibitemOpen
  \bibfield  {author} {\bibinfo {author} {\bibfnamefont {S.}~\bibnamefont
  {Ghosh}}, \bibinfo {author} {\bibfnamefont {S.}~\bibnamefont {Mandal}}, \
  and\ \bibinfo {author} {\bibfnamefont {S.}~\bibnamefont {Chakrabarty}},\
  }\href {\doibase 10.1103/PhysRevC.75.015805} {\bibfield  {journal} {\bibinfo
  {journal} {Phys. Rev. C}\ }\textbf {\bibinfo {volume} {75}},\ \bibinfo
  {pages} {015805} (\bibinfo {year} {2007})}\BibitemShut {NoStop}%
\bibitem [{\citenamefont {Bali}\ \emph {et~al.}(2012)\citenamefont {Bali},
  \citenamefont {Bruckmann}, \citenamefont {Endr\ifmmode~\mbox{\H{o}}\else
  \H{o}\fi{}di}, \citenamefont {Fodor}, \citenamefont {Katz},\ and\
  \citenamefont {Sch\"afer}}]{PhysRevD.86.071502}%
  \BibitemOpen
  \bibfield  {author} {\bibinfo {author} {\bibfnamefont {G.~S.}\ \bibnamefont
  {Bali}}, \bibinfo {author} {\bibfnamefont {F.}~\bibnamefont {Bruckmann}},
  \bibinfo {author} {\bibfnamefont {G.}~\bibnamefont
  {Endr\ifmmode~\mbox{\H{o}}\else \H{o}\fi{}di}}, \bibinfo {author}
  {\bibfnamefont {Z.}~\bibnamefont {Fodor}}, \bibinfo {author} {\bibfnamefont
  {S.~D.}\ \bibnamefont {Katz}}, \ and\ \bibinfo {author} {\bibfnamefont
  {A.}~\bibnamefont {Sch\"afer}},\ }\href {\doibase 10.1103/PhysRevD.86.071502}
  {\bibfield  {journal} {\bibinfo  {journal} {Phys. Rev. D}\ }\textbf {\bibinfo
  {volume} {86}},\ \bibinfo {pages} {071502} (\bibinfo {year}
  {2012})}\BibitemShut {NoStop}%
\bibitem [{\citenamefont {Yamamoto}(2011)}]{Yamamoto:2011gk}%
  \BibitemOpen
  \bibfield  {author} {\bibinfo {author} {\bibfnamefont {A.}~\bibnamefont
  {Yamamoto}},\ }\href {\doibase 10.1103/PhysRevLett.107.031601} {\bibfield
  {journal} {\bibinfo  {journal} {Phys. Rev. Lett.}\ }\textbf {\bibinfo
  {volume} {107}},\ \bibinfo {pages} {031601} (\bibinfo {year} {2011})},\
  \Eprint {http://arxiv.org/abs/1105.0385} {arXiv:1105.0385 [hep-lat]}
  \BibitemShut {NoStop}%
\bibitem [{\citenamefont {Braguta}\ \emph {et~al.}(2016)\citenamefont
  {Braguta}, \citenamefont {Ilgenfritz}, \citenamefont {Kotov}, \citenamefont
  {Petersson},\ and\ \citenamefont {Skinderev}}]{Braguta:2015owi}%
  \BibitemOpen
  \bibfield  {author} {\bibinfo {author} {\bibfnamefont {V.~V.}\ \bibnamefont
  {Braguta}}, \bibinfo {author} {\bibfnamefont {E.~M.}\ \bibnamefont
  {Ilgenfritz}}, \bibinfo {author} {\bibfnamefont {A.~Y.}\ \bibnamefont
  {Kotov}}, \bibinfo {author} {\bibfnamefont {B.}~\bibnamefont {Petersson}}, \
  and\ \bibinfo {author} {\bibfnamefont {S.~A.}\ \bibnamefont {Skinderev}},\
  }\href {\doibase 10.1103/PhysRevD.93.034509} {\bibfield  {journal} {\bibinfo
  {journal} {Phys. Rev. D}\ }\textbf {\bibinfo {volume} {93}},\ \bibinfo
  {pages} {034509} (\bibinfo {year} {2016})},\ \Eprint
  {http://arxiv.org/abs/1512.05873} {arXiv:1512.05873 [hep-lat]} \BibitemShut
  {NoStop}%
\bibitem [{\citenamefont {Braguta}\ \emph {et~al.}(2015)\citenamefont
  {Braguta}, \citenamefont {Goy}, \citenamefont {Ilgenfritz}, \citenamefont
  {Kotov}, \citenamefont {Molochkov}, \citenamefont {Muller-Preussker},\ and\
  \citenamefont {Petersson}}]{Braguta:2015zta}%
  \BibitemOpen
  \bibfield  {author} {\bibinfo {author} {\bibfnamefont {V.~V.}\ \bibnamefont
  {Braguta}}, \bibinfo {author} {\bibfnamefont {V.~A.}\ \bibnamefont {Goy}},
  \bibinfo {author} {\bibfnamefont {E.~M.}\ \bibnamefont {Ilgenfritz}},
  \bibinfo {author} {\bibfnamefont {A.~Y.}\ \bibnamefont {Kotov}}, \bibinfo
  {author} {\bibfnamefont {A.~V.}\ \bibnamefont {Molochkov}}, \bibinfo {author}
  {\bibfnamefont {M.}~\bibnamefont {Muller-Preussker}}, \ and\ \bibinfo
  {author} {\bibfnamefont {B.}~\bibnamefont {Petersson}},\ }\href {\doibase
  10.1007/JHEP06(2015)094} {\bibfield  {journal} {\bibinfo  {journal} {JHEP}\
  }\textbf {\bibinfo {volume} {06}},\ \bibinfo {pages} {094} (\bibinfo {year}
  {2015})},\ \Eprint {http://arxiv.org/abs/1503.06670} {arXiv:1503.06670
  [hep-lat]} \BibitemShut {NoStop}%
\bibitem [{\citenamefont {Alexandru}\ \emph {et~al.}(2016)\citenamefont
  {Alexandru}, \citenamefont {Basar},\ and\ \citenamefont
  {Bedaque}}]{Alexandru:2015xva}%
  \BibitemOpen
  \bibfield  {author} {\bibinfo {author} {\bibfnamefont {A.}~\bibnamefont
  {Alexandru}}, \bibinfo {author} {\bibfnamefont {G.}~\bibnamefont {Basar}}, \
  and\ \bibinfo {author} {\bibfnamefont {P.}~\bibnamefont {Bedaque}},\ }\href
  {\doibase 10.1103/PhysRevD.93.014504} {\bibfield  {journal} {\bibinfo
  {journal} {Phys. Rev. D}\ }\textbf {\bibinfo {volume} {93}},\ \bibinfo
  {pages} {014504} (\bibinfo {year} {2016})},\ \Eprint
  {http://arxiv.org/abs/1510.03258} {arXiv:1510.03258 [hep-lat]} \BibitemShut
  {NoStop}%
\bibitem [{\citenamefont {Scorzato}(2016)}]{Scorzato:2015qts}%
  \BibitemOpen
  \bibfield  {author} {\bibinfo {author} {\bibfnamefont {L.}~\bibnamefont
  {Scorzato}},\ }\href {\doibase 10.22323/1.251.0016} {\bibfield  {journal}
  {\bibinfo  {journal} {PoS}\ }\textbf {\bibinfo {volume} {LATTICE2015}},\
  \bibinfo {pages} {016} (\bibinfo {year} {2016})},\ \Eprint
  {http://arxiv.org/abs/1512.08039} {arXiv:1512.08039 [hep-lat]} \BibitemShut
  {NoStop}%
\bibitem [{\citenamefont {Braguta}\ and\ \citenamefont
  {Kotov}(2016)}]{Braguta:2016aov}%
  \BibitemOpen
  \bibfield  {author} {\bibinfo {author} {\bibfnamefont {V.~V.}\ \bibnamefont
  {Braguta}}\ and\ \bibinfo {author} {\bibfnamefont {A.~Y.}\ \bibnamefont
  {Kotov}},\ }\href {\doibase 10.1103/PhysRevD.93.105025} {\bibfield  {journal}
  {\bibinfo  {journal} {Phys. Rev. D}\ }\textbf {\bibinfo {volume} {93}},\
  \bibinfo {pages} {105025} (\bibinfo {year} {2016})},\ \Eprint
  {http://arxiv.org/abs/1601.04957} {arXiv:1601.04957 [hep-th]} \BibitemShut
  {NoStop}%
\bibitem [{\citenamefont {Kapusta}\ and\ \citenamefont {Gale}(2006)}]{Kapusta}%
  \BibitemOpen
  \bibfield  {author} {\bibinfo {author} {\bibfnamefont {J.~I.}\ \bibnamefont
  {Kapusta}}\ and\ \bibinfo {author} {\bibfnamefont {C.}~\bibnamefont {Gale}},\
  }\href {\doibase 10.1017/CBO9780511535130} {\emph {\bibinfo {title}
  {Finite-Temperature Field Theory: Principles and Applications}}},\ \bibinfo
  {edition} {2nd}\ ed.,\ Cambridge Monographs on Mathematical Physics\
  (\bibinfo  {publisher} {Cambridge University Press},\ \bibinfo {year}
  {2006})\BibitemShut {NoStop}%
\bibitem [{\citenamefont {Feynman}(1982)}]{Feynman}%
  \BibitemOpen
  \bibfield  {author} {\bibinfo {author} {\bibfnamefont {R.~P.}\ \bibnamefont
  {Feynman}},\ }\href {\doibase 10.1007/BF02650179} {\bibfield  {journal}
  {\bibinfo  {journal} {International Journal of Theoretical Physics}\ }\textbf
  {\bibinfo {volume} {21}},\ \bibinfo {pages} {467} (\bibinfo {year}
  {1982})}\BibitemShut {NoStop}%
\bibitem [{\citenamefont {Bauer}\ \emph {et~al.}(2020)\citenamefont {Bauer},
  \citenamefont {Bravyi}, \citenamefont {Motta},\ and\ \citenamefont
  {Chan}}]{Bauer}%
  \BibitemOpen
  \bibfield  {author} {\bibinfo {author} {\bibfnamefont {B.}~\bibnamefont
  {Bauer}}, \bibinfo {author} {\bibfnamefont {S.}~\bibnamefont {Bravyi}},
  \bibinfo {author} {\bibfnamefont {M.}~\bibnamefont {Motta}}, \ and\ \bibinfo
  {author} {\bibfnamefont {G.~K.-L.}\ \bibnamefont {Chan}},\ }\href {\doibase
  10.1021/acs.chemrev.9b00829} {\bibfield  {journal} {\bibinfo  {journal}
  {Chemical Reviews}\ }\textbf {\bibinfo {volume} {120}},\ \bibinfo {pages}
  {12685–12717} (\bibinfo {year} {2020})}\BibitemShut {NoStop}%
\bibitem [{\citenamefont {Terhal}\ and\ \citenamefont
  {DiVincenzo}(2000{\natexlab{a}})}]{PhysRevA.61.022301}%
  \BibitemOpen
  \bibfield  {author} {\bibinfo {author} {\bibfnamefont {B.~M.}\ \bibnamefont
  {Terhal}}\ and\ \bibinfo {author} {\bibfnamefont {D.~P.}\ \bibnamefont
  {DiVincenzo}},\ }\href {\doibase 10.1103/PhysRevA.61.022301} {\bibfield
  {journal} {\bibinfo  {journal} {Phys. Rev. A}\ }\textbf {\bibinfo {volume}
  {61}},\ \bibinfo {pages} {022301} (\bibinfo {year}
  {2000}{\natexlab{a}})}\BibitemShut {NoStop}%
\bibitem [{\citenamefont {Poulin}\ and\ \citenamefont
  {Wocjan}(2009)}]{PhysRevLett.103.220502}%
  \BibitemOpen
  \bibfield  {author} {\bibinfo {author} {\bibfnamefont {D.}~\bibnamefont
  {Poulin}}\ and\ \bibinfo {author} {\bibfnamefont {P.}~\bibnamefont
  {Wocjan}},\ }\href {\doibase 10.1103/PhysRevLett.103.220502} {\bibfield
  {journal} {\bibinfo  {journal} {Phys. Rev. Lett.}\ }\textbf {\bibinfo
  {volume} {103}},\ \bibinfo {pages} {220502} (\bibinfo {year}
  {2009})}\BibitemShut {NoStop}%
\bibitem [{\citenamefont {Riera}\ \emph {et~al.}(2012)\citenamefont {Riera},
  \citenamefont {Gogolin},\ and\ \citenamefont
  {Eisert}}]{PhysRevLett.108.080402}%
  \BibitemOpen
  \bibfield  {author} {\bibinfo {author} {\bibfnamefont {A.}~\bibnamefont
  {Riera}}, \bibinfo {author} {\bibfnamefont {C.}~\bibnamefont {Gogolin}}, \
  and\ \bibinfo {author} {\bibfnamefont {J.}~\bibnamefont {Eisert}},\ }\href
  {\doibase 10.1103/PhysRevLett.108.080402} {\bibfield  {journal} {\bibinfo
  {journal} {Phys. Rev. Lett.}\ }\textbf {\bibinfo {volume} {108}},\ \bibinfo
  {pages} {080402} (\bibinfo {year} {2012})}\BibitemShut {NoStop}%
\bibitem [{\citenamefont {Temme}\ \emph
  {et~al.}(2011{\natexlab{a}})\citenamefont {Temme}, \citenamefont {Osborne},
  \citenamefont {Vollbrecht}, \citenamefont {Poulin},\ and\ \citenamefont
  {Verstraete}}]{Temme2011}%
  \BibitemOpen
  \bibfield  {author} {\bibinfo {author} {\bibfnamefont {K.}~\bibnamefont
  {Temme}}, \bibinfo {author} {\bibfnamefont {T.~J.}\ \bibnamefont {Osborne}},
  \bibinfo {author} {\bibfnamefont {K.~G.}\ \bibnamefont {Vollbrecht}},
  \bibinfo {author} {\bibfnamefont {D.}~\bibnamefont {Poulin}}, \ and\ \bibinfo
  {author} {\bibfnamefont {F.}~\bibnamefont {Verstraete}},\ }\href {\doibase
  10.1038/nature09770} {\bibfield  {journal} {\bibinfo  {journal} {Nature}\
  }\textbf {\bibinfo {volume} {471}},\ \bibinfo {pages} {87} (\bibinfo {year}
  {2011}{\natexlab{a}})}\BibitemShut {NoStop}%
\bibitem [{\citenamefont {Yung}\ and\ \citenamefont
  {Aspuru-Guzik}(2012)}]{Yung754}%
  \BibitemOpen
  \bibfield  {author} {\bibinfo {author} {\bibfnamefont {M.-H.}\ \bibnamefont
  {Yung}}\ and\ \bibinfo {author} {\bibfnamefont {A.}~\bibnamefont
  {Aspuru-Guzik}},\ }\href {\doibase 10.1073/pnas.1111758109} {\bibfield
  {journal} {\bibinfo  {journal} {Proceedings of the National Academy of
  Sciences}\ }\textbf {\bibinfo {volume} {109}},\ \bibinfo {pages} {754}
  (\bibinfo {year} {2012})}\BibitemShut {NoStop}%
\bibitem [{\citenamefont {Li}\ \emph {et~al.}(2021)\citenamefont {Li},
  \citenamefont {Guo}, \citenamefont {Lai}, \citenamefont {Liu}, \citenamefont
  {Wang}, \citenamefont {Xing}, \citenamefont {Zhang},\ and\ \citenamefont
  {Zhu}}]{Li:2021kcs}%
  \BibitemOpen
  \bibfield  {author} {\bibinfo {author} {\bibfnamefont {T.}~\bibnamefont
  {Li}}, \bibinfo {author} {\bibfnamefont {X.}~\bibnamefont {Guo}}, \bibinfo
  {author} {\bibfnamefont {W.~K.}\ \bibnamefont {Lai}}, \bibinfo {author}
  {\bibfnamefont {X.}~\bibnamefont {Liu}}, \bibinfo {author} {\bibfnamefont
  {E.}~\bibnamefont {Wang}}, \bibinfo {author} {\bibfnamefont {H.}~\bibnamefont
  {Xing}}, \bibinfo {author} {\bibfnamefont {D.-B.}\ \bibnamefont {Zhang}}, \
  and\ \bibinfo {author} {\bibfnamefont {S.-L.}\ \bibnamefont {Zhu}},\
  }\href@noop {} {\  (\bibinfo {year} {2021})},\ \Eprint
  {http://arxiv.org/abs/2106.03865} {arXiv:2106.03865 [hep-ph]} \BibitemShut
  {NoStop}%
\bibitem [{\citenamefont {Zhang}\ \emph {et~al.}(2021)\citenamefont {Zhang},
  \citenamefont {Xing}, \citenamefont {Yan}, \citenamefont {Wang},\ and\
  \citenamefont {Zhu}}]{Zhang:2020uqo}%
  \BibitemOpen
  \bibfield  {author} {\bibinfo {author} {\bibfnamefont {D.-B.}\ \bibnamefont
  {Zhang}}, \bibinfo {author} {\bibfnamefont {H.}~\bibnamefont {Xing}},
  \bibinfo {author} {\bibfnamefont {H.}~\bibnamefont {Yan}}, \bibinfo {author}
  {\bibfnamefont {E.}~\bibnamefont {Wang}}, \ and\ \bibinfo {author}
  {\bibfnamefont {S.-L.}\ \bibnamefont {Zhu}},\ }\href {\doibase
  10.1088/1674-1056/abd761} {\bibfield  {journal} {\bibinfo  {journal} {Chin.
  Phys. B}\ }\textbf {\bibinfo {volume} {30}},\ \bibinfo {pages} {020306}
  (\bibinfo {year} {2021})},\ \Eprint {http://arxiv.org/abs/2011.01431}
  {arXiv:2011.01431 [quant-ph]} \BibitemShut {NoStop}%
\bibitem [{\citenamefont {Tomiya}(2022)}]{Tomiya:2022chr}%
  \BibitemOpen
  \bibfield  {author} {\bibinfo {author} {\bibfnamefont {A.}~\bibnamefont
  {Tomiya}},\ }\href@noop {} {\  (\bibinfo {year} {2022})},\ \Eprint
  {http://arxiv.org/abs/2205.08860} {arXiv:2205.08860 [hep-lat]} \BibitemShut
  {NoStop}%
\bibitem [{\citenamefont {Xie}\ \emph {et~al.}(2022)\citenamefont {Xie},
  \citenamefont {Guo}, \citenamefont {Xing}, \citenamefont {Xue}, \citenamefont
  {Zhang},\ and\ \citenamefont {Zhu}}]{Xie:2022jgj}%
  \BibitemOpen
  \bibfield  {author} {\bibinfo {author} {\bibfnamefont {X.-D.}\ \bibnamefont
  {Xie}}, \bibinfo {author} {\bibfnamefont {X.}~\bibnamefont {Guo}}, \bibinfo
  {author} {\bibfnamefont {H.}~\bibnamefont {Xing}}, \bibinfo {author}
  {\bibfnamefont {Z.-Y.}\ \bibnamefont {Xue}}, \bibinfo {author} {\bibfnamefont
  {D.-B.}\ \bibnamefont {Zhang}}, \ and\ \bibinfo {author} {\bibfnamefont
  {S.-L.}\ \bibnamefont {Zhu}} (\bibinfo {collaboration} {QuNu}),\ }\href
  {\doibase 10.1103/PhysRevD.106.054509} {\bibfield  {journal} {\bibinfo
  {journal} {Phys. Rev. D}\ }\textbf {\bibinfo {volume} {106}},\ \bibinfo
  {pages} {054509} (\bibinfo {year} {2022})},\ \Eprint
  {http://arxiv.org/abs/2205.12767} {arXiv:2205.12767 [quant-ph]} \BibitemShut
  {NoStop}%
\bibitem [{\citenamefont {Davoudi}\ \emph {et~al.}(2022)\citenamefont
  {Davoudi}, \citenamefont {Mueller},\ and\ \citenamefont
  {Powers}}]{Davoudi:2022uzo}%
  \BibitemOpen
  \bibfield  {author} {\bibinfo {author} {\bibfnamefont {Z.}~\bibnamefont
  {Davoudi}}, \bibinfo {author} {\bibfnamefont {N.}~\bibnamefont {Mueller}}, \
  and\ \bibinfo {author} {\bibfnamefont {C.}~\bibnamefont {Powers}},\
  }\href@noop {} {\  (\bibinfo {year} {2022})},\ \Eprint
  {http://arxiv.org/abs/2208.13112} {arXiv:2208.13112 [hep-lat]} \BibitemShut
  {NoStop}%
\bibitem [{\citenamefont {Arute}\ \emph {et~al.}(2020)\citenamefont {Arute}
  \emph {et~al.}}]{Arute:2020uxm}%
  \BibitemOpen
  \bibfield  {author} {\bibinfo {author} {\bibfnamefont {F.}~\bibnamefont
  {Arute}} \emph {et~al.},\ }\href {\doibase 10.1126/science.abb9811}
  {\bibfield  {journal} {\bibinfo  {journal} {Science}\ }\textbf {\bibinfo
  {volume} {369}},\ \bibinfo {pages} {1084} (\bibinfo {year} {2020})},\ \Eprint
  {http://arxiv.org/abs/2004.04174} {arXiv:2004.04174 [quant-ph]} \BibitemShut
  {NoStop}%
\bibitem [{\citenamefont {Ma}\ \emph {et~al.}(2020)\citenamefont {Ma},
  \citenamefont {Govoni},\ and\ \citenamefont {Galli}}]{Ma2020}%
  \BibitemOpen
  \bibfield  {author} {\bibinfo {author} {\bibfnamefont {H.}~\bibnamefont
  {Ma}}, \bibinfo {author} {\bibfnamefont {M.}~\bibnamefont {Govoni}}, \ and\
  \bibinfo {author} {\bibfnamefont {G.}~\bibnamefont {Galli}},\ }\href
  {\doibase 10.1038/s41524-020-00353-z} {\bibfield  {journal} {\bibinfo
  {journal} {npj Computational Materials}\ }\textbf {\bibinfo {volume} {6}},\
  \bibinfo {pages} {85} (\bibinfo {year} {2020})}\BibitemShut {NoStop}%
\bibitem [{\citenamefont {Kandala}\ \emph {et~al.}(2019)\citenamefont
  {Kandala}, \citenamefont {Temme}, \citenamefont {C{\'o}rcoles}, \citenamefont
  {Mezzacapo}, \citenamefont {Chow},\ and\ \citenamefont
  {Gambetta}}]{Kandala2019}%
  \BibitemOpen
  \bibfield  {author} {\bibinfo {author} {\bibfnamefont {A.}~\bibnamefont
  {Kandala}}, \bibinfo {author} {\bibfnamefont {K.}~\bibnamefont {Temme}},
  \bibinfo {author} {\bibfnamefont {A.~D.}\ \bibnamefont {C{\'o}rcoles}},
  \bibinfo {author} {\bibfnamefont {A.}~\bibnamefont {Mezzacapo}}, \bibinfo
  {author} {\bibfnamefont {J.~M.}\ \bibnamefont {Chow}}, \ and\ \bibinfo
  {author} {\bibfnamefont {J.~M.}\ \bibnamefont {Gambetta}},\ }\href {\doibase
  10.1038/s41586-019-1040-7} {\bibfield  {journal} {\bibinfo  {journal}
  {Nature}\ }\textbf {\bibinfo {volume} {567}},\ \bibinfo {pages} {491}
  (\bibinfo {year} {2019})}\BibitemShut {NoStop}%
\bibitem [{\citenamefont {O'Malley}\ \emph {et~al.}(2016)\citenamefont
  {O'Malley}, \citenamefont {Babbush}, \citenamefont {Kivlichan}, \citenamefont
  {Romero}, \citenamefont {McClean}, \citenamefont {Barends}, \citenamefont
  {Kelly}, \citenamefont {Roushan}, \citenamefont {Tranter}, \citenamefont
  {Ding}, \citenamefont {Campbell}, \citenamefont {Chen}, \citenamefont {Chen},
  \citenamefont {Chiaro}, \citenamefont {Dunsworth}, \citenamefont {Fowler},
  \citenamefont {Jeffrey}, \citenamefont {Lucero}, \citenamefont {Megrant},
  \citenamefont {Mutus}, \citenamefont {Neeley}, \citenamefont {Neill},
  \citenamefont {Quintana}, \citenamefont {Sank}, \citenamefont {Vainsencher},
  \citenamefont {Wenner}, \citenamefont {White}, \citenamefont {Coveney},
  \citenamefont {Love}, \citenamefont {Neven}, \citenamefont {Aspuru-Guzik},\
  and\ \citenamefont {Martinis}}]{PhysRevX.6.031007}%
  \BibitemOpen
  \bibfield  {author} {\bibinfo {author} {\bibfnamefont {P.~J.~J.}\
  \bibnamefont {O'Malley}}, \bibinfo {author} {\bibfnamefont {R.}~\bibnamefont
  {Babbush}}, \bibinfo {author} {\bibfnamefont {I.~D.}\ \bibnamefont
  {Kivlichan}}, \bibinfo {author} {\bibfnamefont {J.}~\bibnamefont {Romero}},
  \bibinfo {author} {\bibfnamefont {J.~R.}\ \bibnamefont {McClean}}, \bibinfo
  {author} {\bibfnamefont {R.}~\bibnamefont {Barends}}, \bibinfo {author}
  {\bibfnamefont {J.}~\bibnamefont {Kelly}}, \bibinfo {author} {\bibfnamefont
  {P.}~\bibnamefont {Roushan}}, \bibinfo {author} {\bibfnamefont
  {A.}~\bibnamefont {Tranter}}, \bibinfo {author} {\bibfnamefont
  {N.}~\bibnamefont {Ding}}, \bibinfo {author} {\bibfnamefont {B.}~\bibnamefont
  {Campbell}}, \bibinfo {author} {\bibfnamefont {Y.}~\bibnamefont {Chen}},
  \bibinfo {author} {\bibfnamefont {Z.}~\bibnamefont {Chen}}, \bibinfo {author}
  {\bibfnamefont {B.}~\bibnamefont {Chiaro}}, \bibinfo {author} {\bibfnamefont
  {A.}~\bibnamefont {Dunsworth}}, \bibinfo {author} {\bibfnamefont {A.~G.}\
  \bibnamefont {Fowler}}, \bibinfo {author} {\bibfnamefont {E.}~\bibnamefont
  {Jeffrey}}, \bibinfo {author} {\bibfnamefont {E.}~\bibnamefont {Lucero}},
  \bibinfo {author} {\bibfnamefont {A.}~\bibnamefont {Megrant}}, \bibinfo
  {author} {\bibfnamefont {J.~Y.}\ \bibnamefont {Mutus}}, \bibinfo {author}
  {\bibfnamefont {M.}~\bibnamefont {Neeley}}, \bibinfo {author} {\bibfnamefont
  {C.}~\bibnamefont {Neill}}, \bibinfo {author} {\bibfnamefont
  {C.}~\bibnamefont {Quintana}}, \bibinfo {author} {\bibfnamefont
  {D.}~\bibnamefont {Sank}}, \bibinfo {author} {\bibfnamefont {A.}~\bibnamefont
  {Vainsencher}}, \bibinfo {author} {\bibfnamefont {J.}~\bibnamefont {Wenner}},
  \bibinfo {author} {\bibfnamefont {T.~C.}\ \bibnamefont {White}}, \bibinfo
  {author} {\bibfnamefont {P.~V.}\ \bibnamefont {Coveney}}, \bibinfo {author}
  {\bibfnamefont {P.~J.}\ \bibnamefont {Love}}, \bibinfo {author}
  {\bibfnamefont {H.}~\bibnamefont {Neven}}, \bibinfo {author} {\bibfnamefont
  {A.}~\bibnamefont {Aspuru-Guzik}}, \ and\ \bibinfo {author} {\bibfnamefont
  {J.~M.}\ \bibnamefont {Martinis}},\ }\href {\doibase
  10.1103/PhysRevX.6.031007} {\bibfield  {journal} {\bibinfo  {journal} {Phys.
  Rev. X}\ }\textbf {\bibinfo {volume} {6}},\ \bibinfo {pages} {031007}
  (\bibinfo {year} {2016})}\BibitemShut {NoStop}%
\bibitem [{\citenamefont {Kandala}\ \emph {et~al.}(2017)\citenamefont
  {Kandala}, \citenamefont {Mezzacapo}, \citenamefont {Temme}, \citenamefont
  {Takita}, \citenamefont {Brink}, \citenamefont {Chow},\ and\ \citenamefont
  {Gambetta}}]{Kandala2017}%
  \BibitemOpen
  \bibfield  {author} {\bibinfo {author} {\bibfnamefont {A.}~\bibnamefont
  {Kandala}}, \bibinfo {author} {\bibfnamefont {A.}~\bibnamefont {Mezzacapo}},
  \bibinfo {author} {\bibfnamefont {K.}~\bibnamefont {Temme}}, \bibinfo
  {author} {\bibfnamefont {M.}~\bibnamefont {Takita}}, \bibinfo {author}
  {\bibfnamefont {M.}~\bibnamefont {Brink}}, \bibinfo {author} {\bibfnamefont
  {J.~M.}\ \bibnamefont {Chow}}, \ and\ \bibinfo {author} {\bibfnamefont
  {J.~M.}\ \bibnamefont {Gambetta}},\ }\href {\doibase 10.1038/nature23879}
  {\bibfield  {journal} {\bibinfo  {journal} {Nature}\ }\textbf {\bibinfo
  {volume} {549}},\ \bibinfo {pages} {242} (\bibinfo {year}
  {2017})}\BibitemShut {NoStop}%
\bibitem [{\citenamefont {Peruzzo}\ \emph {et~al.}(2014)\citenamefont
  {Peruzzo}, \citenamefont {McClean}, \citenamefont {Shadbolt}, \citenamefont
  {Yung}, \citenamefont {Zhou}, \citenamefont {Love}, \citenamefont
  {Aspuru-Guzik},\ and\ \citenamefont {O'Brien}}]{Peruzzo2014}%
  \BibitemOpen
  \bibfield  {author} {\bibinfo {author} {\bibfnamefont {A.}~\bibnamefont
  {Peruzzo}}, \bibinfo {author} {\bibfnamefont {J.}~\bibnamefont {McClean}},
  \bibinfo {author} {\bibfnamefont {P.}~\bibnamefont {Shadbolt}}, \bibinfo
  {author} {\bibfnamefont {M.-H.}\ \bibnamefont {Yung}}, \bibinfo {author}
  {\bibfnamefont {X.-Q.}\ \bibnamefont {Zhou}}, \bibinfo {author}
  {\bibfnamefont {P.~J.}\ \bibnamefont {Love}}, \bibinfo {author}
  {\bibfnamefont {A.}~\bibnamefont {Aspuru-Guzik}}, \ and\ \bibinfo {author}
  {\bibfnamefont {J.~L.}\ \bibnamefont {O'Brien}},\ }\href {\doibase
  10.1038/ncomms5213} {\bibfield  {journal} {\bibinfo  {journal} {Nature
  Communications}\ }\textbf {\bibinfo {volume} {5}},\ \bibinfo {pages} {4213}
  (\bibinfo {year} {2014})}\BibitemShut {NoStop}%
\bibitem [{\citenamefont {Colless}\ \emph {et~al.}(2018)\citenamefont
  {Colless}, \citenamefont {Ramasesh}, \citenamefont {Dahlen}, \citenamefont
  {Blok}, \citenamefont {Kimchi-Schwartz}, \citenamefont {McClean},
  \citenamefont {Carter}, \citenamefont {de~Jong},\ and\ \citenamefont
  {Siddiqi}}]{PhysRevX.8.011021}%
  \BibitemOpen
  \bibfield  {author} {\bibinfo {author} {\bibfnamefont {J.~I.}\ \bibnamefont
  {Colless}}, \bibinfo {author} {\bibfnamefont {V.~V.}\ \bibnamefont
  {Ramasesh}}, \bibinfo {author} {\bibfnamefont {D.}~\bibnamefont {Dahlen}},
  \bibinfo {author} {\bibfnamefont {M.~S.}\ \bibnamefont {Blok}}, \bibinfo
  {author} {\bibfnamefont {M.~E.}\ \bibnamefont {Kimchi-Schwartz}}, \bibinfo
  {author} {\bibfnamefont {J.~R.}\ \bibnamefont {McClean}}, \bibinfo {author}
  {\bibfnamefont {J.}~\bibnamefont {Carter}}, \bibinfo {author} {\bibfnamefont
  {W.~A.}\ \bibnamefont {de~Jong}}, \ and\ \bibinfo {author} {\bibfnamefont
  {I.}~\bibnamefont {Siddiqi}},\ }\href {\doibase 10.1103/PhysRevX.8.011021}
  {\bibfield  {journal} {\bibinfo  {journal} {Phys. Rev. X}\ }\textbf {\bibinfo
  {volume} {8}},\ \bibinfo {pages} {011021} (\bibinfo {year}
  {2018})}\BibitemShut {NoStop}%
\bibitem [{\citenamefont {Chiesa}\ \emph {et~al.}(2019)\citenamefont {Chiesa},
  \citenamefont {Tacchino}, \citenamefont {Grossi}, \citenamefont {Santini},
  \citenamefont {Tavernelli}, \citenamefont {Gerace},\ and\ \citenamefont
  {Carretta}}]{Chiesa2019}%
  \BibitemOpen
  \bibfield  {author} {\bibinfo {author} {\bibfnamefont {A.}~\bibnamefont
  {Chiesa}}, \bibinfo {author} {\bibfnamefont {F.}~\bibnamefont {Tacchino}},
  \bibinfo {author} {\bibfnamefont {M.}~\bibnamefont {Grossi}}, \bibinfo
  {author} {\bibfnamefont {P.}~\bibnamefont {Santini}}, \bibinfo {author}
  {\bibfnamefont {I.}~\bibnamefont {Tavernelli}}, \bibinfo {author}
  {\bibfnamefont {D.}~\bibnamefont {Gerace}}, \ and\ \bibinfo {author}
  {\bibfnamefont {S.}~\bibnamefont {Carretta}},\ }\href {\doibase
  10.1038/s41567-019-0437-4} {\bibfield  {journal} {\bibinfo  {journal} {Nature
  Physics}\ }\textbf {\bibinfo {volume} {15}},\ \bibinfo {pages} {455}
  (\bibinfo {year} {2019})}\BibitemShut {NoStop}%
\bibitem [{\citenamefont {Smith}\ \emph {et~al.}(2019)\citenamefont {Smith},
  \citenamefont {Kim}, \citenamefont {Pollmann},\ and\ \citenamefont
  {Knolle}}]{Smith2019}%
  \BibitemOpen
  \bibfield  {author} {\bibinfo {author} {\bibfnamefont {A.}~\bibnamefont
  {Smith}}, \bibinfo {author} {\bibfnamefont {M.~S.}\ \bibnamefont {Kim}},
  \bibinfo {author} {\bibfnamefont {F.}~\bibnamefont {Pollmann}}, \ and\
  \bibinfo {author} {\bibfnamefont {J.}~\bibnamefont {Knolle}},\ }\href
  {\doibase 10.1038/s41534-019-0217-0} {\bibfield  {journal} {\bibinfo
  {journal} {npj Quantum Information}\ }\textbf {\bibinfo {volume} {5}},\
  \bibinfo {pages} {106} (\bibinfo {year} {2019})}\BibitemShut {NoStop}%
\bibitem [{\citenamefont {Zhang}\ \emph {et~al.}(2017)\citenamefont {Zhang},
  \citenamefont {Pagano}, \citenamefont {Hess}, \citenamefont {Kyprianidis},
  \citenamefont {Becker}, \citenamefont {Kaplan}, \citenamefont {Gorshkov},
  \citenamefont {Gong},\ and\ \citenamefont {Monroe}}]{Zhang2017}%
  \BibitemOpen
  \bibfield  {author} {\bibinfo {author} {\bibfnamefont {J.}~\bibnamefont
  {Zhang}}, \bibinfo {author} {\bibfnamefont {G.}~\bibnamefont {Pagano}},
  \bibinfo {author} {\bibfnamefont {P.~W.}\ \bibnamefont {Hess}}, \bibinfo
  {author} {\bibfnamefont {A.}~\bibnamefont {Kyprianidis}}, \bibinfo {author}
  {\bibfnamefont {P.}~\bibnamefont {Becker}}, \bibinfo {author} {\bibfnamefont
  {H.}~\bibnamefont {Kaplan}}, \bibinfo {author} {\bibfnamefont {A.~V.}\
  \bibnamefont {Gorshkov}}, \bibinfo {author} {\bibfnamefont {Z.~X.}\
  \bibnamefont {Gong}}, \ and\ \bibinfo {author} {\bibfnamefont
  {C.}~\bibnamefont {Monroe}},\ }\href {\doibase 10.1038/nature24654}
  {\bibfield  {journal} {\bibinfo  {journal} {Nature}\ }\textbf {\bibinfo
  {volume} {551}},\ \bibinfo {pages} {601} (\bibinfo {year}
  {2017})}\BibitemShut {NoStop}%
\bibitem [{\citenamefont {Islam}\ \emph {et~al.}(2013)\citenamefont {Islam},
  \citenamefont {Senko}, \citenamefont {Campbell}, \citenamefont {Korenblit},
  \citenamefont {Smith}, \citenamefont {Lee}, \citenamefont {Edwards},
  \citenamefont {Wang}, \citenamefont {Freericks},\ and\ \citenamefont
  {Monroe}}]{doi:10.1126/science.1232296}%
  \BibitemOpen
  \bibfield  {author} {\bibinfo {author} {\bibfnamefont {R.}~\bibnamefont
  {Islam}}, \bibinfo {author} {\bibfnamefont {C.}~\bibnamefont {Senko}},
  \bibinfo {author} {\bibfnamefont {W.~C.}\ \bibnamefont {Campbell}}, \bibinfo
  {author} {\bibfnamefont {S.}~\bibnamefont {Korenblit}}, \bibinfo {author}
  {\bibfnamefont {J.}~\bibnamefont {Smith}}, \bibinfo {author} {\bibfnamefont
  {A.}~\bibnamefont {Lee}}, \bibinfo {author} {\bibfnamefont {E.~E.}\
  \bibnamefont {Edwards}}, \bibinfo {author} {\bibfnamefont {C.-C.~J.}\
  \bibnamefont {Wang}}, \bibinfo {author} {\bibfnamefont {J.~K.}\ \bibnamefont
  {Freericks}}, \ and\ \bibinfo {author} {\bibfnamefont {C.}~\bibnamefont
  {Monroe}},\ }\href {\doibase 10.1126/science.1232296} {\bibfield  {journal}
  {\bibinfo  {journal} {Science}\ }\textbf {\bibinfo {volume} {340}},\ \bibinfo
  {pages} {583} (\bibinfo {year} {2013})}\BibitemShut {NoStop}%
\bibitem [{\citenamefont {Francis}\ \emph {et~al.}(2020)\citenamefont
  {Francis}, \citenamefont {Freericks},\ and\ \citenamefont
  {Kemper}}]{PhysRevB.101.014411}%
  \BibitemOpen
  \bibfield  {author} {\bibinfo {author} {\bibfnamefont {A.}~\bibnamefont
  {Francis}}, \bibinfo {author} {\bibfnamefont {J.~K.}\ \bibnamefont
  {Freericks}}, \ and\ \bibinfo {author} {\bibfnamefont {A.~F.}\ \bibnamefont
  {Kemper}},\ }\href {\doibase 10.1103/PhysRevB.101.014411} {\bibfield
  {journal} {\bibinfo  {journal} {Phys. Rev. B}\ }\textbf {\bibinfo {volume}
  {101}},\ \bibinfo {pages} {014411} (\bibinfo {year} {2020})}\BibitemShut
  {NoStop}%
\bibitem [{\citenamefont {Lloyd}(1996)}]{doi:10.1126/science.273.5278.1073}%
  \BibitemOpen
  \bibfield  {author} {\bibinfo {author} {\bibfnamefont {S.}~\bibnamefont
  {Lloyd}},\ }\href {\doibase 10.1126/science.273.5278.1073} {\bibfield
  {journal} {\bibinfo  {journal} {Science}\ }\textbf {\bibinfo {volume}
  {273}},\ \bibinfo {pages} {1073} (\bibinfo {year} {1996})}\BibitemShut
  {NoStop}%
\bibitem [{\citenamefont {De~Jong}\ \emph {et~al.}(2021)\citenamefont
  {De~Jong}, \citenamefont {Metcalf}, \citenamefont {Mulligan}, \citenamefont
  {P\l{}osko\'n}, \citenamefont {Ringer},\ and\ \citenamefont
  {Yao}}]{DeJong:2020riy}%
  \BibitemOpen
  \bibfield  {author} {\bibinfo {author} {\bibfnamefont {W.~A.}\ \bibnamefont
  {De~Jong}}, \bibinfo {author} {\bibfnamefont {M.}~\bibnamefont {Metcalf}},
  \bibinfo {author} {\bibfnamefont {J.}~\bibnamefont {Mulligan}}, \bibinfo
  {author} {\bibfnamefont {M.}~\bibnamefont {P\l{}osko\'n}}, \bibinfo {author}
  {\bibfnamefont {F.}~\bibnamefont {Ringer}}, \ and\ \bibinfo {author}
  {\bibfnamefont {X.}~\bibnamefont {Yao}},\ }\href {\doibase
  10.1103/PhysRevD.104.L051501} {\bibfield  {journal} {\bibinfo  {journal}
  {Phys. Rev. D}\ }\textbf {\bibinfo {volume} {104}},\ \bibinfo {pages}
  {051501} (\bibinfo {year} {2021})},\ \Eprint
  {http://arxiv.org/abs/2010.03571} {arXiv:2010.03571 [hep-ph]} \BibitemShut
  {NoStop}%
\bibitem [{\citenamefont {de~Jong}\ \emph {et~al.}(2021)\citenamefont
  {de~Jong}, \citenamefont {Lee}, \citenamefont {Mulligan}, \citenamefont
  {P\l{}osko\'n}, \citenamefont {Ringer},\ and\ \citenamefont
  {Yao}}]{deJong:2021wsd}%
  \BibitemOpen
  \bibfield  {author} {\bibinfo {author} {\bibfnamefont {W.~A.}\ \bibnamefont
  {de~Jong}}, \bibinfo {author} {\bibfnamefont {K.}~\bibnamefont {Lee}},
  \bibinfo {author} {\bibfnamefont {J.}~\bibnamefont {Mulligan}}, \bibinfo
  {author} {\bibfnamefont {M.}~\bibnamefont {P\l{}osko\'n}}, \bibinfo {author}
  {\bibfnamefont {F.}~\bibnamefont {Ringer}}, \ and\ \bibinfo {author}
  {\bibfnamefont {X.}~\bibnamefont {Yao}},\ }\href@noop {} {\  (\bibinfo {year}
  {2021})},\ \Eprint {http://arxiv.org/abs/2106.08394} {arXiv:2106.08394
  [quant-ph]} \BibitemShut {NoStop}%
\bibitem [{\citenamefont {Wallraff}\ \emph {et~al.}(2004)\citenamefont
  {Wallraff}, \citenamefont {Schuster}, \citenamefont {Blais}, \citenamefont
  {Frunzio}, \citenamefont {Huang}, \citenamefont {Majer}, \citenamefont
  {Kumar}, \citenamefont {Girvin},\ and\ \citenamefont
  {Schoelkopf}}]{wallraff_strong_2004}%
  \BibitemOpen
  \bibfield  {author} {\bibinfo {author} {\bibfnamefont {A.}~\bibnamefont
  {Wallraff}}, \bibinfo {author} {\bibfnamefont {D.~I.}\ \bibnamefont
  {Schuster}}, \bibinfo {author} {\bibfnamefont {A.}~\bibnamefont {Blais}},
  \bibinfo {author} {\bibfnamefont {L.}~\bibnamefont {Frunzio}}, \bibinfo
  {author} {\bibfnamefont {R.-S.}\ \bibnamefont {Huang}}, \bibinfo {author}
  {\bibfnamefont {J.}~\bibnamefont {Majer}}, \bibinfo {author} {\bibfnamefont
  {S.}~\bibnamefont {Kumar}}, \bibinfo {author} {\bibfnamefont {S.~M.}\
  \bibnamefont {Girvin}}, \ and\ \bibinfo {author} {\bibfnamefont {R.~J.}\
  \bibnamefont {Schoelkopf}},\ }\href {\doibase 10.1038/nature02851} {\bibfield
   {journal} {\bibinfo  {journal} {Nature}\ }\textbf {\bibinfo {volume}
  {431}},\ \bibinfo {pages} {162} (\bibinfo {year} {2004})},\ \bibinfo {note}
  {bandiera\_abtest: a Cg\_type: Nature Research Journals Number: 7005
  Primary\_atype: Research Publisher: Nature Publishing Group}\BibitemShut
  {NoStop}%
\bibitem [{\citenamefont {Majer}\ \emph {et~al.}(2007)\citenamefont {Majer},
  \citenamefont {Chow}, \citenamefont {Gambetta}, \citenamefont {Koch},
  \citenamefont {Johnson}, \citenamefont {Schreier}, \citenamefont {Frunzio},
  \citenamefont {Schuster}, \citenamefont {Houck}, \citenamefont {Wallraff},
  \citenamefont {Blais}, \citenamefont {Devoret}, \citenamefont {Girvin},\ and\
  \citenamefont {Schoelkopf}}]{majer_coupling_2007}%
  \BibitemOpen
  \bibfield  {author} {\bibinfo {author} {\bibfnamefont {J.}~\bibnamefont
  {Majer}}, \bibinfo {author} {\bibfnamefont {J.~M.}\ \bibnamefont {Chow}},
  \bibinfo {author} {\bibfnamefont {J.~M.}\ \bibnamefont {Gambetta}}, \bibinfo
  {author} {\bibfnamefont {J.}~\bibnamefont {Koch}}, \bibinfo {author}
  {\bibfnamefont {B.~R.}\ \bibnamefont {Johnson}}, \bibinfo {author}
  {\bibfnamefont {J.~A.}\ \bibnamefont {Schreier}}, \bibinfo {author}
  {\bibfnamefont {L.}~\bibnamefont {Frunzio}}, \bibinfo {author} {\bibfnamefont
  {D.~I.}\ \bibnamefont {Schuster}}, \bibinfo {author} {\bibfnamefont {A.~A.}\
  \bibnamefont {Houck}}, \bibinfo {author} {\bibfnamefont {A.}~\bibnamefont
  {Wallraff}}, \bibinfo {author} {\bibfnamefont {A.}~\bibnamefont {Blais}},
  \bibinfo {author} {\bibfnamefont {M.~H.}\ \bibnamefont {Devoret}}, \bibinfo
  {author} {\bibfnamefont {S.~M.}\ \bibnamefont {Girvin}}, \ and\ \bibinfo
  {author} {\bibfnamefont {R.~J.}\ \bibnamefont {Schoelkopf}},\ }\href
  {\doibase 10.1038/nature06184} {\bibfield  {journal} {\bibinfo  {journal}
  {Nature}\ }\textbf {\bibinfo {volume} {449}},\ \bibinfo {pages} {443}
  (\bibinfo {year} {2007})},\ \bibinfo {note} {bandiera\_abtest: a Cg\_type:
  Nature Research Journals Number: 7161 Primary\_atype: Research Publisher:
  Nature Publishing Group}\BibitemShut {NoStop}%
\bibitem [{\citenamefont {Jordan}\ \emph {et~al.}(2012)\citenamefont {Jordan},
  \citenamefont {Lee},\ and\ \citenamefont {Preskill}}]{jordan_quantum_2012}%
  \BibitemOpen
  \bibfield  {author} {\bibinfo {author} {\bibfnamefont {S.~P.}\ \bibnamefont
  {Jordan}}, \bibinfo {author} {\bibfnamefont {K.~S.~M.}\ \bibnamefont {Lee}},
  \ and\ \bibinfo {author} {\bibfnamefont {J.}~\bibnamefont {Preskill}},\
  }\href {\doibase 10.1126/science.1217069} {\bibfield  {journal} {\bibinfo
  {journal} {Science}\ }\textbf {\bibinfo {volume} {336}},\ \bibinfo {pages}
  {1130} (\bibinfo {year} {2012})},\ \bibinfo {note} {arXiv:
  1111.3633}\BibitemShut {NoStop}%
\bibitem [{\citenamefont {Zohar}\ \emph {et~al.}(2012)\citenamefont {Zohar},
  \citenamefont {Cirac},\ and\ \citenamefont {Reznik}}]{zohar_simulating_2012}%
  \BibitemOpen
  \bibfield  {author} {\bibinfo {author} {\bibfnamefont {E.}~\bibnamefont
  {Zohar}}, \bibinfo {author} {\bibfnamefont {J.~I.}\ \bibnamefont {Cirac}}, \
  and\ \bibinfo {author} {\bibfnamefont {B.}~\bibnamefont {Reznik}},\ }\href
  {\doibase 10.1103/PhysRevLett.109.125302} {\bibfield  {journal} {\bibinfo
  {journal} {Physical Review Letters}\ }\textbf {\bibinfo {volume} {109}},\
  \bibinfo {pages} {125302} (\bibinfo {year} {2012})},\ \bibinfo {note}
  {publisher: American Physical Society}\BibitemShut {NoStop}%
\bibitem [{\citenamefont {Zohar}\ \emph {et~al.}(2013)\citenamefont {Zohar},
  \citenamefont {Cirac},\ and\ \citenamefont {Reznik}}]{zohar_cold-atom_2013}%
  \BibitemOpen
  \bibfield  {author} {\bibinfo {author} {\bibfnamefont {E.}~\bibnamefont
  {Zohar}}, \bibinfo {author} {\bibfnamefont {J.~I.}\ \bibnamefont {Cirac}}, \
  and\ \bibinfo {author} {\bibfnamefont {B.}~\bibnamefont {Reznik}},\ }\href
  {\doibase 10.1103/PhysRevLett.110.125304} {\bibfield  {journal} {\bibinfo
  {journal} {Physical Review Letters}\ }\textbf {\bibinfo {volume} {110}},\
  \bibinfo {pages} {125304} (\bibinfo {year} {2013})},\ \bibinfo {note}
  {publisher: American Physical Society}\BibitemShut {NoStop}%
\bibitem [{\citenamefont {Banerjee}\ \emph {et~al.}(2013)\citenamefont
  {Banerjee}, \citenamefont {B{\"o}gli}, \citenamefont {Dalmonte},
  \citenamefont {Rico}, \citenamefont {Stebler}, \citenamefont {Wiese},\ and\
  \citenamefont {Zoller}}]{banerjee_atomic_2013}%
  \BibitemOpen
  \bibfield  {author} {\bibinfo {author} {\bibfnamefont {D.}~\bibnamefont
  {Banerjee}}, \bibinfo {author} {\bibfnamefont {M.}~\bibnamefont {B{\"o}gli}},
  \bibinfo {author} {\bibfnamefont {M.}~\bibnamefont {Dalmonte}}, \bibinfo
  {author} {\bibfnamefont {E.}~\bibnamefont {Rico}}, \bibinfo {author}
  {\bibfnamefont {P.}~\bibnamefont {Stebler}}, \bibinfo {author} {\bibfnamefont
  {U.-J.}\ \bibnamefont {Wiese}}, \ and\ \bibinfo {author} {\bibfnamefont
  {P.}~\bibnamefont {Zoller}},\ }\href {\doibase
  10.1103/PhysRevLett.110.125303} {\bibfield  {journal} {\bibinfo  {journal}
  {Physical Review Letters}\ }\textbf {\bibinfo {volume} {110}},\ \bibinfo
  {pages} {125303} (\bibinfo {year} {2013})},\ \bibinfo {note} {publisher:
  American Physical Society}\BibitemShut {NoStop}%
\bibitem [{\citenamefont {Banerjee}\ \emph {et~al.}(2012)\citenamefont
  {Banerjee}, \citenamefont {Dalmonte}, \citenamefont {M{\"u}ller},
  \citenamefont {Rico}, \citenamefont {Stebler}, \citenamefont {Wiese},\ and\
  \citenamefont {Zoller}}]{banerjee_atomic_2012}%
  \BibitemOpen
  \bibfield  {author} {\bibinfo {author} {\bibfnamefont {D.}~\bibnamefont
  {Banerjee}}, \bibinfo {author} {\bibfnamefont {M.}~\bibnamefont {Dalmonte}},
  \bibinfo {author} {\bibfnamefont {M.}~\bibnamefont {M{\"u}ller}}, \bibinfo
  {author} {\bibfnamefont {E.}~\bibnamefont {Rico}}, \bibinfo {author}
  {\bibfnamefont {P.}~\bibnamefont {Stebler}}, \bibinfo {author} {\bibfnamefont
  {U.-J.}\ \bibnamefont {Wiese}}, \ and\ \bibinfo {author} {\bibfnamefont
  {P.}~\bibnamefont {Zoller}},\ }\href {\doibase
  10.1103/PhysRevLett.109.175302} {\bibfield  {journal} {\bibinfo  {journal}
  {Physical Review Letters}\ }\textbf {\bibinfo {volume} {109}},\ \bibinfo
  {pages} {175302} (\bibinfo {year} {2012})},\ \bibinfo {note} {arXiv:
  1205.6366}\BibitemShut {NoStop}%
\bibitem [{\citenamefont {Wiese}(2013)}]{wiese_ultracold_2013}%
  \BibitemOpen
  \bibfield  {author} {\bibinfo {author} {\bibfnamefont {U.-J.}\ \bibnamefont
  {Wiese}},\ }\href {\doibase 10.1002/andp.201300104} {\bibfield  {journal}
  {\bibinfo  {journal} {Annalen der Physik}\ }\textbf {\bibinfo {volume}
  {525}},\ \bibinfo {pages} {777} (\bibinfo {year} {2013})},\ \bibinfo {note}
  {\_eprint:
  https://onlinelibrary.wiley.com/doi/pdf/10.1002/andp.201300104}\BibitemShut
  {NoStop}%
\bibitem [{\citenamefont {Wiese}(2014)}]{wiese_towards_2014}%
  \BibitemOpen
  \bibfield  {author} {\bibinfo {author} {\bibfnamefont {U.-J.}\ \bibnamefont
  {Wiese}},\ }\href {\doibase 10.1016/j.nuclphysa.2014.09.102} {\bibfield
  {journal} {\bibinfo  {journal} {Nuclear Physics A}\ }\bibinfo {series}
  {{QUARK} {MATTER} 2014},\ \textbf {\bibinfo {volume} {931}},\ \bibinfo
  {pages} {246} (\bibinfo {year} {2014})}\BibitemShut {NoStop}%
\bibitem [{\citenamefont {Jordan}\ \emph {et~al.}(2014)\citenamefont {Jordan},
  \citenamefont {Lee},\ and\ \citenamefont {Preskill}}]{jordan_quantum_2014}%
  \BibitemOpen
  \bibfield  {author} {\bibinfo {author} {\bibfnamefont {S.~P.}\ \bibnamefont
  {Jordan}}, \bibinfo {author} {\bibfnamefont {K.~S.~M.}\ \bibnamefont {Lee}},
  \ and\ \bibinfo {author} {\bibfnamefont {J.}~\bibnamefont {Preskill}},\
  }\href {http://arxiv.org/abs/1404.7115} {\bibfield  {journal} {\bibinfo
  {journal} {arXiv:1404.7115 [hep-th, physics:quant-ph]}\ } (\bibinfo {year}
  {2014})},\ \bibinfo {note} {arXiv: 1404.7115}\BibitemShut {NoStop}%
\bibitem [{\citenamefont {Garc{\'i}a-{\'A}lvarez}\ \emph
  {et~al.}(2015)\citenamefont {Garc{\'i}a-{\'A}lvarez}, \citenamefont
  {Casanova}, \citenamefont {Mezzacapo}, \citenamefont {Egusquiza},
  \citenamefont {Lamata}, \citenamefont {Romero},\ and\ \citenamefont
  {Solano}}]{garcia-alvarez_fermion-fermion_2015}%
  \BibitemOpen
  \bibfield  {author} {\bibinfo {author} {\bibfnamefont {L.}~\bibnamefont
  {Garc{\'i}a-{\'A}lvarez}}, \bibinfo {author} {\bibfnamefont {J.}~\bibnamefont
  {Casanova}}, \bibinfo {author} {\bibfnamefont {A.}~\bibnamefont {Mezzacapo}},
  \bibinfo {author} {\bibfnamefont {I.~L.}\ \bibnamefont {Egusquiza}}, \bibinfo
  {author} {\bibfnamefont {L.}~\bibnamefont {Lamata}}, \bibinfo {author}
  {\bibfnamefont {G.}~\bibnamefont {Romero}}, \ and\ \bibinfo {author}
  {\bibfnamefont {E.}~\bibnamefont {Solano}},\ }\href {\doibase
  10.1103/PhysRevLett.114.070502} {\bibfield  {journal} {\bibinfo  {journal}
  {Physical Review Letters}\ }\textbf {\bibinfo {volume} {114}},\ \bibinfo
  {pages} {070502} (\bibinfo {year} {2015})},\ \bibinfo {note} {publisher:
  American Physical Society}\BibitemShut {NoStop}%
\bibitem [{\citenamefont {Marcos}\ \emph {et~al.}(2014)\citenamefont {Marcos},
  \citenamefont {Widmer}, \citenamefont {Rico}, \citenamefont {Hafezi},
  \citenamefont {Rabl}, \citenamefont {Wiese},\ and\ \citenamefont
  {Zoller}}]{marcos_two-dimensional_2014}%
  \BibitemOpen
  \bibfield  {author} {\bibinfo {author} {\bibfnamefont {D.}~\bibnamefont
  {Marcos}}, \bibinfo {author} {\bibfnamefont {P.}~\bibnamefont {Widmer}},
  \bibinfo {author} {\bibfnamefont {E.}~\bibnamefont {Rico}}, \bibinfo {author}
  {\bibfnamefont {M.}~\bibnamefont {Hafezi}}, \bibinfo {author} {\bibfnamefont
  {P.}~\bibnamefont {Rabl}}, \bibinfo {author} {\bibfnamefont {U.~J.}\
  \bibnamefont {Wiese}}, \ and\ \bibinfo {author} {\bibfnamefont
  {P.}~\bibnamefont {Zoller}},\ }\href {\doibase 10.1016/j.aop.2014.09.011}
  {\bibfield  {journal} {\bibinfo  {journal} {Annals of Physics}\ }\textbf
  {\bibinfo {volume} {351}},\ \bibinfo {pages} {634} (\bibinfo {year}
  {2014})}\BibitemShut {NoStop}%
\bibitem [{\citenamefont {Bazavov}\ \emph {et~al.}(2015)\citenamefont
  {Bazavov}, \citenamefont {Meurice}, \citenamefont {Tsai}, \citenamefont
  {Unmuth-Yockey},\ and\ \citenamefont {Zhang}}]{bazavov_gauge-invariant_2015}%
  \BibitemOpen
  \bibfield  {author} {\bibinfo {author} {\bibfnamefont {A.}~\bibnamefont
  {Bazavov}}, \bibinfo {author} {\bibfnamefont {Y.}~\bibnamefont {Meurice}},
  \bibinfo {author} {\bibfnamefont {S.-W.}\ \bibnamefont {Tsai}}, \bibinfo
  {author} {\bibfnamefont {J.}~\bibnamefont {Unmuth-Yockey}}, \ and\ \bibinfo
  {author} {\bibfnamefont {J.}~\bibnamefont {Zhang}},\ }\href {\doibase
  10.1103/PhysRevD.92.076003} {\bibfield  {journal} {\bibinfo  {journal}
  {Physical Review D}\ }\textbf {\bibinfo {volume} {92}},\ \bibinfo {pages}
  {076003} (\bibinfo {year} {2015})},\ \bibinfo {note} {publisher: American
  Physical Society}\BibitemShut {NoStop}%
\bibitem [{\citenamefont {Zohar}\ \emph {et~al.}(2015)\citenamefont {Zohar},
  \citenamefont {Cirac},\ and\ \citenamefont {Reznik}}]{zohar_quantum_2015}%
  \BibitemOpen
  \bibfield  {author} {\bibinfo {author} {\bibfnamefont {E.}~\bibnamefont
  {Zohar}}, \bibinfo {author} {\bibfnamefont {J.~I.}\ \bibnamefont {Cirac}}, \
  and\ \bibinfo {author} {\bibfnamefont {B.}~\bibnamefont {Reznik}},\ }\href
  {\doibase 10.1088/0034-4885/79/1/014401} {\bibfield  {journal} {\bibinfo
  {journal} {Reports on Progress in Physics}\ }\textbf {\bibinfo {volume}
  {79}},\ \bibinfo {pages} {014401} (\bibinfo {year} {2015})},\ \bibinfo {note}
  {publisher: IOP Publishing}\BibitemShut {NoStop}%
\bibitem [{\citenamefont {Mezzacapo}\ \emph {et~al.}(2015)\citenamefont
  {Mezzacapo}, \citenamefont {Rico}, \citenamefont {Sab{\'i}n}, \citenamefont
  {Egusquiza}, \citenamefont {Lamata},\ and\ \citenamefont
  {Solano}}]{mezzacapo_non-abelian_2015}%
  \BibitemOpen
  \bibfield  {author} {\bibinfo {author} {\bibfnamefont {A.}~\bibnamefont
  {Mezzacapo}}, \bibinfo {author} {\bibfnamefont {E.}~\bibnamefont {Rico}},
  \bibinfo {author} {\bibfnamefont {C.}~\bibnamefont {Sab{\'i}n}}, \bibinfo
  {author} {\bibfnamefont {I.~L.}\ \bibnamefont {Egusquiza}}, \bibinfo {author}
  {\bibfnamefont {L.}~\bibnamefont {Lamata}}, \ and\ \bibinfo {author}
  {\bibfnamefont {E.}~\bibnamefont {Solano}},\ }\href {\doibase
  10.1103/PhysRevLett.115.240502} {\bibfield  {journal} {\bibinfo  {journal}
  {Physical Review Letters}\ }\textbf {\bibinfo {volume} {115}},\ \bibinfo
  {pages} {240502} (\bibinfo {year} {2015})},\ \bibinfo {note} {publisher:
  American Physical Society}\BibitemShut {NoStop}%
\bibitem [{\citenamefont {Dalmonte}\ and\ \citenamefont
  {Montangero}(2016)}]{dalmonte_lattice_2016}%
  \BibitemOpen
  \bibfield  {author} {\bibinfo {author} {\bibfnamefont {M.}~\bibnamefont
  {Dalmonte}}\ and\ \bibinfo {author} {\bibfnamefont {S.}~\bibnamefont
  {Montangero}},\ }\href {\doibase 10.1080/00107514.2016.1151199} {\bibfield
  {journal} {\bibinfo  {journal} {Contemporary Physics}\ }\textbf {\bibinfo
  {volume} {57}},\ \bibinfo {pages} {388} (\bibinfo {year} {2016})},\ \bibinfo
  {note} {arXiv: 1602.03776}\BibitemShut {NoStop}%
\bibitem [{\citenamefont {Zohar}\ \emph {et~al.}(2017)\citenamefont {Zohar},
  \citenamefont {Farace}, \citenamefont {Reznik},\ and\ \citenamefont
  {Cirac}}]{zohar_digital_2017}%
  \BibitemOpen
  \bibfield  {author} {\bibinfo {author} {\bibfnamefont {E.}~\bibnamefont
  {Zohar}}, \bibinfo {author} {\bibfnamefont {A.}~\bibnamefont {Farace}},
  \bibinfo {author} {\bibfnamefont {B.}~\bibnamefont {Reznik}}, \ and\ \bibinfo
  {author} {\bibfnamefont {J.~I.}\ \bibnamefont {Cirac}},\ }\href {\doibase
  10.1103/PhysRevA.95.023604} {\bibfield  {journal} {\bibinfo  {journal}
  {Physical Review A}\ }\textbf {\bibinfo {volume} {95}},\ \bibinfo {pages}
  {023604} (\bibinfo {year} {2017})},\ \bibinfo {note} {arXiv:
  1607.08121}\BibitemShut {NoStop}%
\bibitem [{\citenamefont {Martinez}\ \emph {et~al.}(2016)\citenamefont
  {Martinez}, \citenamefont {Muschik}, \citenamefont {Schindler}, \citenamefont
  {Nigg}, \citenamefont {Erhard}, \citenamefont {Heyl}, \citenamefont {Hauke},
  \citenamefont {Dalmonte}, \citenamefont {Monz}, \citenamefont {Zoller},\ and\
  \citenamefont {Blatt}}]{martinez_real-time_2016}%
  \BibitemOpen
  \bibfield  {author} {\bibinfo {author} {\bibfnamefont {E.~A.}\ \bibnamefont
  {Martinez}}, \bibinfo {author} {\bibfnamefont {C.~A.}\ \bibnamefont
  {Muschik}}, \bibinfo {author} {\bibfnamefont {P.}~\bibnamefont {Schindler}},
  \bibinfo {author} {\bibfnamefont {D.}~\bibnamefont {Nigg}}, \bibinfo {author}
  {\bibfnamefont {A.}~\bibnamefont {Erhard}}, \bibinfo {author} {\bibfnamefont
  {M.}~\bibnamefont {Heyl}}, \bibinfo {author} {\bibfnamefont {P.}~\bibnamefont
  {Hauke}}, \bibinfo {author} {\bibfnamefont {M.}~\bibnamefont {Dalmonte}},
  \bibinfo {author} {\bibfnamefont {T.}~\bibnamefont {Monz}}, \bibinfo {author}
  {\bibfnamefont {P.}~\bibnamefont {Zoller}}, \ and\ \bibinfo {author}
  {\bibfnamefont {R.}~\bibnamefont {Blatt}},\ }\href {\doibase
  10.1038/nature18318} {\bibfield  {journal} {\bibinfo  {journal} {Nature}\
  }\textbf {\bibinfo {volume} {534}},\ \bibinfo {pages} {516} (\bibinfo {year}
  {2016})},\ \bibinfo {note} {arXiv: 1605.04570}\BibitemShut {NoStop}%
\bibitem [{\citenamefont {Bermudez}\ \emph {et~al.}(2017)\citenamefont
  {Bermudez}, \citenamefont {Aarts},\ and\ \citenamefont
  {M{\"u}ller}}]{bermudez_quantum_2017}%
  \BibitemOpen
  \bibfield  {author} {\bibinfo {author} {\bibfnamefont {A.}~\bibnamefont
  {Bermudez}}, \bibinfo {author} {\bibfnamefont {G.}~\bibnamefont {Aarts}}, \
  and\ \bibinfo {author} {\bibfnamefont {M.}~\bibnamefont {M{\"u}ller}},\
  }\href {\doibase 10.1103/PhysRevX.7.041012} {\bibfield  {journal} {\bibinfo
  {journal} {Physical Review X}\ }\textbf {\bibinfo {volume} {7}},\ \bibinfo
  {pages} {041012} (\bibinfo {year} {2017})},\ \bibinfo {note} {publisher:
  American Physical Society}\BibitemShut {NoStop}%
\bibitem [{\citenamefont {Gambetta}\ \emph {et~al.}(2017)\citenamefont
  {Gambetta}, \citenamefont {Chow},\ and\ \citenamefont
  {Steffen}}]{gambetta_building_2017}%
  \BibitemOpen
  \bibfield  {author} {\bibinfo {author} {\bibfnamefont {J.~M.}\ \bibnamefont
  {Gambetta}}, \bibinfo {author} {\bibfnamefont {J.~M.}\ \bibnamefont {Chow}},
  \ and\ \bibinfo {author} {\bibfnamefont {M.}~\bibnamefont {Steffen}},\ }\href
  {\doibase 10.1038/s41534-016-0004-0} {\bibfield  {journal} {\bibinfo
  {journal} {npj Quantum Information}\ }\textbf {\bibinfo {volume} {3}},\
  \bibinfo {pages} {1} (\bibinfo {year} {2017})},\ \bibinfo {note}
  {bandiera\_abtest: a Cc\_license\_type: cc\_by Cg\_type: Nature Research
  Journals Number: 1 Primary\_atype: Reviews Publisher: Nature Publishing Group
  Subject\_term: Quantum information;Qubits Subject\_term\_id:
  quantum-information;qubits}\BibitemShut {NoStop}%
\bibitem [{\citenamefont {Krinner}\ \emph {et~al.}(2018)\citenamefont
  {Krinner}, \citenamefont {Stewart}, \citenamefont {Pazmi{\~n}o},
  \citenamefont {Kwon},\ and\ \citenamefont
  {Schneble}}]{krinner_spontaneous_2018}%
  \BibitemOpen
  \bibfield  {author} {\bibinfo {author} {\bibfnamefont {L.}~\bibnamefont
  {Krinner}}, \bibinfo {author} {\bibfnamefont {M.}~\bibnamefont {Stewart}},
  \bibinfo {author} {\bibfnamefont {A.}~\bibnamefont {Pazmi{\~n}o}}, \bibinfo
  {author} {\bibfnamefont {J.}~\bibnamefont {Kwon}}, \ and\ \bibinfo {author}
  {\bibfnamefont {D.}~\bibnamefont {Schneble}},\ }\href {\doibase
  10.1038/s41586-018-0348-z} {\bibfield  {journal} {\bibinfo  {journal}
  {Nature}\ }\textbf {\bibinfo {volume} {559}},\ \bibinfo {pages} {589}
  (\bibinfo {year} {2018})},\ \bibinfo {note} {bandiera\_abtest: a Cg\_type:
  Nature Research Journals Number: 7715 Primary\_atype: Research Publisher:
  Nature Publishing Group Subject\_term: Matter waves and particle
  beams;Quantum simulation;Single photons and quantum effects;Ultracold gases
  Subject\_term\_id:
  matter-waves-and-particle-beams;quantum-simulation;single-photons-and-quantum-effects;ultracold-gases}\BibitemShut
  {NoStop}%
\bibitem [{\citenamefont {Macridin}\ \emph {et~al.}(2018)\citenamefont
  {Macridin}, \citenamefont {Spentzouris}, \citenamefont {Amundson},\ and\
  \citenamefont {Harnik}}]{macridin_electron-phonon_2018}%
  \BibitemOpen
  \bibfield  {author} {\bibinfo {author} {\bibfnamefont {A.}~\bibnamefont
  {Macridin}}, \bibinfo {author} {\bibfnamefont {P.}~\bibnamefont
  {Spentzouris}}, \bibinfo {author} {\bibfnamefont {J.}~\bibnamefont
  {Amundson}}, \ and\ \bibinfo {author} {\bibfnamefont {R.}~\bibnamefont
  {Harnik}},\ }\href {\doibase 10.1103/PhysRevLett.121.110504} {\bibfield
  {journal} {\bibinfo  {journal} {Physical Review Letters}\ }\textbf {\bibinfo
  {volume} {121}},\ \bibinfo {pages} {110504} (\bibinfo {year} {2018})},\
  \bibinfo {note} {publisher: American Physical Society}\BibitemShut {NoStop}%
\bibitem [{\citenamefont {Zache}\ \emph {et~al.}(2018)\citenamefont {Zache},
  \citenamefont {Hebenstreit}, \citenamefont {Jendrzejewski}, \citenamefont
  {Oberthaler}, \citenamefont {Berges},\ and\ \citenamefont
  {Hauke}}]{zache_quantum_2018}%
  \BibitemOpen
  \bibfield  {author} {\bibinfo {author} {\bibfnamefont {T.~V.}\ \bibnamefont
  {Zache}}, \bibinfo {author} {\bibfnamefont {F.}~\bibnamefont {Hebenstreit}},
  \bibinfo {author} {\bibfnamefont {F.}~\bibnamefont {Jendrzejewski}}, \bibinfo
  {author} {\bibfnamefont {M.~K.}\ \bibnamefont {Oberthaler}}, \bibinfo
  {author} {\bibfnamefont {J.}~\bibnamefont {Berges}}, \ and\ \bibinfo {author}
  {\bibfnamefont {P.}~\bibnamefont {Hauke}},\ }\href {\doibase
  10.1088/2058-9565/aac33b} {\bibfield  {journal} {\bibinfo  {journal} {Quantum
  Science and Technology}\ }\textbf {\bibinfo {volume} {3}},\ \bibinfo {pages}
  {034010} (\bibinfo {year} {2018})},\ \bibinfo {note} {publisher: IOP
  Publishing}\BibitemShut {NoStop}%
\bibitem [{\citenamefont {Zhang}\ \emph {et~al.}(2018)\citenamefont {Zhang},
  \citenamefont {Unmuth-Yockey}, \citenamefont {Zeiher}, \citenamefont
  {Bazavov}, \citenamefont {Tsai},\ and\ \citenamefont
  {Meurice}}]{zhang_quantum_2018}%
  \BibitemOpen
  \bibfield  {author} {\bibinfo {author} {\bibfnamefont {J.}~\bibnamefont
  {Zhang}}, \bibinfo {author} {\bibfnamefont {J.}~\bibnamefont
  {Unmuth-Yockey}}, \bibinfo {author} {\bibfnamefont {J.}~\bibnamefont
  {Zeiher}}, \bibinfo {author} {\bibfnamefont {A.}~\bibnamefont {Bazavov}},
  \bibinfo {author} {\bibfnamefont {S.-W.}\ \bibnamefont {Tsai}}, \ and\
  \bibinfo {author} {\bibfnamefont {Y.}~\bibnamefont {Meurice}},\ }\href
  {\doibase 10.1103/PhysRevLett.121.223201} {\bibfield  {journal} {\bibinfo
  {journal} {Physical Review Letters}\ }\textbf {\bibinfo {volume} {121}},\
  \bibinfo {pages} {223201} (\bibinfo {year} {2018})},\ \bibinfo {note}
  {publisher: American Physical Society}\BibitemShut {NoStop}%
\bibitem [{\citenamefont {Klco}\ \emph {et~al.}(2018)\citenamefont {Klco},
  \citenamefont {Dumitrescu}, \citenamefont {McCaskey}, \citenamefont {Morris},
  \citenamefont {Pooser}, \citenamefont {Sanz}, \citenamefont {Solano},
  \citenamefont {Lougovski},\ and\ \citenamefont
  {Savage}}]{klco_quantum-classical_2018}%
  \BibitemOpen
  \bibfield  {author} {\bibinfo {author} {\bibfnamefont {N.}~\bibnamefont
  {Klco}}, \bibinfo {author} {\bibfnamefont {E.~F.}\ \bibnamefont
  {Dumitrescu}}, \bibinfo {author} {\bibfnamefont {A.~J.}\ \bibnamefont
  {McCaskey}}, \bibinfo {author} {\bibfnamefont {T.~D.}\ \bibnamefont
  {Morris}}, \bibinfo {author} {\bibfnamefont {R.~C.}\ \bibnamefont {Pooser}},
  \bibinfo {author} {\bibfnamefont {M.}~\bibnamefont {Sanz}}, \bibinfo {author}
  {\bibfnamefont {E.}~\bibnamefont {Solano}}, \bibinfo {author} {\bibfnamefont
  {P.}~\bibnamefont {Lougovski}}, \ and\ \bibinfo {author} {\bibfnamefont
  {M.~J.}\ \bibnamefont {Savage}},\ }\href {\doibase
  10.1103/PhysRevA.98.032331} {\bibfield  {journal} {\bibinfo  {journal}
  {Physical Review A}\ }\textbf {\bibinfo {volume} {98}},\ \bibinfo {pages}
  {032331} (\bibinfo {year} {2018})},\ \bibinfo {note} {arXiv:
  1803.03326}\BibitemShut {NoStop}%
\bibitem [{\citenamefont {Klco}\ and\ \citenamefont
  {Savage}(2019)}]{klco_digitization_2019}%
  \BibitemOpen
  \bibfield  {author} {\bibinfo {author} {\bibfnamefont {N.}~\bibnamefont
  {Klco}}\ and\ \bibinfo {author} {\bibfnamefont {M.~J.}\ \bibnamefont
  {Savage}},\ }\href {\doibase 10.1103/PhysRevA.99.052335} {\bibfield
  {journal} {\bibinfo  {journal} {Physical Review A}\ }\textbf {\bibinfo
  {volume} {99}},\ \bibinfo {pages} {052335} (\bibinfo {year} {2019})},\
  \bibinfo {note} {arXiv: 1808.10378}\BibitemShut {NoStop}%
\bibitem [{\citenamefont {Gustafson}\ \emph {et~al.}(2019)\citenamefont
  {Gustafson}, \citenamefont {Meurice},\ and\ \citenamefont
  {Unmuth-Yockey}}]{gustafson_quantum_2019}%
  \BibitemOpen
  \bibfield  {author} {\bibinfo {author} {\bibfnamefont {E.}~\bibnamefont
  {Gustafson}}, \bibinfo {author} {\bibfnamefont {Y.}~\bibnamefont {Meurice}},
  \ and\ \bibinfo {author} {\bibfnamefont {J.}~\bibnamefont {Unmuth-Yockey}},\
  }\href {\doibase 10.1103/PhysRevD.99.094503} {\bibfield  {journal} {\bibinfo
  {journal} {Physical Review D}\ }\textbf {\bibinfo {volume} {99}},\ \bibinfo
  {pages} {094503} (\bibinfo {year} {2019})},\ \bibinfo {note} {publisher:
  American Physical Society}\BibitemShut {NoStop}%
\bibitem [{\citenamefont {{NuQS Collaboration}}\ \emph
  {et~al.}(2019)\citenamefont {{NuQS Collaboration}}, \citenamefont
  {Alexandru}, \citenamefont {Bedaque}, \citenamefont {Lamm},\ and\
  \citenamefont {Lawrence}}]{nuqs_collaboration_ensuremathsigma_2019}%
  \BibitemOpen
  \bibfield  {author} {\bibinfo {author} {\bibnamefont {{NuQS Collaboration}}},
  \bibinfo {author} {\bibfnamefont {A.}~\bibnamefont {Alexandru}}, \bibinfo
  {author} {\bibfnamefont {P.~F.}\ \bibnamefont {Bedaque}}, \bibinfo {author}
  {\bibfnamefont {H.}~\bibnamefont {Lamm}}, \ and\ \bibinfo {author}
  {\bibfnamefont {S.}~\bibnamefont {Lawrence}},\ }\href {\doibase
  10.1103/PhysRevLett.123.090501} {\bibfield  {journal} {\bibinfo  {journal}
  {Physical Review Letters}\ }\textbf {\bibinfo {volume} {123}},\ \bibinfo
  {pages} {090501} (\bibinfo {year} {2019})},\ \bibinfo {note} {publisher:
  American Physical Society}\BibitemShut {NoStop}%
\bibitem [{\citenamefont {Magnifico}\ \emph {et~al.}(2020)\citenamefont
  {Magnifico}, \citenamefont {Dalmonte}, \citenamefont {Facchi}, \citenamefont
  {Pascazio}, \citenamefont {Pepe},\ and\ \citenamefont
  {Ercolessi}}]{magnifico_real_2020}%
  \BibitemOpen
  \bibfield  {author} {\bibinfo {author} {\bibfnamefont {G.}~\bibnamefont
  {Magnifico}}, \bibinfo {author} {\bibfnamefont {M.}~\bibnamefont {Dalmonte}},
  \bibinfo {author} {\bibfnamefont {P.}~\bibnamefont {Facchi}}, \bibinfo
  {author} {\bibfnamefont {S.}~\bibnamefont {Pascazio}}, \bibinfo {author}
  {\bibfnamefont {F.~V.}\ \bibnamefont {Pepe}}, \ and\ \bibinfo {author}
  {\bibfnamefont {E.}~\bibnamefont {Ercolessi}},\ }\href {\doibase
  10.22331/q-2020-06-15-281} {\bibfield  {journal} {\bibinfo  {journal}
  {Quantum}\ }\textbf {\bibinfo {volume} {4}},\ \bibinfo {pages} {281}
  (\bibinfo {year} {2020})},\ \bibinfo {note} {arXiv: 1909.04821}\BibitemShut
  {NoStop}%
\bibitem [{\citenamefont {Jordan}\ \emph {et~al.}(2019)\citenamefont {Jordan},
  \citenamefont {Lee},\ and\ \citenamefont {Preskill}}]{jordan_quantum_2019}%
  \BibitemOpen
  \bibfield  {author} {\bibinfo {author} {\bibfnamefont {S.~P.}\ \bibnamefont
  {Jordan}}, \bibinfo {author} {\bibfnamefont {K.~S.~M.}\ \bibnamefont {Lee}},
  \ and\ \bibinfo {author} {\bibfnamefont {J.}~\bibnamefont {Preskill}},\
  }\href {http://arxiv.org/abs/1112.4833} {\bibfield  {journal} {\bibinfo
  {journal} {arXiv:1112.4833 [hep-th, physics:quant-ph]}\ } (\bibinfo {year}
  {2019})},\ \bibinfo {note} {arXiv: 1112.4833}\BibitemShut {NoStop}%
\bibitem [{\citenamefont {Lu}\ \emph {et~al.}(2019)\citenamefont {Lu},
  \citenamefont {Klco}, \citenamefont {Lukens}, \citenamefont {Morris},
  \citenamefont {Bansal}, \citenamefont {Ekstr{\"o}m}, \citenamefont {Hagen},
  \citenamefont {Papenbrock}, \citenamefont {Weiner}, \citenamefont {Savage},\
  and\ \citenamefont {Lougovski}}]{lu_simulations_2019}%
  \BibitemOpen
  \bibfield  {author} {\bibinfo {author} {\bibfnamefont {H.-H.}\ \bibnamefont
  {Lu}}, \bibinfo {author} {\bibfnamefont {N.}~\bibnamefont {Klco}}, \bibinfo
  {author} {\bibfnamefont {J.~M.}\ \bibnamefont {Lukens}}, \bibinfo {author}
  {\bibfnamefont {T.~D.}\ \bibnamefont {Morris}}, \bibinfo {author}
  {\bibfnamefont {A.}~\bibnamefont {Bansal}}, \bibinfo {author} {\bibfnamefont
  {A.}~\bibnamefont {Ekstr{\"o}m}}, \bibinfo {author} {\bibfnamefont
  {G.}~\bibnamefont {Hagen}}, \bibinfo {author} {\bibfnamefont
  {T.}~\bibnamefont {Papenbrock}}, \bibinfo {author} {\bibfnamefont {A.~M.}\
  \bibnamefont {Weiner}}, \bibinfo {author} {\bibfnamefont {M.~J.}\
  \bibnamefont {Savage}}, \ and\ \bibinfo {author} {\bibfnamefont
  {P.}~\bibnamefont {Lougovski}},\ }\href {\doibase
  10.1103/PhysRevA.100.012320} {\bibfield  {journal} {\bibinfo  {journal}
  {Physical Review A}\ }\textbf {\bibinfo {volume} {100}},\ \bibinfo {pages}
  {012320} (\bibinfo {year} {2019})},\ \bibinfo {note} {arXiv:
  1810.03959}\BibitemShut {NoStop}%
\bibitem [{\citenamefont {Klco}\ and\ \citenamefont
  {Savage}(2020)}]{klco_minimally-entangled_2020}%
  \BibitemOpen
  \bibfield  {author} {\bibinfo {author} {\bibfnamefont {N.}~\bibnamefont
  {Klco}}\ and\ \bibinfo {author} {\bibfnamefont {M.~J.}\ \bibnamefont
  {Savage}},\ }\href {\doibase 10.1103/PhysRevA.102.012612} {\bibfield
  {journal} {\bibinfo  {journal} {Physical Review A}\ }\textbf {\bibinfo
  {volume} {102}},\ \bibinfo {pages} {012612} (\bibinfo {year} {2020})},\
  \bibinfo {note} {arXiv: 1904.10440}\BibitemShut {NoStop}%
\bibitem [{\citenamefont {Lamm}\ and\ \citenamefont
  {Lawrence}(2018)}]{lamm_simulation_2018}%
  \BibitemOpen
  \bibfield  {author} {\bibinfo {author} {\bibfnamefont {H.}~\bibnamefont
  {Lamm}}\ and\ \bibinfo {author} {\bibfnamefont {S.}~\bibnamefont
  {Lawrence}},\ }\href {\doibase 10.1103/PhysRevLett.121.170501} {\bibfield
  {journal} {\bibinfo  {journal} {Physical Review Letters}\ }\textbf {\bibinfo
  {volume} {121}},\ \bibinfo {pages} {170501} (\bibinfo {year} {2018})},\
  \bibinfo {note} {arXiv: 1806.06649}\BibitemShut {NoStop}%
\bibitem [{\citenamefont {Klco}\ \emph {et~al.}(2020)\citenamefont {Klco},
  \citenamefont {Stryker},\ and\ \citenamefont {Savage}}]{klco_su2_2020}%
  \BibitemOpen
  \bibfield  {author} {\bibinfo {author} {\bibfnamefont {N.}~\bibnamefont
  {Klco}}, \bibinfo {author} {\bibfnamefont {J.~R.}\ \bibnamefont {Stryker}}, \
  and\ \bibinfo {author} {\bibfnamefont {M.~J.}\ \bibnamefont {Savage}},\
  }\href {\doibase 10.1103/PhysRevD.101.074512} {\bibfield  {journal} {\bibinfo
   {journal} {Physical Review D}\ }\textbf {\bibinfo {volume} {101}},\ \bibinfo
  {pages} {074512} (\bibinfo {year} {2020})},\ \bibinfo {note} {arXiv:
  1908.06935}\BibitemShut {NoStop}%
\bibitem [{\citenamefont {Alexandru}\ \emph {et~al.}(2019)\citenamefont
  {Alexandru}, \citenamefont {Bedaque}, \citenamefont {Harmalkar},
  \citenamefont {Lamm}, \citenamefont {Lawrence},\ and\ \citenamefont
  {Warrington}}]{alexandru_gluon_2019}%
  \BibitemOpen
  \bibfield  {author} {\bibinfo {author} {\bibfnamefont {A.}~\bibnamefont
  {Alexandru}}, \bibinfo {author} {\bibfnamefont {P.~F.}\ \bibnamefont
  {Bedaque}}, \bibinfo {author} {\bibfnamefont {S.}~\bibnamefont {Harmalkar}},
  \bibinfo {author} {\bibfnamefont {H.}~\bibnamefont {Lamm}}, \bibinfo {author}
  {\bibfnamefont {S.}~\bibnamefont {Lawrence}}, \ and\ \bibinfo {author}
  {\bibfnamefont {N.~C.}\ \bibnamefont {Warrington}},\ }\href {\doibase
  10.1103/PhysRevD.100.114501} {\bibfield  {journal} {\bibinfo  {journal}
  {Physical Review D}\ }\textbf {\bibinfo {volume} {100}},\ \bibinfo {pages}
  {114501} (\bibinfo {year} {2019})},\ \bibinfo {note} {arXiv:
  1906.11213}\BibitemShut {NoStop}%
\bibitem [{\citenamefont {Mueller}\ \emph {et~al.}(2020)\citenamefont
  {Mueller}, \citenamefont {Tarasov},\ and\ \citenamefont
  {Venugopalan}}]{mueller_deeply_2020}%
  \BibitemOpen
  \bibfield  {author} {\bibinfo {author} {\bibfnamefont {N.}~\bibnamefont
  {Mueller}}, \bibinfo {author} {\bibfnamefont {A.}~\bibnamefont {Tarasov}}, \
  and\ \bibinfo {author} {\bibfnamefont {R.}~\bibnamefont {Venugopalan}},\
  }\href {\doibase 10.1103/PhysRevD.102.016007} {\bibfield  {journal} {\bibinfo
   {journal} {Physical Review D}\ }\textbf {\bibinfo {volume} {102}},\ \bibinfo
  {pages} {016007} (\bibinfo {year} {2020})},\ \bibinfo {note} {arXiv:
  1908.07051}\BibitemShut {NoStop}%
\bibitem [{\citenamefont {Lamm}\ \emph {et~al.}(2020)\citenamefont {Lamm},
  \citenamefont {Lawrence},\ and\ \citenamefont {Yamauchi}}]{lamm_parton_2020}%
  \BibitemOpen
  \bibfield  {author} {\bibinfo {author} {\bibfnamefont {H.}~\bibnamefont
  {Lamm}}, \bibinfo {author} {\bibfnamefont {S.}~\bibnamefont {Lawrence}}, \
  and\ \bibinfo {author} {\bibfnamefont {Y.}~\bibnamefont {Yamauchi}},\ }\href
  {\doibase 10.1103/PhysRevResearch.2.013272} {\bibfield  {journal} {\bibinfo
  {journal} {Physical Review Research}\ }\textbf {\bibinfo {volume} {2}},\
  \bibinfo {pages} {013272} (\bibinfo {year} {2020})},\ \bibinfo {note} {arXiv:
  1908.10439}\BibitemShut {NoStop}%
\bibitem [{\citenamefont {Chakraborty}\ \emph {et~al.}(2020)\citenamefont
  {Chakraborty}, \citenamefont {Honda}, \citenamefont {Izubuchi}, \citenamefont
  {Kikuchi},\ and\ \citenamefont {Tomiya}}]{chakraborty_digital_2020}%
  \BibitemOpen
  \bibfield  {author} {\bibinfo {author} {\bibfnamefont {B.}~\bibnamefont
  {Chakraborty}}, \bibinfo {author} {\bibfnamefont {M.}~\bibnamefont {Honda}},
  \bibinfo {author} {\bibfnamefont {T.}~\bibnamefont {Izubuchi}}, \bibinfo
  {author} {\bibfnamefont {Y.}~\bibnamefont {Kikuchi}}, \ and\ \bibinfo
  {author} {\bibfnamefont {A.}~\bibnamefont {Tomiya}},\ }\href
  {http://arxiv.org/abs/2001.00485} {\bibfield  {journal} {\bibinfo  {journal}
  {arXiv:2001.00485 [cond-mat, physics:hep-lat, physics:hep-ph, physics:hep-th,
  physics:quant-ph]}\ } (\bibinfo {year} {2020})},\ \bibinfo {note} {arXiv:
  2001.00485}\BibitemShut {NoStop}%
\bibitem [{\citenamefont {Bermudez}\ \emph {et~al.}(2018)\citenamefont
  {Bermudez}, \citenamefont {Tirrito}, \citenamefont {Rizzi}, \citenamefont
  {Lewenstein},\ and\ \citenamefont {Hands}}]{Bermudez:2018eyh}%
  \BibitemOpen
  \bibfield  {author} {\bibinfo {author} {\bibfnamefont {A.}~\bibnamefont
  {Bermudez}}, \bibinfo {author} {\bibfnamefont {E.}~\bibnamefont {Tirrito}},
  \bibinfo {author} {\bibfnamefont {M.}~\bibnamefont {Rizzi}}, \bibinfo
  {author} {\bibfnamefont {M.}~\bibnamefont {Lewenstein}}, \ and\ \bibinfo
  {author} {\bibfnamefont {S.}~\bibnamefont {Hands}},\ }\href {\doibase
  10.1016/j.aop.2018.10.007} {\bibfield  {journal} {\bibinfo  {journal} {Annals
  Phys.}\ }\textbf {\bibinfo {volume} {399}},\ \bibinfo {pages} {149} (\bibinfo
  {year} {2018})},\ \Eprint {http://arxiv.org/abs/1807.03202} {arXiv:1807.03202
  [cond-mat.quant-gas]} \BibitemShut {NoStop}%
\bibitem [{\citenamefont {Ziegler}\ \emph {et~al.}(2020)\citenamefont
  {Ziegler}, \citenamefont {Tirrito}, \citenamefont {Lewenstein}, \citenamefont
  {Hands},\ and\ \citenamefont {Bermudez}}]{Ziegler:2020zkq}%
  \BibitemOpen
  \bibfield  {author} {\bibinfo {author} {\bibfnamefont {L.}~\bibnamefont
  {Ziegler}}, \bibinfo {author} {\bibfnamefont {E.}~\bibnamefont {Tirrito}},
  \bibinfo {author} {\bibfnamefont {M.}~\bibnamefont {Lewenstein}}, \bibinfo
  {author} {\bibfnamefont {S.}~\bibnamefont {Hands}}, \ and\ \bibinfo {author}
  {\bibfnamefont {A.}~\bibnamefont {Bermudez}},\ }\href@noop {} {\  (\bibinfo
  {year} {2020})},\ \Eprint {http://arxiv.org/abs/2011.08744} {arXiv:2011.08744
  [cond-mat.quant-gas]} \BibitemShut {NoStop}%
\bibitem [{\citenamefont {Ziegler}\ \emph {et~al.}(2021)\citenamefont
  {Ziegler}, \citenamefont {Tirrito}, \citenamefont {Lewenstein}, \citenamefont
  {Hands},\ and\ \citenamefont {Bermudez}}]{Ziegler:2021yua}%
  \BibitemOpen
  \bibfield  {author} {\bibinfo {author} {\bibfnamefont {L.}~\bibnamefont
  {Ziegler}}, \bibinfo {author} {\bibfnamefont {E.}~\bibnamefont {Tirrito}},
  \bibinfo {author} {\bibfnamefont {M.}~\bibnamefont {Lewenstein}}, \bibinfo
  {author} {\bibfnamefont {S.}~\bibnamefont {Hands}}, \ and\ \bibinfo {author}
  {\bibfnamefont {A.}~\bibnamefont {Bermudez}},\ }\href@noop {} {\  (\bibinfo
  {year} {2021})},\ \Eprint {http://arxiv.org/abs/2111.04485} {arXiv:2111.04485
  [cond-mat.quant-gas]} \BibitemShut {NoStop}%
\bibitem [{\citenamefont {Sun}\ \emph {et~al.}(2021)\citenamefont {Sun},
  \citenamefont {Motta}, \citenamefont {Tazhigulov}, \citenamefont {Tan},
  \citenamefont {Chan},\ and\ \citenamefont {Minnich}}]{PRXQuantum.2.010317}%
  \BibitemOpen
  \bibfield  {author} {\bibinfo {author} {\bibfnamefont {S.-N.}\ \bibnamefont
  {Sun}}, \bibinfo {author} {\bibfnamefont {M.}~\bibnamefont {Motta}}, \bibinfo
  {author} {\bibfnamefont {R.~N.}\ \bibnamefont {Tazhigulov}}, \bibinfo
  {author} {\bibfnamefont {A.~T.}\ \bibnamefont {Tan}}, \bibinfo {author}
  {\bibfnamefont {G.~K.-L.}\ \bibnamefont {Chan}}, \ and\ \bibinfo {author}
  {\bibfnamefont {A.~J.}\ \bibnamefont {Minnich}},\ }\href {\doibase
  10.1103/PRXQuantum.2.010317} {\bibfield  {journal} {\bibinfo  {journal} {PRX
  Quantum}\ }\textbf {\bibinfo {volume} {2}},\ \bibinfo {pages} {010317}
  (\bibinfo {year} {2021})}\BibitemShut {NoStop}%
\bibitem [{\citenamefont {Ville}\ \emph {et~al.}(2021)\citenamefont {Ville}
  \emph {et~al.}}]{Ville:2021hrl}%
  \BibitemOpen
  \bibfield  {author} {\bibinfo {author} {\bibfnamefont {J.-L.}\ \bibnamefont
  {Ville}} \emph {et~al.},\ }\href@noop {} {\  (\bibinfo {year} {2021})},\
  \Eprint {http://arxiv.org/abs/2104.08785} {arXiv:2104.08785 [quant-ph]}
  \BibitemShut {NoStop}%
\bibitem [{\citenamefont {Czajka}\ \emph {et~al.}(2022)\citenamefont {Czajka},
  \citenamefont {Kang}, \citenamefont {Ma},\ and\ \citenamefont
  {Zhao}}]{Czajka:2021yll}%
  \BibitemOpen
  \bibfield  {author} {\bibinfo {author} {\bibfnamefont {A.~M.}\ \bibnamefont
  {Czajka}}, \bibinfo {author} {\bibfnamefont {Z.-B.}\ \bibnamefont {Kang}},
  \bibinfo {author} {\bibfnamefont {H.}~\bibnamefont {Ma}}, \ and\ \bibinfo
  {author} {\bibfnamefont {F.}~\bibnamefont {Zhao}},\ }\href {\doibase
  10.1007/JHEP08(2022)209} {\bibfield  {journal} {\bibinfo  {journal} {JHEP}\
  }\textbf {\bibinfo {volume} {08}},\ \bibinfo {pages} {209} (\bibinfo {year}
  {2022})},\ \Eprint {http://arxiv.org/abs/2112.03944} {arXiv:2112.03944
  [hep-ph]} \BibitemShut {NoStop}%
\bibitem [{\citenamefont {Gross}\ and\ \citenamefont
  {Neveu}(1974)}]{Gross:1974jv}%
  \BibitemOpen
  \bibfield  {author} {\bibinfo {author} {\bibfnamefont {D.~J.}\ \bibnamefont
  {Gross}}\ and\ \bibinfo {author} {\bibfnamefont {A.}~\bibnamefont {Neveu}},\
  }\href {\doibase 10.1103/PhysRevD.10.3235} {\bibfield  {journal} {\bibinfo
  {journal} {Phys. Rev. D}\ }\textbf {\bibinfo {volume} {10}},\ \bibinfo
  {pages} {3235} (\bibinfo {year} {1974})}\BibitemShut {NoStop}%
\bibitem [{\citenamefont {Borsanyi}\ \emph {et~al.}(2010)\citenamefont
  {Borsanyi}, \citenamefont {Endrodi}, \citenamefont {Fodor}, \citenamefont
  {Jakovac}, \citenamefont {Katz}, \citenamefont {Krieg}, \citenamefont
  {Ratti},\ and\ \citenamefont {Szabo}}]{Borsanyi:2010cj}%
  \BibitemOpen
  \bibfield  {author} {\bibinfo {author} {\bibfnamefont {S.}~\bibnamefont
  {Borsanyi}}, \bibinfo {author} {\bibfnamefont {G.}~\bibnamefont {Endrodi}},
  \bibinfo {author} {\bibfnamefont {Z.}~\bibnamefont {Fodor}}, \bibinfo
  {author} {\bibfnamefont {A.}~\bibnamefont {Jakovac}}, \bibinfo {author}
  {\bibfnamefont {S.~D.}\ \bibnamefont {Katz}}, \bibinfo {author}
  {\bibfnamefont {S.}~\bibnamefont {Krieg}}, \bibinfo {author} {\bibfnamefont
  {C.}~\bibnamefont {Ratti}}, \ and\ \bibinfo {author} {\bibfnamefont {K.~K.}\
  \bibnamefont {Szabo}},\ }\href {\doibase 10.1007/JHEP11(2010)077} {\bibfield
  {journal} {\bibinfo  {journal} {JHEP}\ }\textbf {\bibinfo {volume} {11}},\
  \bibinfo {pages} {077} (\bibinfo {year} {2010})},\ \Eprint
  {http://arxiv.org/abs/1007.2580} {arXiv:1007.2580 [hep-lat]} \BibitemShut
  {NoStop}%
\bibitem [{\citenamefont {Kogut}\ and\ \citenamefont
  {Susskind}(1975)}]{Staggered}%
  \BibitemOpen
  \bibfield  {author} {\bibinfo {author} {\bibfnamefont {J.}~\bibnamefont
  {Kogut}}\ and\ \bibinfo {author} {\bibfnamefont {L.}~\bibnamefont
  {Susskind}},\ }\href {\doibase 10.1103/PhysRevD.11.395} {\bibfield  {journal}
  {\bibinfo  {journal} {Phys. Rev. D}\ }\textbf {\bibinfo {volume} {11}},\
  \bibinfo {pages} {395} (\bibinfo {year} {1975})}\BibitemShut {NoStop}%
\bibitem [{\citenamefont {Borsanyi}\ \emph {et~al.}(2014)\citenamefont
  {Borsanyi}, \citenamefont {Fodor}, \citenamefont {Hoelbling}, \citenamefont
  {Katz}, \citenamefont {Krieg},\ and\ \citenamefont
  {Szabo}}]{Borsanyi:2013bia}%
  \BibitemOpen
  \bibfield  {author} {\bibinfo {author} {\bibfnamefont {S.}~\bibnamefont
  {Borsanyi}}, \bibinfo {author} {\bibfnamefont {Z.}~\bibnamefont {Fodor}},
  \bibinfo {author} {\bibfnamefont {C.}~\bibnamefont {Hoelbling}}, \bibinfo
  {author} {\bibfnamefont {S.~D.}\ \bibnamefont {Katz}}, \bibinfo {author}
  {\bibfnamefont {S.}~\bibnamefont {Krieg}}, \ and\ \bibinfo {author}
  {\bibfnamefont {K.~K.}\ \bibnamefont {Szabo}},\ }\href {\doibase
  10.1016/j.physletb.2014.01.007} {\bibfield  {journal} {\bibinfo  {journal}
  {Phys. Lett. B}\ }\textbf {\bibinfo {volume} {730}},\ \bibinfo {pages} {99}
  (\bibinfo {year} {2014})},\ \Eprint {http://arxiv.org/abs/1309.5258}
  {arXiv:1309.5258 [hep-lat]} \BibitemShut {NoStop}%
\bibitem [{\citenamefont {Aoki}\ \emph {et~al.}(2006)\citenamefont {Aoki},
  \citenamefont {Fodor}, \citenamefont {Katz},\ and\ \citenamefont
  {Szabo}}]{Aoki:2005vt}%
  \BibitemOpen
  \bibfield  {author} {\bibinfo {author} {\bibfnamefont {Y.}~\bibnamefont
  {Aoki}}, \bibinfo {author} {\bibfnamefont {Z.}~\bibnamefont {Fodor}},
  \bibinfo {author} {\bibfnamefont {S.~D.}\ \bibnamefont {Katz}}, \ and\
  \bibinfo {author} {\bibfnamefont {K.~K.}\ \bibnamefont {Szabo}},\ }\href
  {\doibase 10.1088/1126-6708/2006/01/089} {\bibfield  {journal} {\bibinfo
  {journal} {JHEP}\ }\textbf {\bibinfo {volume} {01}},\ \bibinfo {pages} {089}
  (\bibinfo {year} {2006})},\ \Eprint {http://arxiv.org/abs/hep-lat/0510084}
  {arXiv:hep-lat/0510084} \BibitemShut {NoStop}%
\bibitem [{\citenamefont {Bazavov}\ \emph {et~al.}(2014)\citenamefont {Bazavov}
  \emph {et~al.}}]{HotQCD:2014kol}%
  \BibitemOpen
  \bibfield  {author} {\bibinfo {author} {\bibfnamefont {A.}~\bibnamefont
  {Bazavov}} \emph {et~al.} (\bibinfo {collaboration} {HotQCD}),\ }\href
  {\doibase 10.1103/PhysRevD.90.094503} {\bibfield  {journal} {\bibinfo
  {journal} {Phys. Rev. D}\ }\textbf {\bibinfo {volume} {90}},\ \bibinfo
  {pages} {094503} (\bibinfo {year} {2014})},\ \Eprint
  {http://arxiv.org/abs/1407.6387} {arXiv:1407.6387 [hep-lat]} \BibitemShut
  {NoStop}%
\bibitem [{\citenamefont {Aubin}\ \emph {et~al.}(2020)\citenamefont {Aubin},
  \citenamefont {Blum}, \citenamefont {Tu}, \citenamefont {Golterman},
  \citenamefont {Jung},\ and\ \citenamefont {Peris}}]{Aubin:2019usy}%
  \BibitemOpen
  \bibfield  {author} {\bibinfo {author} {\bibfnamefont {C.}~\bibnamefont
  {Aubin}}, \bibinfo {author} {\bibfnamefont {T.}~\bibnamefont {Blum}},
  \bibinfo {author} {\bibfnamefont {C.}~\bibnamefont {Tu}}, \bibinfo {author}
  {\bibfnamefont {M.}~\bibnamefont {Golterman}}, \bibinfo {author}
  {\bibfnamefont {C.}~\bibnamefont {Jung}}, \ and\ \bibinfo {author}
  {\bibfnamefont {S.}~\bibnamefont {Peris}},\ }\href {\doibase
  10.1103/PhysRevD.101.014503} {\bibfield  {journal} {\bibinfo  {journal}
  {Phys. Rev. D}\ }\textbf {\bibinfo {volume} {101}},\ \bibinfo {pages}
  {014503} (\bibinfo {year} {2020})},\ \Eprint
  {http://arxiv.org/abs/1905.09307} {arXiv:1905.09307 [hep-lat]} \BibitemShut
  {NoStop}%
\bibitem [{\citenamefont {Jordan}\ and\ \citenamefont
  {Wigner}(1928)}]{Jordan1928}%
  \BibitemOpen
  \bibfield  {author} {\bibinfo {author} {\bibfnamefont {P.}~\bibnamefont
  {Jordan}}\ and\ \bibinfo {author} {\bibfnamefont {E.}~\bibnamefont
  {Wigner}},\ }\href {\doibase 10.1007/BF01331938} {\bibfield  {journal}
  {\bibinfo  {journal} {Zeitschrift f{\"u}r Physik}\ }\textbf {\bibinfo
  {volume} {47}},\ \bibinfo {pages} {631} (\bibinfo {year} {1928})}\BibitemShut
  {NoStop}%
\bibitem [{\citenamefont {Trotter}(1959)}]{trotter1959product}%
  \BibitemOpen
  \bibfield  {author} {\bibinfo {author} {\bibfnamefont {H.~F.}\ \bibnamefont
  {Trotter}},\ }\href@noop {} {\bibfield  {journal} {\bibinfo  {journal}
  {Proceedings of the American Mathematical Society}\ }\textbf {\bibinfo
  {volume} {10}},\ \bibinfo {pages} {545} (\bibinfo {year} {1959})}\BibitemShut
  {NoStop}%
\bibitem [{\citenamefont {Suzuki}(1976)}]{Suzuki}%
  \BibitemOpen
  \bibfield  {author} {\bibinfo {author} {\bibfnamefont {M.}~\bibnamefont
  {Suzuki}},\ }\href {\doibase 10.1007/BF01609348} {\bibfield  {journal}
  {\bibinfo  {journal} {Communications in Mathematical Physics}\ }\textbf
  {\bibinfo {volume} {51}},\ \bibinfo {pages} {183} (\bibinfo {year}
  {1976})}\BibitemShut {NoStop}%
\bibitem [{\citenamefont {Motta}\ \emph {et~al.}(2020)\citenamefont {Motta},
  \citenamefont {Sun}, \citenamefont {Tan}, \citenamefont {O'Rourke},
  \citenamefont {Ye}, \citenamefont {Minnich}, \citenamefont {Brand{\~a}o},\
  and\ \citenamefont {Chan}}]{Motta}%
  \BibitemOpen
  \bibfield  {author} {\bibinfo {author} {\bibfnamefont {M.}~\bibnamefont
  {Motta}}, \bibinfo {author} {\bibfnamefont {C.}~\bibnamefont {Sun}}, \bibinfo
  {author} {\bibfnamefont {A.~T.~K.}\ \bibnamefont {Tan}}, \bibinfo {author}
  {\bibfnamefont {M.~J.}\ \bibnamefont {O'Rourke}}, \bibinfo {author}
  {\bibfnamefont {E.}~\bibnamefont {Ye}}, \bibinfo {author} {\bibfnamefont
  {A.~J.}\ \bibnamefont {Minnich}}, \bibinfo {author} {\bibfnamefont {F.~G.
  S.~L.}\ \bibnamefont {Brand{\~a}o}}, \ and\ \bibinfo {author} {\bibfnamefont
  {G.~K.-L.}\ \bibnamefont {Chan}},\ }\href {\doibase
  10.1038/s41567-019-0704-4} {\bibfield  {journal} {\bibinfo  {journal} {Nature
  Physics}\ }\textbf {\bibinfo {volume} {16}},\ \bibinfo {pages} {205}
  (\bibinfo {year} {2020})}\BibitemShut {NoStop}%
\bibitem [{\citenamefont {Terhal}\ and\ \citenamefont
  {DiVincenzo}(2000{\natexlab{b}})}]{Terhal:1998yh}%
  \BibitemOpen
  \bibfield  {author} {\bibinfo {author} {\bibfnamefont {B.~M.}\ \bibnamefont
  {Terhal}}\ and\ \bibinfo {author} {\bibfnamefont {D.~P.}\ \bibnamefont
  {DiVincenzo}},\ }\href {\doibase 10.1103/PhysRevA.61.22301} {\bibfield
  {journal} {\bibinfo  {journal} {Phys. Rev. A}\ }\textbf {\bibinfo {volume}
  {61}},\ \bibinfo {pages} {22301} (\bibinfo {year} {2000}{\natexlab{b}})},\
  \Eprint {http://arxiv.org/abs/quant-ph/9810063} {arXiv:quant-ph/9810063}
  \BibitemShut {NoStop}%
\bibitem [{\citenamefont {Temme}\ \emph
  {et~al.}(2011{\natexlab{b}})\citenamefont {Temme}, \citenamefont {Osborne},
  \citenamefont {Vollbrecht}, \citenamefont {Poulin},\ and\ \citenamefont
  {Verstraete}}]{Temme:2009wa}%
  \BibitemOpen
  \bibfield  {author} {\bibinfo {author} {\bibfnamefont {K.}~\bibnamefont
  {Temme}}, \bibinfo {author} {\bibfnamefont {T.~J.}\ \bibnamefont {Osborne}},
  \bibinfo {author} {\bibfnamefont {K.~G.}\ \bibnamefont {Vollbrecht}},
  \bibinfo {author} {\bibfnamefont {D.}~\bibnamefont {Poulin}}, \ and\ \bibinfo
  {author} {\bibfnamefont {F.}~\bibnamefont {Verstraete}},\ }\href {\doibase
  10.1038/nature09770} {\bibfield  {journal} {\bibinfo  {journal} {Nature}\
  }\textbf {\bibinfo {volume} {471}},\ \bibinfo {pages} {87} (\bibinfo {year}
  {2011}{\natexlab{b}})},\ \Eprint {http://arxiv.org/abs/0911.3635}
  {arXiv:0911.3635 [quant-ph]} \BibitemShut {NoStop}%
\bibitem [{\citenamefont {Chowdhury}\ and\ \citenamefont
  {Somma}(2016)}]{https://doi.org/10.48550/arxiv.1603.02940}%
  \BibitemOpen
  \bibfield  {author} {\bibinfo {author} {\bibfnamefont {A.~N.}\ \bibnamefont
  {Chowdhury}}\ and\ \bibinfo {author} {\bibfnamefont {R.~D.}\ \bibnamefont
  {Somma}},\ }\href {\doibase 10.48550/ARXIV.1603.02940} {\  (\bibinfo {year}
  {2016}),\ 10.48550/ARXIV.1603.02940}\BibitemShut {NoStop}%
\bibitem [{\citenamefont {Brand\~ao}\ and\ \citenamefont
  {Kastoryano}(2019)}]{Brandao:2016mfe}%
  \BibitemOpen
  \bibfield  {author} {\bibinfo {author} {\bibfnamefont {F.~G. S.~L.}\
  \bibnamefont {Brand\~ao}}\ and\ \bibinfo {author} {\bibfnamefont {M.~J.}\
  \bibnamefont {Kastoryano}},\ }\href {\doibase 10.1007/s00220-018-3150-8}
  {\bibfield  {journal} {\bibinfo  {journal} {Commun. Math. Phys.}\ }\textbf
  {\bibinfo {volume} {365}},\ \bibinfo {pages} {1} (\bibinfo {year} {2019})},\
  \Eprint {http://arxiv.org/abs/1609.07877} {arXiv:1609.07877 [quant-ph]}
  \BibitemShut {NoStop}%
\bibitem [{\citenamefont {Barison}\ \emph {et~al.}(2022)\citenamefont
  {Barison}, \citenamefont {Vicentini}, \citenamefont {Cirac},\ and\
  \citenamefont {Carleo}}]{Barison:2022drt}%
  \BibitemOpen
  \bibfield  {author} {\bibinfo {author} {\bibfnamefont {S.}~\bibnamefont
  {Barison}}, \bibinfo {author} {\bibfnamefont {F.}~\bibnamefont {Vicentini}},
  \bibinfo {author} {\bibfnamefont {I.}~\bibnamefont {Cirac}}, \ and\ \bibinfo
  {author} {\bibfnamefont {G.}~\bibnamefont {Carleo}},\ }\href@noop {} {\
  (\bibinfo {year} {2022})},\ \Eprint {http://arxiv.org/abs/2204.03454}
  {arXiv:2204.03454 [quant-ph]} \BibitemShut {NoStop}%
\bibitem [{\citenamefont {Johnson}\ \emph {et~al.}(2022)\citenamefont
  {Johnson}, \citenamefont {Kunitsa}, \citenamefont {Gonthier}, \citenamefont
  {Radin}, \citenamefont {Buda}, \citenamefont {Doskocil}, \citenamefont
  {Abuan},\ and\ \citenamefont {Romero}}]{Johnson:2022lpl}%
  \BibitemOpen
  \bibfield  {author} {\bibinfo {author} {\bibfnamefont {P.~D.}\ \bibnamefont
  {Johnson}}, \bibinfo {author} {\bibfnamefont {A.~A.}\ \bibnamefont
  {Kunitsa}}, \bibinfo {author} {\bibfnamefont {J.~F.}\ \bibnamefont
  {Gonthier}}, \bibinfo {author} {\bibfnamefont {M.~D.}\ \bibnamefont {Radin}},
  \bibinfo {author} {\bibfnamefont {C.}~\bibnamefont {Buda}}, \bibinfo {author}
  {\bibfnamefont {E.~J.}\ \bibnamefont {Doskocil}}, \bibinfo {author}
  {\bibfnamefont {C.~M.}\ \bibnamefont {Abuan}}, \ and\ \bibinfo {author}
  {\bibfnamefont {J.}~\bibnamefont {Romero}},\ }\href@noop {} {\  (\bibinfo
  {year} {2022})},\ \Eprint {http://arxiv.org/abs/2203.07275} {arXiv:2203.07275
  [quant-ph]} \BibitemShut {NoStop}%
\bibitem [{\citenamefont {Cao}\ \emph {et~al.}(2021)\citenamefont {Cao},
  \citenamefont {Yu}, \citenamefont {Wu}, \citenamefont {Shannon},
  \citenamefont {Zeng},\ and\ \citenamefont {Joynt}}]{Cao:2021uls}%
  \BibitemOpen
  \bibfield  {author} {\bibinfo {author} {\bibfnamefont {C.}~\bibnamefont
  {Cao}}, \bibinfo {author} {\bibfnamefont {Y.}~\bibnamefont {Yu}}, \bibinfo
  {author} {\bibfnamefont {Z.}~\bibnamefont {Wu}}, \bibinfo {author}
  {\bibfnamefont {N.}~\bibnamefont {Shannon}}, \bibinfo {author} {\bibfnamefont
  {B.}~\bibnamefont {Zeng}}, \ and\ \bibinfo {author} {\bibfnamefont
  {R.}~\bibnamefont {Joynt}},\ }\href@noop {} {\  (\bibinfo {year} {2021})},\
  \Eprint {http://arxiv.org/abs/2109.08132} {arXiv:2109.08132 [quant-ph]}
  \BibitemShut {NoStop}%
\bibitem [{\citenamefont {Omiya*}\ \emph {et~al.}(2022)\citenamefont {Omiya*},
  \citenamefont {Nakagawa*}, \citenamefont {Koh}, \citenamefont {Mizukami},
  \citenamefont {Gao},\ and\ \citenamefont {Kobayashi}}]{Omiya:2021vol}%
  \BibitemOpen
  \bibfield  {author} {\bibinfo {author} {\bibfnamefont {K.}~\bibnamefont
  {Omiya*}}, \bibinfo {author} {\bibfnamefont {Y.~O.}\ \bibnamefont
  {Nakagawa*}}, \bibinfo {author} {\bibfnamefont {S.}~\bibnamefont {Koh}},
  \bibinfo {author} {\bibfnamefont {W.}~\bibnamefont {Mizukami}}, \bibinfo
  {author} {\bibfnamefont {Q.}~\bibnamefont {Gao}}, \ and\ \bibinfo {author}
  {\bibfnamefont {T.}~\bibnamefont {Kobayashi}},\ }\href {\doibase
  10.1021/acs.jctc.1c00877} {\bibfield  {journal} {\bibinfo  {journal} {J.
  Chem. Theor. Comput.}\ }\textbf {\bibinfo {volume} {18}},\ \bibinfo {pages}
  {741} (\bibinfo {year} {2022})},\ \Eprint {http://arxiv.org/abs/2107.12705}
  {arXiv:2107.12705 [physics.chem-ph]} \BibitemShut {NoStop}%
\bibitem [{\citenamefont {Stair}\ and\ \citenamefont
  {Evangelista}(2021)}]{stair2021qforte}%
  \BibitemOpen
  \bibfield  {author} {\bibinfo {author} {\bibfnamefont {N.~H.}\ \bibnamefont
  {Stair}}\ and\ \bibinfo {author} {\bibfnamefont {F.~A.}\ \bibnamefont
  {Evangelista}},\ }\href@noop {} {\enquote {\bibinfo {title} {Qforte: an
  efficient state simulator and quantum algorithms library for molecular
  electronic structure},}\ } (\bibinfo {year} {2021}),\ \Eprint
  {http://arxiv.org/abs/2108.04413} {arXiv:2108.04413 [quant-ph]} \BibitemShut
  {NoStop}%
\bibitem [{\citenamefont {ANIS}\ \emph {et~al.}(2021)\citenamefont {ANIS},
  \citenamefont {Abby-Mitchell}, \citenamefont {Abraham}, \citenamefont
  {AduOffei}, \citenamefont {Agarwal}, \citenamefont {Agliardi}, \citenamefont
  {Aharoni}, \citenamefont {Akhalwaya}, \citenamefont {Aleksandrowicz},
  \citenamefont {Alexander}, \citenamefont {Amy}, \citenamefont {Anagolum},
  \citenamefont {Anthony-Gandon}, \citenamefont {Arbel}, \citenamefont {Asfaw},
  \citenamefont {Athalye}, \citenamefont {Avkhadiev}, \citenamefont {Azaustre},
  \citenamefont {BHOLE}, \citenamefont {Banerjee}, \citenamefont {Banerjee},
  \citenamefont {Bang}, \citenamefont {Bansal}, \citenamefont {Barkoutsos},
  \citenamefont {Barnawal}, \citenamefont {Barron}, \citenamefont {Barron},
  \citenamefont {Bello}, \citenamefont {Ben-Haim}, \citenamefont {Bennett},
  \citenamefont {Bevenius}, \citenamefont {Bhatnagar}, \citenamefont {Bhobe},
  \citenamefont {Bianchini}, \citenamefont {Bishop}, \citenamefont {Blank},
  \citenamefont {Bolos}, \citenamefont {Bopardikar}, \citenamefont {Bosch},
  \citenamefont {Brandhofer}, \citenamefont {Brandon}, \citenamefont {Bravyi},
  \citenamefont {Bronn}, \citenamefont {Bryce-Fuller}, \citenamefont {Bucher},
  \citenamefont {Burov}, \citenamefont {Cabrera}, \citenamefont {Calpin},
  \citenamefont {Capelluto}, \citenamefont {Carballo}, \citenamefont
  {Carrascal}, \citenamefont {Carriker}, \citenamefont {Carvalho},
  \citenamefont {Chen}, \citenamefont {Chen}, \citenamefont {Chen},
  \citenamefont {Chen}, \citenamefont {Chen}, \citenamefont {Chevallier},
  \citenamefont {Chinda}, \citenamefont {Cholarajan}, \citenamefont {Chow},
  \citenamefont {Churchill}, \citenamefont {CisterMoke}, \citenamefont {Claus},
  \citenamefont {Clauss}, \citenamefont {Clothier}, \citenamefont {Cocking},
  \citenamefont {Cocuzzo}, \citenamefont {Connor}, \citenamefont {Correa},
  \citenamefont {Crockett}, \citenamefont {Cross}, \citenamefont {Cross},
  \citenamefont {Cross}, \citenamefont {Cruz-Benito}, \citenamefont {Culver},
  \citenamefont {C{\'o}rcoles-Gonzales}, \citenamefont {D}, \citenamefont
  {Dague}, \citenamefont {Dandachi}, \citenamefont {Dangwal}, \citenamefont
  {Daniel}, \citenamefont {Daniels}, \citenamefont {Dartiailh}, \citenamefont
  {Davila}, \citenamefont {Debouni}, \citenamefont {Dekusar}, \citenamefont
  {Deshmukh}, \citenamefont {Deshpande}, \citenamefont {Ding}, \citenamefont
  {Doi}, \citenamefont {Dow}, \citenamefont {Drechsler}, \citenamefont
  {Dumitrescu}, \citenamefont {Dumon}, \citenamefont {Duran}, \citenamefont
  {EL-Safty}, \citenamefont {Eastman}, \citenamefont {Eberle}, \citenamefont
  {Ebrahimi}, \citenamefont {Eendebak}, \citenamefont {Egger}, \citenamefont
  {ElePT}, \citenamefont {Emilio}, \citenamefont {Espiricueta}, \citenamefont
  {Everitt}, \citenamefont {Facoetti}, \citenamefont {Farida}, \citenamefont
  {Fern{\'a}ndez}, \citenamefont {Ferracin}, \citenamefont {Ferrari},
  \citenamefont {Ferrera}, \citenamefont {Fouilland}, \citenamefont {Frisch},
  \citenamefont {Fuhrer}, \citenamefont {Fuller}, \citenamefont {GEORGE},
  \citenamefont {Gacon}, \citenamefont {Gago}, \citenamefont {Gambella},
  \citenamefont {Gambetta}, \citenamefont {Gammanpila}, \citenamefont {Garcia},
  \citenamefont {Garg}, \citenamefont {Garion}, \citenamefont {Garrison},
  \citenamefont {Gates}, \citenamefont {Gil}, \citenamefont {Gilliam},
  \citenamefont {Giridharan}, \citenamefont {Gomez-Mosquera}, \citenamefont
  {Gonzalo}, \citenamefont {de~la Puente~Gonz{\'a}lez}, \citenamefont
  {Gorzinski}, \citenamefont {Gould}, \citenamefont {Greenberg}, \citenamefont
  {Grinko}, \citenamefont {Guan}, \citenamefont {Guijo}, \citenamefont
  {Gunnels}, \citenamefont {Gupta}, \citenamefont {Gupta}, \citenamefont
  {G{\"u}nther}, \citenamefont {Haglund}, \citenamefont {Haide}, \citenamefont
  {Hamamura}, \citenamefont {Hamido}, \citenamefont {Harkins}, \citenamefont
  {Hartman}, \citenamefont {Hasan}, \citenamefont {Havlicek}, \citenamefont
  {Hellmers}, \citenamefont {Herok}, \citenamefont {Hillmich}, \citenamefont
  {Horii}, \citenamefont {Howington}, \citenamefont {Hu}, \citenamefont {Hu},
  \citenamefont {Huang}, \citenamefont {Huisman}, \citenamefont {Imai},
  \citenamefont {Imamichi}, \citenamefont {Ishizaki}, \citenamefont {Ishwor},
  \citenamefont {Iten}, \citenamefont {Itoko}, \citenamefont {Ivrii},
  \citenamefont {Javadi}, \citenamefont {Javadi-Abhari}, \citenamefont {Javed},
  \citenamefont {Jianhua}, \citenamefont {Jivrajani}, \citenamefont {Johns},
  \citenamefont {Johnstun}, \citenamefont {Jonathan-Shoemaker}, \citenamefont
  {JosDenmark}, \citenamefont {JoshDumo}, \citenamefont {Judge}, \citenamefont
  {Kachmann}, \citenamefont {Kale}, \citenamefont {Kanazawa}, \citenamefont
  {Kane}, \citenamefont {Kang-Bae}, \citenamefont {Kapila}, \citenamefont
  {Karazeev}, \citenamefont {Kassebaum}, \citenamefont {Kelso}, \citenamefont
  {Kelso}, \citenamefont {Khanderao}, \citenamefont {King}, \citenamefont
  {Kobayashi}, \citenamefont {Kovi11Day}, \citenamefont {Kovyrshin},
  \citenamefont {Krishnakumar}, \citenamefont {Krishnan}, \citenamefont
  {Krsulich}, \citenamefont {Kumkar}, \citenamefont {Kus}, \citenamefont
  {LaRose}, \citenamefont {Lacal}, \citenamefont {Lambert}, \citenamefont
  {Landa}, \citenamefont {Lapeyre}, \citenamefont {Latone}, \citenamefont
  {Lawrence}, \citenamefont {Lee}, \citenamefont {Li}, \citenamefont {Lishman},
  \citenamefont {Liu}, \citenamefont {Liu}, \citenamefont {Lolcroc},
  \citenamefont {M}, \citenamefont {Madden}, \citenamefont {Maeng},
  \citenamefont {Maheshkar}, \citenamefont {Majmudar}, \citenamefont
  {Malyshev}, \citenamefont {Mandouh}, \citenamefont {Manela}, \citenamefont
  {Manjula}, \citenamefont {Marecek}, \citenamefont {Marques}, \citenamefont
  {Marwaha}, \citenamefont {Maslov}, \citenamefont {Maszota}, \citenamefont
  {Mathews}, \citenamefont {Matsuo}, \citenamefont {Mazhandu}, \citenamefont
  {McClure}, \citenamefont {McElaney}, \citenamefont {McGarry}, \citenamefont
  {McKay}, \citenamefont {McPherson}, \citenamefont {Meesala}, \citenamefont
  {Meirom}, \citenamefont {Mendell}, \citenamefont {Metcalfe}, \citenamefont
  {Mevissen}, \citenamefont {Meyer}, \citenamefont {Mezzacapo}, \citenamefont
  {Midha}, \citenamefont {Miller}, \citenamefont {Minev}, \citenamefont
  {Mitchell}, \citenamefont {Moll}, \citenamefont {Montanez}, \citenamefont
  {Monteiro}, \citenamefont {Mooring}, \citenamefont {Morales}, \citenamefont
  {Moran}, \citenamefont {Morcuende}, \citenamefont {Mostafa}, \citenamefont
  {Motta}, \citenamefont {Moyard}, \citenamefont {Murali}, \citenamefont
  {M{\"u}ggenburg}, \citenamefont {NEMOZ}, \citenamefont {Nadlinger},
  \citenamefont {Nakanishi}, \citenamefont {Nannicini}, \citenamefont {Nation},
  \citenamefont {Navarro}, \citenamefont {Naveh}, \citenamefont {Neagle},
  \citenamefont {Neuweiler}, \citenamefont {Ngoueya}, \citenamefont {Nicander},
  \citenamefont {Nick-Singstock}, \citenamefont {Niroula}, \citenamefont
  {Norlen}, \citenamefont {NuoWenLei}, \citenamefont {O'Riordan}, \citenamefont
  {Ogunbayo}, \citenamefont {Ollitrault}, \citenamefont {Onodera},
  \citenamefont {Otaolea}, \citenamefont {Oud}, \citenamefont {Padilha},
  \citenamefont {Paik}, \citenamefont {Pal}, \citenamefont {Pang},
  \citenamefont {Panigrahi}, \citenamefont {Pascuzzi}, \citenamefont
  {Perriello}, \citenamefont {Peterson}, \citenamefont {Phan}, \citenamefont
  {Pilch}, \citenamefont {Piro}, \citenamefont {Pistoia}, \citenamefont
  {Piveteau}, \citenamefont {Plewa}, \citenamefont {Pocreau}, \citenamefont
  {Pozas-Kerstjens}, \citenamefont {Pracht}, \citenamefont {Prokop},
  \citenamefont {Prutyanov}, \citenamefont {Puri}, \citenamefont {Puzzuoli},
  \citenamefont {P{\'e}rez}, \citenamefont {Quant02}, \citenamefont {Quintiii},
  \citenamefont {Rahman}, \citenamefont {Raja}, \citenamefont {Rajeev},
  \citenamefont {Rajput}, \citenamefont {Ramagiri}, \citenamefont {Rao},
  \citenamefont {Raymond}, \citenamefont {Reardon-Smith}, \citenamefont
  {Redondo}, \citenamefont {Reuter}, \citenamefont {Rice}, \citenamefont
  {Riedemann}, \citenamefont {Rietesh}, \citenamefont {Risinger}, \citenamefont
  {Rocca}, \citenamefont {Rodr{\'\i}guez}, \citenamefont {RohithKarur},
  \citenamefont {Rosand}, \citenamefont {Rossmannek}, \citenamefont {Ryu},
  \citenamefont {SAPV}, \citenamefont {Sa}, \citenamefont {Saha}, \citenamefont
  {Ash-Saki}, \citenamefont {Sanand}, \citenamefont {Sandberg}, \citenamefont
  {Sandesara}, \citenamefont {Sapra}, \citenamefont {Sargsyan}, \citenamefont
  {Sarkar}, \citenamefont {Sathaye}, \citenamefont {Schmitt}, \citenamefont
  {Schnabel}, \citenamefont {Schoenfeld}, \citenamefont {Scholten},
  \citenamefont {Schoute}, \citenamefont {Schulterbrandt}, \citenamefont
  {Schwarm}, \citenamefont {Seaward}, \citenamefont {Sergi}, \citenamefont
  {Sertage}, \citenamefont {Setia}, \citenamefont {Shah}, \citenamefont
  {Shammah}, \citenamefont {Sharma}, \citenamefont {Shi}, \citenamefont
  {Shoemaker}, \citenamefont {Silva}, \citenamefont {Simonetto}, \citenamefont
  {Singh}, \citenamefont {Singh}, \citenamefont {Singh}, \citenamefont
  {Singkanipa}, \citenamefont {Siraichi}, \citenamefont {Siri}, \citenamefont
  {Sistos}, \citenamefont {Sitdikov}, \citenamefont {Sivarajah}, \citenamefont
  {Sletfjerding}, \citenamefont {Smolin}, \citenamefont {Soeken}, \citenamefont
  {Sokolov}, \citenamefont {Sokolov}, \citenamefont {Soloviev}, \citenamefont
  {SooluThomas}, \citenamefont {Starfish}, \citenamefont {Steenken},
  \citenamefont {Stypulkoski}, \citenamefont {Suau}, \citenamefont {Sun},
  \citenamefont {Sung}, \citenamefont {Suwama}, \citenamefont {S{\l}owik},
  \citenamefont {Takahashi}, \citenamefont {Takawale}, \citenamefont
  {Tavernelli}, \citenamefont {Taylor}, \citenamefont {Taylour}, \citenamefont
  {Thomas}, \citenamefont {Tian}, \citenamefont {Tillet}, \citenamefont {Tod},
  \citenamefont {Tomasik}, \citenamefont {Tornow}, \citenamefont {de~la Torre},
  \citenamefont {Toural}, \citenamefont {Trabing}, \citenamefont {Treinish},
  \citenamefont {Trenev}, \citenamefont {TrishaPe}, \citenamefont {Truger},
  \citenamefont {Tsilimigkounakis}, \citenamefont {Tulsi}, \citenamefont
  {Turner}, \citenamefont {Vaknin}, \citenamefont {Valcarce}, \citenamefont
  {Varchon}, \citenamefont {Vartak}, \citenamefont {Vazquez}, \citenamefont
  {Vijaywargiya}, \citenamefont {Villar}, \citenamefont {Vishnu}, \citenamefont
  {Vogt-Lee}, \citenamefont {Vuillot}, \citenamefont {Weaver}, \citenamefont
  {Weidenfeller}, \citenamefont {Wieczorek}, \citenamefont {Wildstrom},
  \citenamefont {Wilson}, \citenamefont {Winston}, \citenamefont
  {WinterSoldier}, \citenamefont {Woehr}, \citenamefont {Woerner},
  \citenamefont {Woo}, \citenamefont {Wood}, \citenamefont {Wood},
  \citenamefont {Wood}, \citenamefont {Wootton}, \citenamefont {Wright},
  \citenamefont {Xing}, \citenamefont {YU}, \citenamefont {Yang}, \citenamefont
  {Yang}, \citenamefont {Yao}, \citenamefont {Yeralin}, \citenamefont
  {Yonekura}, \citenamefont {Yonge-Mallo}, \citenamefont {Yoshida},
  \citenamefont {Young}, \citenamefont {Yu}, \citenamefont {Yu}, \citenamefont
  {Zachow}, \citenamefont {Zdanski}, \citenamefont {Zhang}, \citenamefont
  {Zidaru}, \citenamefont {Zoufal}, \citenamefont {aeddins ibm}, \citenamefont
  {alexzhang13}, \citenamefont {b63}, \citenamefont {bartek bartlomiej},
  \citenamefont {bcamorrison}, \citenamefont {brandhsn}, \citenamefont
  {charmerDark}, \citenamefont {deeplokhande}, \citenamefont {dekel.meirom},
  \citenamefont {dime10}, \citenamefont {dlasecki}, \citenamefont {ehchen},
  \citenamefont {fanizzamarco}, \citenamefont {fs1132429}, \citenamefont
  {gadial}, \citenamefont {galeinston}, \citenamefont {georgezhou20},
  \citenamefont {georgios ts}, \citenamefont {gruu}, \citenamefont {hhorii},
  \citenamefont {hykavitha}, \citenamefont {itoko}, \citenamefont
  {jeppevinkel}, \citenamefont {jessica angel7}, \citenamefont {jezerjojo14},
  \citenamefont {jliu45}, \citenamefont {jscott2}, \citenamefont {klinvill},
  \citenamefont {krutik2966}, \citenamefont {ma5x}, \citenamefont
  {michelle4654}, \citenamefont {msuwama}, \citenamefont {nico lgrs},
  \citenamefont {ntgiwsvp}, \citenamefont {ordmoj}, \citenamefont {sagar
  pahwa}, \citenamefont {pritamsinha2304}, \citenamefont {ryancocuzzo},
  \citenamefont {saktar unr}, \citenamefont {saswati qiskit}, \citenamefont
  {septembrr}, \citenamefont {sethmerkel}, \citenamefont {sg495}, \citenamefont
  {shaashwat}, \citenamefont {smturro2}, \citenamefont {sternparky},
  \citenamefont {strickroman}, \citenamefont {tigerjack}, \citenamefont {tsura
  crisaldo}, \citenamefont {vadebayo49}, \citenamefont {welien}, \citenamefont
  {willhbang}, \citenamefont {wmurphy collabstar}, \citenamefont {yang.luh},\
  and\ \citenamefont {{\v{C}}epulkovskis}}]{Qiskit}%
  \BibitemOpen
  \bibfield  {author} {\bibinfo {author} {\bibfnamefont {M.~S.}\ \bibnamefont
  {ANIS}}, \bibinfo {author} {\bibnamefont {Abby-Mitchell}}, \bibinfo {author}
  {\bibfnamefont {H.}~\bibnamefont {Abraham}}, \bibinfo {author} {\bibnamefont
  {AduOffei}}, \bibinfo {author} {\bibfnamefont {R.}~\bibnamefont {Agarwal}},
  \bibinfo {author} {\bibfnamefont {G.}~\bibnamefont {Agliardi}}, \bibinfo
  {author} {\bibfnamefont {M.}~\bibnamefont {Aharoni}}, \bibinfo {author}
  {\bibfnamefont {I.~Y.}\ \bibnamefont {Akhalwaya}}, \bibinfo {author}
  {\bibfnamefont {G.}~\bibnamefont {Aleksandrowicz}}, \bibinfo {author}
  {\bibfnamefont {T.}~\bibnamefont {Alexander}}, \bibinfo {author}
  {\bibfnamefont {M.}~\bibnamefont {Amy}}, \bibinfo {author} {\bibfnamefont
  {S.}~\bibnamefont {Anagolum}}, \bibinfo {author} {\bibnamefont
  {Anthony-Gandon}}, \bibinfo {author} {\bibfnamefont {E.}~\bibnamefont
  {Arbel}}, \bibinfo {author} {\bibfnamefont {A.}~\bibnamefont {Asfaw}},
  \bibinfo {author} {\bibfnamefont {A.}~\bibnamefont {Athalye}}, \bibinfo
  {author} {\bibfnamefont {A.}~\bibnamefont {Avkhadiev}}, \bibinfo {author}
  {\bibfnamefont {C.}~\bibnamefont {Azaustre}}, \bibinfo {author}
  {\bibfnamefont {P.}~\bibnamefont {BHOLE}}, \bibinfo {author} {\bibfnamefont
  {A.}~\bibnamefont {Banerjee}}, \bibinfo {author} {\bibfnamefont
  {S.}~\bibnamefont {Banerjee}}, \bibinfo {author} {\bibfnamefont
  {W.}~\bibnamefont {Bang}}, \bibinfo {author} {\bibfnamefont {A.}~\bibnamefont
  {Bansal}}, \bibinfo {author} {\bibfnamefont {P.}~\bibnamefont {Barkoutsos}},
  \bibinfo {author} {\bibfnamefont {A.}~\bibnamefont {Barnawal}}, \bibinfo
  {author} {\bibfnamefont {G.}~\bibnamefont {Barron}}, \bibinfo {author}
  {\bibfnamefont {G.~S.}\ \bibnamefont {Barron}}, \bibinfo {author}
  {\bibfnamefont {L.}~\bibnamefont {Bello}}, \bibinfo {author} {\bibfnamefont
  {Y.}~\bibnamefont {Ben-Haim}}, \bibinfo {author} {\bibfnamefont {M.~C.}\
  \bibnamefont {Bennett}}, \bibinfo {author} {\bibfnamefont {D.}~\bibnamefont
  {Bevenius}}, \bibinfo {author} {\bibfnamefont {D.}~\bibnamefont {Bhatnagar}},
  \bibinfo {author} {\bibfnamefont {A.}~\bibnamefont {Bhobe}}, \bibinfo
  {author} {\bibfnamefont {P.}~\bibnamefont {Bianchini}}, \bibinfo {author}
  {\bibfnamefont {L.~S.}\ \bibnamefont {Bishop}}, \bibinfo {author}
  {\bibfnamefont {C.}~\bibnamefont {Blank}}, \bibinfo {author} {\bibfnamefont
  {S.}~\bibnamefont {Bolos}}, \bibinfo {author} {\bibfnamefont
  {S.}~\bibnamefont {Bopardikar}}, \bibinfo {author} {\bibfnamefont
  {S.}~\bibnamefont {Bosch}}, \bibinfo {author} {\bibfnamefont
  {S.}~\bibnamefont {Brandhofer}}, \bibinfo {author} {\bibnamefont {Brandon}},
  \bibinfo {author} {\bibfnamefont {S.}~\bibnamefont {Bravyi}}, \bibinfo
  {author} {\bibfnamefont {N.}~\bibnamefont {Bronn}}, \bibinfo {author}
  {\bibnamefont {Bryce-Fuller}}, \bibinfo {author} {\bibfnamefont
  {D.}~\bibnamefont {Bucher}}, \bibinfo {author} {\bibfnamefont
  {A.}~\bibnamefont {Burov}}, \bibinfo {author} {\bibfnamefont
  {F.}~\bibnamefont {Cabrera}}, \bibinfo {author} {\bibfnamefont
  {P.}~\bibnamefont {Calpin}}, \bibinfo {author} {\bibfnamefont
  {L.}~\bibnamefont {Capelluto}}, \bibinfo {author} {\bibfnamefont
  {J.}~\bibnamefont {Carballo}}, \bibinfo {author} {\bibfnamefont
  {G.}~\bibnamefont {Carrascal}}, \bibinfo {author} {\bibfnamefont
  {A.}~\bibnamefont {Carriker}}, \bibinfo {author} {\bibfnamefont
  {I.}~\bibnamefont {Carvalho}}, \bibinfo {author} {\bibfnamefont
  {A.}~\bibnamefont {Chen}}, \bibinfo {author} {\bibfnamefont {C.-F.}\
  \bibnamefont {Chen}}, \bibinfo {author} {\bibfnamefont {E.}~\bibnamefont
  {Chen}}, \bibinfo {author} {\bibfnamefont {J.~C.}\ \bibnamefont {Chen}},
  \bibinfo {author} {\bibfnamefont {R.}~\bibnamefont {Chen}}, \bibinfo {author}
  {\bibfnamefont {F.}~\bibnamefont {Chevallier}}, \bibinfo {author}
  {\bibfnamefont {K.}~\bibnamefont {Chinda}}, \bibinfo {author} {\bibfnamefont
  {R.}~\bibnamefont {Cholarajan}}, \bibinfo {author} {\bibfnamefont {J.~M.}\
  \bibnamefont {Chow}}, \bibinfo {author} {\bibfnamefont {S.}~\bibnamefont
  {Churchill}}, \bibinfo {author} {\bibnamefont {CisterMoke}}, \bibinfo
  {author} {\bibfnamefont {C.}~\bibnamefont {Claus}}, \bibinfo {author}
  {\bibfnamefont {C.}~\bibnamefont {Clauss}}, \bibinfo {author} {\bibfnamefont
  {C.}~\bibnamefont {Clothier}}, \bibinfo {author} {\bibfnamefont
  {R.}~\bibnamefont {Cocking}}, \bibinfo {author} {\bibfnamefont
  {R.}~\bibnamefont {Cocuzzo}}, \bibinfo {author} {\bibfnamefont
  {J.}~\bibnamefont {Connor}}, \bibinfo {author} {\bibfnamefont
  {F.}~\bibnamefont {Correa}}, \bibinfo {author} {\bibfnamefont
  {Z.}~\bibnamefont {Crockett}}, \bibinfo {author} {\bibfnamefont {A.~J.}\
  \bibnamefont {Cross}}, \bibinfo {author} {\bibfnamefont {A.~W.}\ \bibnamefont
  {Cross}}, \bibinfo {author} {\bibfnamefont {S.}~\bibnamefont {Cross}},
  \bibinfo {author} {\bibfnamefont {J.}~\bibnamefont {Cruz-Benito}}, \bibinfo
  {author} {\bibfnamefont {C.}~\bibnamefont {Culver}}, \bibinfo {author}
  {\bibfnamefont {A.~D.}\ \bibnamefont {C{\'o}rcoles-Gonzales}}, \bibinfo
  {author} {\bibfnamefont {N.}~\bibnamefont {D}}, \bibinfo {author}
  {\bibfnamefont {S.}~\bibnamefont {Dague}}, \bibinfo {author} {\bibfnamefont
  {T.~E.}\ \bibnamefont {Dandachi}}, \bibinfo {author} {\bibfnamefont {A.~N.}\
  \bibnamefont {Dangwal}}, \bibinfo {author} {\bibfnamefont {J.}~\bibnamefont
  {Daniel}}, \bibinfo {author} {\bibfnamefont {M.}~\bibnamefont {Daniels}},
  \bibinfo {author} {\bibfnamefont {M.}~\bibnamefont {Dartiailh}}, \bibinfo
  {author} {\bibfnamefont {A.~R.}\ \bibnamefont {Davila}}, \bibinfo {author}
  {\bibfnamefont {F.}~\bibnamefont {Debouni}}, \bibinfo {author} {\bibfnamefont
  {A.}~\bibnamefont {Dekusar}}, \bibinfo {author} {\bibfnamefont
  {A.}~\bibnamefont {Deshmukh}}, \bibinfo {author} {\bibfnamefont
  {M.}~\bibnamefont {Deshpande}}, \bibinfo {author} {\bibfnamefont
  {D.}~\bibnamefont {Ding}}, \bibinfo {author} {\bibfnamefont {J.}~\bibnamefont
  {Doi}}, \bibinfo {author} {\bibfnamefont {E.~M.}\ \bibnamefont {Dow}},
  \bibinfo {author} {\bibfnamefont {E.}~\bibnamefont {Drechsler}}, \bibinfo
  {author} {\bibfnamefont {E.}~\bibnamefont {Dumitrescu}}, \bibinfo {author}
  {\bibfnamefont {K.}~\bibnamefont {Dumon}}, \bibinfo {author} {\bibfnamefont
  {I.}~\bibnamefont {Duran}}, \bibinfo {author} {\bibfnamefont
  {K.}~\bibnamefont {EL-Safty}}, \bibinfo {author} {\bibfnamefont
  {E.}~\bibnamefont {Eastman}}, \bibinfo {author} {\bibfnamefont
  {G.}~\bibnamefont {Eberle}}, \bibinfo {author} {\bibfnamefont
  {A.}~\bibnamefont {Ebrahimi}}, \bibinfo {author} {\bibfnamefont
  {P.}~\bibnamefont {Eendebak}}, \bibinfo {author} {\bibfnamefont
  {D.}~\bibnamefont {Egger}}, \bibinfo {author} {\bibnamefont {ElePT}},
  \bibinfo {author} {\bibnamefont {Emilio}}, \bibinfo {author} {\bibfnamefont
  {A.}~\bibnamefont {Espiricueta}}, \bibinfo {author} {\bibfnamefont
  {M.}~\bibnamefont {Everitt}}, \bibinfo {author} {\bibfnamefont
  {D.}~\bibnamefont {Facoetti}}, \bibinfo {author} {\bibnamefont {Farida}},
  \bibinfo {author} {\bibfnamefont {P.~M.}\ \bibnamefont {Fern{\'a}ndez}},
  \bibinfo {author} {\bibfnamefont {S.}~\bibnamefont {Ferracin}}, \bibinfo
  {author} {\bibfnamefont {D.}~\bibnamefont {Ferrari}}, \bibinfo {author}
  {\bibfnamefont {A.~H.}\ \bibnamefont {Ferrera}}, \bibinfo {author}
  {\bibfnamefont {R.}~\bibnamefont {Fouilland}}, \bibinfo {author}
  {\bibfnamefont {A.}~\bibnamefont {Frisch}}, \bibinfo {author} {\bibfnamefont
  {A.}~\bibnamefont {Fuhrer}}, \bibinfo {author} {\bibfnamefont
  {B.}~\bibnamefont {Fuller}}, \bibinfo {author} {\bibfnamefont
  {M.}~\bibnamefont {GEORGE}}, \bibinfo {author} {\bibfnamefont
  {J.}~\bibnamefont {Gacon}}, \bibinfo {author} {\bibfnamefont {B.~G.}\
  \bibnamefont {Gago}}, \bibinfo {author} {\bibfnamefont {C.}~\bibnamefont
  {Gambella}}, \bibinfo {author} {\bibfnamefont {J.~M.}\ \bibnamefont
  {Gambetta}}, \bibinfo {author} {\bibfnamefont {A.}~\bibnamefont
  {Gammanpila}}, \bibinfo {author} {\bibfnamefont {L.}~\bibnamefont {Garcia}},
  \bibinfo {author} {\bibfnamefont {T.}~\bibnamefont {Garg}}, \bibinfo {author}
  {\bibfnamefont {S.}~\bibnamefont {Garion}}, \bibinfo {author} {\bibfnamefont
  {J.~R.}\ \bibnamefont {Garrison}}, \bibinfo {author} {\bibfnamefont
  {T.}~\bibnamefont {Gates}}, \bibinfo {author} {\bibfnamefont
  {L.}~\bibnamefont {Gil}}, \bibinfo {author} {\bibfnamefont {A.}~\bibnamefont
  {Gilliam}}, \bibinfo {author} {\bibfnamefont {A.}~\bibnamefont {Giridharan}},
  \bibinfo {author} {\bibfnamefont {J.}~\bibnamefont {Gomez-Mosquera}},
  \bibinfo {author} {\bibnamefont {Gonzalo}}, \bibinfo {author} {\bibfnamefont
  {S.}~\bibnamefont {de~la Puente~Gonz{\'a}lez}}, \bibinfo {author}
  {\bibfnamefont {J.}~\bibnamefont {Gorzinski}}, \bibinfo {author}
  {\bibfnamefont {I.}~\bibnamefont {Gould}}, \bibinfo {author} {\bibfnamefont
  {D.}~\bibnamefont {Greenberg}}, \bibinfo {author} {\bibfnamefont
  {D.}~\bibnamefont {Grinko}}, \bibinfo {author} {\bibfnamefont
  {W.}~\bibnamefont {Guan}}, \bibinfo {author} {\bibfnamefont {D.}~\bibnamefont
  {Guijo}}, \bibinfo {author} {\bibfnamefont {J.~A.}\ \bibnamefont {Gunnels}},
  \bibinfo {author} {\bibfnamefont {H.}~\bibnamefont {Gupta}}, \bibinfo
  {author} {\bibfnamefont {N.}~\bibnamefont {Gupta}}, \bibinfo {author}
  {\bibfnamefont {J.~M.}\ \bibnamefont {G{\"u}nther}}, \bibinfo {author}
  {\bibfnamefont {M.}~\bibnamefont {Haglund}}, \bibinfo {author} {\bibfnamefont
  {I.}~\bibnamefont {Haide}}, \bibinfo {author} {\bibfnamefont
  {I.}~\bibnamefont {Hamamura}}, \bibinfo {author} {\bibfnamefont {O.~C.}\
  \bibnamefont {Hamido}}, \bibinfo {author} {\bibfnamefont {F.}~\bibnamefont
  {Harkins}}, \bibinfo {author} {\bibfnamefont {K.}~\bibnamefont {Hartman}},
  \bibinfo {author} {\bibfnamefont {A.}~\bibnamefont {Hasan}}, \bibinfo
  {author} {\bibfnamefont {V.}~\bibnamefont {Havlicek}}, \bibinfo {author}
  {\bibfnamefont {J.}~\bibnamefont {Hellmers}}, \bibinfo {author}
  {\bibfnamefont {{\L}.}~\bibnamefont {Herok}}, \bibinfo {author}
  {\bibfnamefont {S.}~\bibnamefont {Hillmich}}, \bibinfo {author}
  {\bibfnamefont {H.}~\bibnamefont {Horii}}, \bibinfo {author} {\bibfnamefont
  {C.}~\bibnamefont {Howington}}, \bibinfo {author} {\bibfnamefont
  {S.}~\bibnamefont {Hu}}, \bibinfo {author} {\bibfnamefont {W.}~\bibnamefont
  {Hu}}, \bibinfo {author} {\bibfnamefont {J.}~\bibnamefont {Huang}}, \bibinfo
  {author} {\bibfnamefont {R.}~\bibnamefont {Huisman}}, \bibinfo {author}
  {\bibfnamefont {H.}~\bibnamefont {Imai}}, \bibinfo {author} {\bibfnamefont
  {T.}~\bibnamefont {Imamichi}}, \bibinfo {author} {\bibfnamefont
  {K.}~\bibnamefont {Ishizaki}}, \bibinfo {author} {\bibnamefont {Ishwor}},
  \bibinfo {author} {\bibfnamefont {R.}~\bibnamefont {Iten}}, \bibinfo {author}
  {\bibfnamefont {T.}~\bibnamefont {Itoko}}, \bibinfo {author} {\bibfnamefont
  {A.}~\bibnamefont {Ivrii}}, \bibinfo {author} {\bibfnamefont
  {A.}~\bibnamefont {Javadi}}, \bibinfo {author} {\bibfnamefont
  {A.}~\bibnamefont {Javadi-Abhari}}, \bibinfo {author} {\bibfnamefont
  {W.}~\bibnamefont {Javed}}, \bibinfo {author} {\bibfnamefont
  {Q.}~\bibnamefont {Jianhua}}, \bibinfo {author} {\bibfnamefont
  {M.}~\bibnamefont {Jivrajani}}, \bibinfo {author} {\bibfnamefont
  {K.}~\bibnamefont {Johns}}, \bibinfo {author} {\bibfnamefont
  {S.}~\bibnamefont {Johnstun}}, \bibinfo {author} {\bibnamefont
  {Jonathan-Shoemaker}}, \bibinfo {author} {\bibnamefont {JosDenmark}},
  \bibinfo {author} {\bibnamefont {JoshDumo}}, \bibinfo {author} {\bibfnamefont
  {J.}~\bibnamefont {Judge}}, \bibinfo {author} {\bibfnamefont
  {T.}~\bibnamefont {Kachmann}}, \bibinfo {author} {\bibfnamefont
  {A.}~\bibnamefont {Kale}}, \bibinfo {author} {\bibfnamefont {N.}~\bibnamefont
  {Kanazawa}}, \bibinfo {author} {\bibfnamefont {J.}~\bibnamefont {Kane}},
  \bibinfo {author} {\bibnamefont {Kang-Bae}}, \bibinfo {author} {\bibfnamefont
  {A.}~\bibnamefont {Kapila}}, \bibinfo {author} {\bibfnamefont
  {A.}~\bibnamefont {Karazeev}}, \bibinfo {author} {\bibfnamefont
  {P.}~\bibnamefont {Kassebaum}}, \bibinfo {author} {\bibfnamefont
  {J.}~\bibnamefont {Kelso}}, \bibinfo {author} {\bibfnamefont
  {S.}~\bibnamefont {Kelso}}, \bibinfo {author} {\bibfnamefont
  {V.}~\bibnamefont {Khanderao}}, \bibinfo {author} {\bibfnamefont
  {S.}~\bibnamefont {King}}, \bibinfo {author} {\bibfnamefont {Y.}~\bibnamefont
  {Kobayashi}}, \bibinfo {author} {\bibnamefont {Kovi11Day}}, \bibinfo {author}
  {\bibfnamefont {A.}~\bibnamefont {Kovyrshin}}, \bibinfo {author}
  {\bibfnamefont {R.}~\bibnamefont {Krishnakumar}}, \bibinfo {author}
  {\bibfnamefont {V.}~\bibnamefont {Krishnan}}, \bibinfo {author}
  {\bibfnamefont {K.}~\bibnamefont {Krsulich}}, \bibinfo {author}
  {\bibfnamefont {P.}~\bibnamefont {Kumkar}}, \bibinfo {author} {\bibfnamefont
  {G.}~\bibnamefont {Kus}}, \bibinfo {author} {\bibfnamefont {R.}~\bibnamefont
  {LaRose}}, \bibinfo {author} {\bibfnamefont {E.}~\bibnamefont {Lacal}},
  \bibinfo {author} {\bibfnamefont {R.}~\bibnamefont {Lambert}}, \bibinfo
  {author} {\bibfnamefont {H.}~\bibnamefont {Landa}}, \bibinfo {author}
  {\bibfnamefont {J.}~\bibnamefont {Lapeyre}}, \bibinfo {author} {\bibfnamefont
  {J.}~\bibnamefont {Latone}}, \bibinfo {author} {\bibfnamefont
  {S.}~\bibnamefont {Lawrence}}, \bibinfo {author} {\bibfnamefont
  {C.}~\bibnamefont {Lee}}, \bibinfo {author} {\bibfnamefont {G.}~\bibnamefont
  {Li}}, \bibinfo {author} {\bibfnamefont {J.}~\bibnamefont {Lishman}},
  \bibinfo {author} {\bibfnamefont {D.}~\bibnamefont {Liu}}, \bibinfo {author}
  {\bibfnamefont {P.}~\bibnamefont {Liu}}, \bibinfo {author} {\bibnamefont
  {Lolcroc}}, \bibinfo {author} {\bibfnamefont {A.~K.}\ \bibnamefont {M}},
  \bibinfo {author} {\bibfnamefont {L.}~\bibnamefont {Madden}}, \bibinfo
  {author} {\bibfnamefont {Y.}~\bibnamefont {Maeng}}, \bibinfo {author}
  {\bibfnamefont {S.}~\bibnamefont {Maheshkar}}, \bibinfo {author}
  {\bibfnamefont {K.}~\bibnamefont {Majmudar}}, \bibinfo {author}
  {\bibfnamefont {A.}~\bibnamefont {Malyshev}}, \bibinfo {author}
  {\bibfnamefont {M.~E.}\ \bibnamefont {Mandouh}}, \bibinfo {author}
  {\bibfnamefont {J.}~\bibnamefont {Manela}}, \bibinfo {author} {\bibnamefont
  {Manjula}}, \bibinfo {author} {\bibfnamefont {J.}~\bibnamefont {Marecek}},
  \bibinfo {author} {\bibfnamefont {M.}~\bibnamefont {Marques}}, \bibinfo
  {author} {\bibfnamefont {K.}~\bibnamefont {Marwaha}}, \bibinfo {author}
  {\bibfnamefont {D.}~\bibnamefont {Maslov}}, \bibinfo {author} {\bibfnamefont
  {P.}~\bibnamefont {Maszota}}, \bibinfo {author} {\bibfnamefont
  {D.}~\bibnamefont {Mathews}}, \bibinfo {author} {\bibfnamefont
  {A.}~\bibnamefont {Matsuo}}, \bibinfo {author} {\bibfnamefont
  {F.}~\bibnamefont {Mazhandu}}, \bibinfo {author} {\bibfnamefont
  {D.}~\bibnamefont {McClure}}, \bibinfo {author} {\bibfnamefont
  {M.}~\bibnamefont {McElaney}}, \bibinfo {author} {\bibfnamefont
  {C.}~\bibnamefont {McGarry}}, \bibinfo {author} {\bibfnamefont
  {D.}~\bibnamefont {McKay}}, \bibinfo {author} {\bibfnamefont
  {D.}~\bibnamefont {McPherson}}, \bibinfo {author} {\bibfnamefont
  {S.}~\bibnamefont {Meesala}}, \bibinfo {author} {\bibfnamefont
  {D.}~\bibnamefont {Meirom}}, \bibinfo {author} {\bibfnamefont
  {C.}~\bibnamefont {Mendell}}, \bibinfo {author} {\bibfnamefont
  {T.}~\bibnamefont {Metcalfe}}, \bibinfo {author} {\bibfnamefont
  {M.}~\bibnamefont {Mevissen}}, \bibinfo {author} {\bibfnamefont
  {A.}~\bibnamefont {Meyer}}, \bibinfo {author} {\bibfnamefont
  {A.}~\bibnamefont {Mezzacapo}}, \bibinfo {author} {\bibfnamefont
  {R.}~\bibnamefont {Midha}}, \bibinfo {author} {\bibfnamefont
  {D.}~\bibnamefont {Miller}}, \bibinfo {author} {\bibfnamefont
  {Z.}~\bibnamefont {Minev}}, \bibinfo {author} {\bibfnamefont
  {A.}~\bibnamefont {Mitchell}}, \bibinfo {author} {\bibfnamefont
  {N.}~\bibnamefont {Moll}}, \bibinfo {author} {\bibfnamefont {A.}~\bibnamefont
  {Montanez}}, \bibinfo {author} {\bibfnamefont {G.}~\bibnamefont {Monteiro}},
  \bibinfo {author} {\bibfnamefont {M.~D.}\ \bibnamefont {Mooring}}, \bibinfo
  {author} {\bibfnamefont {R.}~\bibnamefont {Morales}}, \bibinfo {author}
  {\bibfnamefont {N.}~\bibnamefont {Moran}}, \bibinfo {author} {\bibfnamefont
  {D.}~\bibnamefont {Morcuende}}, \bibinfo {author} {\bibfnamefont
  {S.}~\bibnamefont {Mostafa}}, \bibinfo {author} {\bibfnamefont
  {M.}~\bibnamefont {Motta}}, \bibinfo {author} {\bibfnamefont
  {R.}~\bibnamefont {Moyard}}, \bibinfo {author} {\bibfnamefont
  {P.}~\bibnamefont {Murali}}, \bibinfo {author} {\bibfnamefont
  {J.}~\bibnamefont {M{\"u}ggenburg}}, \bibinfo {author} {\bibfnamefont
  {T.}~\bibnamefont {NEMOZ}}, \bibinfo {author} {\bibfnamefont
  {D.}~\bibnamefont {Nadlinger}}, \bibinfo {author} {\bibfnamefont
  {K.}~\bibnamefont {Nakanishi}}, \bibinfo {author} {\bibfnamefont
  {G.}~\bibnamefont {Nannicini}}, \bibinfo {author} {\bibfnamefont
  {P.}~\bibnamefont {Nation}}, \bibinfo {author} {\bibfnamefont
  {E.}~\bibnamefont {Navarro}}, \bibinfo {author} {\bibfnamefont
  {Y.}~\bibnamefont {Naveh}}, \bibinfo {author} {\bibfnamefont {S.~W.}\
  \bibnamefont {Neagle}}, \bibinfo {author} {\bibfnamefont {P.}~\bibnamefont
  {Neuweiler}}, \bibinfo {author} {\bibfnamefont {A.}~\bibnamefont {Ngoueya}},
  \bibinfo {author} {\bibfnamefont {J.}~\bibnamefont {Nicander}}, \bibinfo
  {author} {\bibnamefont {Nick-Singstock}}, \bibinfo {author} {\bibfnamefont
  {P.}~\bibnamefont {Niroula}}, \bibinfo {author} {\bibfnamefont
  {H.}~\bibnamefont {Norlen}}, \bibinfo {author} {\bibnamefont {NuoWenLei}},
  \bibinfo {author} {\bibfnamefont {L.~J.}\ \bibnamefont {O'Riordan}}, \bibinfo
  {author} {\bibfnamefont {O.}~\bibnamefont {Ogunbayo}}, \bibinfo {author}
  {\bibfnamefont {P.}~\bibnamefont {Ollitrault}}, \bibinfo {author}
  {\bibfnamefont {T.}~\bibnamefont {Onodera}}, \bibinfo {author} {\bibfnamefont
  {R.}~\bibnamefont {Otaolea}}, \bibinfo {author} {\bibfnamefont
  {S.}~\bibnamefont {Oud}}, \bibinfo {author} {\bibfnamefont {D.}~\bibnamefont
  {Padilha}}, \bibinfo {author} {\bibfnamefont {H.}~\bibnamefont {Paik}},
  \bibinfo {author} {\bibfnamefont {S.}~\bibnamefont {Pal}}, \bibinfo {author}
  {\bibfnamefont {Y.}~\bibnamefont {Pang}}, \bibinfo {author} {\bibfnamefont
  {A.}~\bibnamefont {Panigrahi}}, \bibinfo {author} {\bibfnamefont {V.~R.}\
  \bibnamefont {Pascuzzi}}, \bibinfo {author} {\bibfnamefont {S.}~\bibnamefont
  {Perriello}}, \bibinfo {author} {\bibfnamefont {E.}~\bibnamefont {Peterson}},
  \bibinfo {author} {\bibfnamefont {A.}~\bibnamefont {Phan}}, \bibinfo {author}
  {\bibfnamefont {K.}~\bibnamefont {Pilch}}, \bibinfo {author} {\bibfnamefont
  {F.}~\bibnamefont {Piro}}, \bibinfo {author} {\bibfnamefont {M.}~\bibnamefont
  {Pistoia}}, \bibinfo {author} {\bibfnamefont {C.}~\bibnamefont {Piveteau}},
  \bibinfo {author} {\bibfnamefont {J.}~\bibnamefont {Plewa}}, \bibinfo
  {author} {\bibfnamefont {P.}~\bibnamefont {Pocreau}}, \bibinfo {author}
  {\bibfnamefont {A.}~\bibnamefont {Pozas-Kerstjens}}, \bibinfo {author}
  {\bibfnamefont {R.}~\bibnamefont {Pracht}}, \bibinfo {author} {\bibfnamefont
  {M.}~\bibnamefont {Prokop}}, \bibinfo {author} {\bibfnamefont
  {V.}~\bibnamefont {Prutyanov}}, \bibinfo {author} {\bibfnamefont
  {S.}~\bibnamefont {Puri}}, \bibinfo {author} {\bibfnamefont {D.}~\bibnamefont
  {Puzzuoli}}, \bibinfo {author} {\bibfnamefont {J.}~\bibnamefont {P{\'e}rez}},
  \bibinfo {author} {\bibnamefont {Quant02}}, \bibinfo {author} {\bibnamefont
  {Quintiii}}, \bibinfo {author} {\bibfnamefont {R.~I.}\ \bibnamefont
  {Rahman}}, \bibinfo {author} {\bibfnamefont {A.}~\bibnamefont {Raja}},
  \bibinfo {author} {\bibfnamefont {R.}~\bibnamefont {Rajeev}}, \bibinfo
  {author} {\bibfnamefont {I.}~\bibnamefont {Rajput}}, \bibinfo {author}
  {\bibfnamefont {N.}~\bibnamefont {Ramagiri}}, \bibinfo {author}
  {\bibfnamefont {A.}~\bibnamefont {Rao}}, \bibinfo {author} {\bibfnamefont
  {R.}~\bibnamefont {Raymond}}, \bibinfo {author} {\bibfnamefont
  {O.}~\bibnamefont {Reardon-Smith}}, \bibinfo {author} {\bibfnamefont
  {R.~M.-C.}\ \bibnamefont {Redondo}}, \bibinfo {author} {\bibfnamefont
  {M.}~\bibnamefont {Reuter}}, \bibinfo {author} {\bibfnamefont
  {J.}~\bibnamefont {Rice}}, \bibinfo {author} {\bibfnamefont {M.}~\bibnamefont
  {Riedemann}}, \bibinfo {author} {\bibnamefont {Rietesh}}, \bibinfo {author}
  {\bibfnamefont {D.}~\bibnamefont {Risinger}}, \bibinfo {author}
  {\bibfnamefont {M.~L.}\ \bibnamefont {Rocca}}, \bibinfo {author}
  {\bibfnamefont {D.~M.}\ \bibnamefont {Rodr{\'\i}guez}}, \bibinfo {author}
  {\bibnamefont {RohithKarur}}, \bibinfo {author} {\bibfnamefont
  {B.}~\bibnamefont {Rosand}}, \bibinfo {author} {\bibfnamefont
  {M.}~\bibnamefont {Rossmannek}}, \bibinfo {author} {\bibfnamefont
  {M.}~\bibnamefont {Ryu}}, \bibinfo {author} {\bibfnamefont {T.}~\bibnamefont
  {SAPV}}, \bibinfo {author} {\bibfnamefont {N.~R.~C.}\ \bibnamefont {Sa}},
  \bibinfo {author} {\bibfnamefont {A.}~\bibnamefont {Saha}}, \bibinfo {author}
  {\bibfnamefont {A.}~\bibnamefont {Ash-Saki}}, \bibinfo {author}
  {\bibfnamefont {S.}~\bibnamefont {Sanand}}, \bibinfo {author} {\bibfnamefont
  {M.}~\bibnamefont {Sandberg}}, \bibinfo {author} {\bibfnamefont
  {H.}~\bibnamefont {Sandesara}}, \bibinfo {author} {\bibfnamefont
  {R.}~\bibnamefont {Sapra}}, \bibinfo {author} {\bibfnamefont
  {H.}~\bibnamefont {Sargsyan}}, \bibinfo {author} {\bibfnamefont
  {A.}~\bibnamefont {Sarkar}}, \bibinfo {author} {\bibfnamefont
  {N.}~\bibnamefont {Sathaye}}, \bibinfo {author} {\bibfnamefont
  {B.}~\bibnamefont {Schmitt}}, \bibinfo {author} {\bibfnamefont
  {C.}~\bibnamefont {Schnabel}}, \bibinfo {author} {\bibfnamefont
  {Z.}~\bibnamefont {Schoenfeld}}, \bibinfo {author} {\bibfnamefont {T.~L.}\
  \bibnamefont {Scholten}}, \bibinfo {author} {\bibfnamefont {E.}~\bibnamefont
  {Schoute}}, \bibinfo {author} {\bibfnamefont {M.}~\bibnamefont
  {Schulterbrandt}}, \bibinfo {author} {\bibfnamefont {J.}~\bibnamefont
  {Schwarm}}, \bibinfo {author} {\bibfnamefont {J.}~\bibnamefont {Seaward}},
  \bibinfo {author} {\bibnamefont {Sergi}}, \bibinfo {author} {\bibfnamefont
  {I.~F.}\ \bibnamefont {Sertage}}, \bibinfo {author} {\bibfnamefont
  {K.}~\bibnamefont {Setia}}, \bibinfo {author} {\bibfnamefont
  {F.}~\bibnamefont {Shah}}, \bibinfo {author} {\bibfnamefont {N.}~\bibnamefont
  {Shammah}}, \bibinfo {author} {\bibfnamefont {R.}~\bibnamefont {Sharma}},
  \bibinfo {author} {\bibfnamefont {Y.}~\bibnamefont {Shi}}, \bibinfo {author}
  {\bibfnamefont {J.}~\bibnamefont {Shoemaker}}, \bibinfo {author}
  {\bibfnamefont {A.}~\bibnamefont {Silva}}, \bibinfo {author} {\bibfnamefont
  {A.}~\bibnamefont {Simonetto}}, \bibinfo {author} {\bibfnamefont
  {D.}~\bibnamefont {Singh}}, \bibinfo {author} {\bibfnamefont
  {D.}~\bibnamefont {Singh}}, \bibinfo {author} {\bibfnamefont
  {P.}~\bibnamefont {Singh}}, \bibinfo {author} {\bibfnamefont
  {P.}~\bibnamefont {Singkanipa}}, \bibinfo {author} {\bibfnamefont
  {Y.}~\bibnamefont {Siraichi}}, \bibinfo {author} {\bibnamefont {Siri}},
  \bibinfo {author} {\bibfnamefont {J.}~\bibnamefont {Sistos}}, \bibinfo
  {author} {\bibfnamefont {I.}~\bibnamefont {Sitdikov}}, \bibinfo {author}
  {\bibfnamefont {S.}~\bibnamefont {Sivarajah}}, \bibinfo {author}
  {\bibfnamefont {M.~B.}\ \bibnamefont {Sletfjerding}}, \bibinfo {author}
  {\bibfnamefont {J.~A.}\ \bibnamefont {Smolin}}, \bibinfo {author}
  {\bibfnamefont {M.}~\bibnamefont {Soeken}}, \bibinfo {author} {\bibfnamefont
  {I.~O.}\ \bibnamefont {Sokolov}}, \bibinfo {author} {\bibfnamefont
  {I.}~\bibnamefont {Sokolov}}, \bibinfo {author} {\bibfnamefont {V.~P.}\
  \bibnamefont {Soloviev}}, \bibinfo {author} {\bibnamefont {SooluThomas}},
  \bibinfo {author} {\bibnamefont {Starfish}}, \bibinfo {author} {\bibfnamefont
  {D.}~\bibnamefont {Steenken}}, \bibinfo {author} {\bibfnamefont
  {M.}~\bibnamefont {Stypulkoski}}, \bibinfo {author} {\bibfnamefont
  {A.}~\bibnamefont {Suau}}, \bibinfo {author} {\bibfnamefont {S.}~\bibnamefont
  {Sun}}, \bibinfo {author} {\bibfnamefont {K.~J.}\ \bibnamefont {Sung}},
  \bibinfo {author} {\bibfnamefont {M.}~\bibnamefont {Suwama}}, \bibinfo
  {author} {\bibfnamefont {O.}~\bibnamefont {S{\l}owik}}, \bibinfo {author}
  {\bibfnamefont {H.}~\bibnamefont {Takahashi}}, \bibinfo {author}
  {\bibfnamefont {T.}~\bibnamefont {Takawale}}, \bibinfo {author}
  {\bibfnamefont {I.}~\bibnamefont {Tavernelli}}, \bibinfo {author}
  {\bibfnamefont {C.}~\bibnamefont {Taylor}}, \bibinfo {author} {\bibfnamefont
  {P.}~\bibnamefont {Taylour}}, \bibinfo {author} {\bibfnamefont
  {S.}~\bibnamefont {Thomas}}, \bibinfo {author} {\bibfnamefont
  {K.}~\bibnamefont {Tian}}, \bibinfo {author} {\bibfnamefont {M.}~\bibnamefont
  {Tillet}}, \bibinfo {author} {\bibfnamefont {M.}~\bibnamefont {Tod}},
  \bibinfo {author} {\bibfnamefont {M.}~\bibnamefont {Tomasik}}, \bibinfo
  {author} {\bibfnamefont {C.}~\bibnamefont {Tornow}}, \bibinfo {author}
  {\bibfnamefont {E.}~\bibnamefont {de~la Torre}}, \bibinfo {author}
  {\bibfnamefont {J.~L.~S.}\ \bibnamefont {Toural}}, \bibinfo {author}
  {\bibfnamefont {K.}~\bibnamefont {Trabing}}, \bibinfo {author} {\bibfnamefont
  {M.}~\bibnamefont {Treinish}}, \bibinfo {author} {\bibfnamefont
  {D.}~\bibnamefont {Trenev}}, \bibinfo {author} {\bibnamefont {TrishaPe}},
  \bibinfo {author} {\bibfnamefont {F.}~\bibnamefont {Truger}}, \bibinfo
  {author} {\bibfnamefont {G.}~\bibnamefont {Tsilimigkounakis}}, \bibinfo
  {author} {\bibfnamefont {D.}~\bibnamefont {Tulsi}}, \bibinfo {author}
  {\bibfnamefont {W.}~\bibnamefont {Turner}}, \bibinfo {author} {\bibfnamefont
  {Y.}~\bibnamefont {Vaknin}}, \bibinfo {author} {\bibfnamefont {C.~R.}\
  \bibnamefont {Valcarce}}, \bibinfo {author} {\bibfnamefont {F.}~\bibnamefont
  {Varchon}}, \bibinfo {author} {\bibfnamefont {A.}~\bibnamefont {Vartak}},
  \bibinfo {author} {\bibfnamefont {A.~C.}\ \bibnamefont {Vazquez}}, \bibinfo
  {author} {\bibfnamefont {P.}~\bibnamefont {Vijaywargiya}}, \bibinfo {author}
  {\bibfnamefont {V.}~\bibnamefont {Villar}}, \bibinfo {author} {\bibfnamefont
  {B.}~\bibnamefont {Vishnu}}, \bibinfo {author} {\bibfnamefont
  {D.}~\bibnamefont {Vogt-Lee}}, \bibinfo {author} {\bibfnamefont
  {C.}~\bibnamefont {Vuillot}}, \bibinfo {author} {\bibfnamefont
  {J.}~\bibnamefont {Weaver}}, \bibinfo {author} {\bibfnamefont
  {J.}~\bibnamefont {Weidenfeller}}, \bibinfo {author} {\bibfnamefont
  {R.}~\bibnamefont {Wieczorek}}, \bibinfo {author} {\bibfnamefont {J.~A.}\
  \bibnamefont {Wildstrom}}, \bibinfo {author} {\bibfnamefont {J.}~\bibnamefont
  {Wilson}}, \bibinfo {author} {\bibfnamefont {E.}~\bibnamefont {Winston}},
  \bibinfo {author} {\bibnamefont {WinterSoldier}}, \bibinfo {author}
  {\bibfnamefont {J.~J.}\ \bibnamefont {Woehr}}, \bibinfo {author}
  {\bibfnamefont {S.}~\bibnamefont {Woerner}}, \bibinfo {author} {\bibfnamefont
  {R.}~\bibnamefont {Woo}}, \bibinfo {author} {\bibfnamefont {C.~J.}\
  \bibnamefont {Wood}}, \bibinfo {author} {\bibfnamefont {R.}~\bibnamefont
  {Wood}}, \bibinfo {author} {\bibfnamefont {S.}~\bibnamefont {Wood}}, \bibinfo
  {author} {\bibfnamefont {J.}~\bibnamefont {Wootton}}, \bibinfo {author}
  {\bibfnamefont {M.}~\bibnamefont {Wright}}, \bibinfo {author} {\bibfnamefont
  {L.}~\bibnamefont {Xing}}, \bibinfo {author} {\bibfnamefont {J.}~\bibnamefont
  {YU}}, \bibinfo {author} {\bibfnamefont {B.}~\bibnamefont {Yang}}, \bibinfo
  {author} {\bibfnamefont {U.}~\bibnamefont {Yang}}, \bibinfo {author}
  {\bibfnamefont {J.}~\bibnamefont {Yao}}, \bibinfo {author} {\bibfnamefont
  {D.}~\bibnamefont {Yeralin}}, \bibinfo {author} {\bibfnamefont
  {R.}~\bibnamefont {Yonekura}}, \bibinfo {author} {\bibfnamefont
  {D.}~\bibnamefont {Yonge-Mallo}}, \bibinfo {author} {\bibfnamefont
  {R.}~\bibnamefont {Yoshida}}, \bibinfo {author} {\bibfnamefont
  {R.}~\bibnamefont {Young}}, \bibinfo {author} {\bibfnamefont
  {J.}~\bibnamefont {Yu}}, \bibinfo {author} {\bibfnamefont {L.}~\bibnamefont
  {Yu}}, \bibinfo {author} {\bibfnamefont {C.}~\bibnamefont {Zachow}}, \bibinfo
  {author} {\bibfnamefont {L.}~\bibnamefont {Zdanski}}, \bibinfo {author}
  {\bibfnamefont {H.}~\bibnamefont {Zhang}}, \bibinfo {author} {\bibfnamefont
  {I.}~\bibnamefont {Zidaru}}, \bibinfo {author} {\bibfnamefont
  {C.}~\bibnamefont {Zoufal}}, \bibinfo {author} {\bibnamefont {aeddins ibm}},
  \bibinfo {author} {\bibnamefont {alexzhang13}}, \bibinfo {author}
  {\bibnamefont {b63}}, \bibinfo {author} {\bibnamefont {bartek bartlomiej}},
  \bibinfo {author} {\bibnamefont {bcamorrison}}, \bibinfo {author}
  {\bibnamefont {brandhsn}}, \bibinfo {author} {\bibnamefont {charmerDark}},
  \bibinfo {author} {\bibnamefont {deeplokhande}}, \bibinfo {author}
  {\bibnamefont {dekel.meirom}}, \bibinfo {author} {\bibnamefont {dime10}},
  \bibinfo {author} {\bibnamefont {dlasecki}}, \bibinfo {author} {\bibnamefont
  {ehchen}}, \bibinfo {author} {\bibnamefont {fanizzamarco}}, \bibinfo {author}
  {\bibnamefont {fs1132429}}, \bibinfo {author} {\bibnamefont {gadial}},
  \bibinfo {author} {\bibnamefont {galeinston}}, \bibinfo {author}
  {\bibnamefont {georgezhou20}}, \bibinfo {author} {\bibnamefont {georgios
  ts}}, \bibinfo {author} {\bibnamefont {gruu}}, \bibinfo {author}
  {\bibnamefont {hhorii}}, \bibinfo {author} {\bibnamefont {hykavitha}},
  \bibinfo {author} {\bibnamefont {itoko}}, \bibinfo {author} {\bibnamefont
  {jeppevinkel}}, \bibinfo {author} {\bibnamefont {jessica angel7}}, \bibinfo
  {author} {\bibnamefont {jezerjojo14}}, \bibinfo {author} {\bibnamefont
  {jliu45}}, \bibinfo {author} {\bibnamefont {jscott2}}, \bibinfo {author}
  {\bibnamefont {klinvill}}, \bibinfo {author} {\bibnamefont {krutik2966}},
  \bibinfo {author} {\bibnamefont {ma5x}}, \bibinfo {author} {\bibnamefont
  {michelle4654}}, \bibinfo {author} {\bibnamefont {msuwama}}, \bibinfo
  {author} {\bibnamefont {nico lgrs}}, \bibinfo {author} {\bibnamefont
  {ntgiwsvp}}, \bibinfo {author} {\bibnamefont {ordmoj}}, \bibinfo {author}
  {\bibnamefont {sagar pahwa}}, \bibinfo {author} {\bibnamefont
  {pritamsinha2304}}, \bibinfo {author} {\bibnamefont {ryancocuzzo}}, \bibinfo
  {author} {\bibnamefont {saktar unr}}, \bibinfo {author} {\bibnamefont
  {saswati qiskit}}, \bibinfo {author} {\bibnamefont {septembrr}}, \bibinfo
  {author} {\bibnamefont {sethmerkel}}, \bibinfo {author} {\bibnamefont
  {sg495}}, \bibinfo {author} {\bibnamefont {shaashwat}}, \bibinfo {author}
  {\bibnamefont {smturro2}}, \bibinfo {author} {\bibnamefont {sternparky}},
  \bibinfo {author} {\bibnamefont {strickroman}}, \bibinfo {author}
  {\bibnamefont {tigerjack}}, \bibinfo {author} {\bibnamefont {tsura
  crisaldo}}, \bibinfo {author} {\bibnamefont {vadebayo49}}, \bibinfo {author}
  {\bibnamefont {welien}}, \bibinfo {author} {\bibnamefont {willhbang}},
  \bibinfo {author} {\bibnamefont {wmurphy collabstar}}, \bibinfo {author}
  {\bibnamefont {yang.luh}}, \ and\ \bibinfo {author} {\bibfnamefont
  {M.}~\bibnamefont {{\v{C}}epulkovskis}},\ }\href {\doibase
  10.5281/zenodo.2573505} {\enquote {\bibinfo {title} {Qiskit: An open-source
  framework for quantum computing},}\ } (\bibinfo {year} {2021})\BibitemShut
  {NoStop}%
\bibitem [{ibm()}]{ibm}%
  \BibitemOpen
  \href {https://quantum-computing.ibm.com/} {\enquote {\bibinfo {title} {Ibm
  quantum},}\ }\BibitemShut {NoStop}%
\bibitem [{\citenamefont {Gross}\ \emph {et~al.}(1981)\citenamefont {Gross},
  \citenamefont {Pisarski},\ and\ \citenamefont {Yaffe}}]{RevModPhys.53.43}%
  \BibitemOpen
  \bibfield  {author} {\bibinfo {author} {\bibfnamefont {D.~J.}\ \bibnamefont
  {Gross}}, \bibinfo {author} {\bibfnamefont {R.~D.}\ \bibnamefont {Pisarski}},
  \ and\ \bibinfo {author} {\bibfnamefont {L.~G.}\ \bibnamefont {Yaffe}},\
  }\href {\doibase 10.1103/RevModPhys.53.43} {\bibfield  {journal} {\bibinfo
  {journal} {Rev. Mod. Phys.}\ }\textbf {\bibinfo {volume} {53}},\ \bibinfo
  {pages} {43} (\bibinfo {year} {1981})}\BibitemShut {NoStop}%
\bibitem [{\citenamefont {Nambu}\ and\ \citenamefont
  {Jona-Lasinio}(1961{\natexlab{c}})}]{PhysRev.122.345}%
  \BibitemOpen
  \bibfield  {author} {\bibinfo {author} {\bibfnamefont {Y.}~\bibnamefont
  {Nambu}}\ and\ \bibinfo {author} {\bibfnamefont {G.}~\bibnamefont
  {Jona-Lasinio}},\ }\href {\doibase 10.1103/PhysRev.122.345} {\bibfield
  {journal} {\bibinfo  {journal} {Phys. Rev.}\ }\textbf {\bibinfo {volume}
  {122}},\ \bibinfo {pages} {345} (\bibinfo {year}
  {1961}{\natexlab{c}})}\BibitemShut {NoStop}%
\bibitem [{\citenamefont {McLerran}\ and\ \citenamefont
  {Svetitsky}(1981)}]{PhysRevD.24.450}%
  \BibitemOpen
  \bibfield  {author} {\bibinfo {author} {\bibfnamefont {L.~D.}\ \bibnamefont
  {McLerran}}\ and\ \bibinfo {author} {\bibfnamefont {B.}~\bibnamefont
  {Svetitsky}},\ }\href {\doibase 10.1103/PhysRevD.24.450} {\bibfield
  {journal} {\bibinfo  {journal} {Phys. Rev. D}\ }\textbf {\bibinfo {volume}
  {24}},\ \bibinfo {pages} {450} (\bibinfo {year} {1981})}\BibitemShut
  {NoStop}%
\bibitem [{\citenamefont {Polyakov}(1978)}]{POLYAKOV1978477}%
  \BibitemOpen
  \bibfield  {author} {\bibinfo {author} {\bibfnamefont {A.}~\bibnamefont
  {Polyakov}},\ }\href {\doibase https://doi.org/10.1016/0370-2693(78)90737-2}
  {\bibfield  {journal} {\bibinfo  {journal} {Physics Letters B}\ }\textbf
  {\bibinfo {volume} {72}},\ \bibinfo {pages} {477} (\bibinfo {year}
  {1978})}\BibitemShut {NoStop}%
\bibitem [{\citenamefont {Fang}\ \emph {et~al.}(2018)\citenamefont {Fang},
  \citenamefont {Wu},\ and\ \citenamefont {Zhang}}]{Fang:2018vkp}%
  \BibitemOpen
  \bibfield  {author} {\bibinfo {author} {\bibfnamefont {Z.}~\bibnamefont
  {Fang}}, \bibinfo {author} {\bibfnamefont {Y.-L.}\ \bibnamefont {Wu}}, \ and\
  \bibinfo {author} {\bibfnamefont {L.}~\bibnamefont {Zhang}},\ }\href
  {\doibase 10.1103/PhysRevD.98.114003} {\bibfield  {journal} {\bibinfo
  {journal} {Phys. Rev. D}\ }\textbf {\bibinfo {volume} {98}},\ \bibinfo
  {pages} {114003} (\bibinfo {year} {2018})},\ \Eprint
  {http://arxiv.org/abs/1805.05019} {arXiv:1805.05019 [hep-ph]} \BibitemShut
  {NoStop}%
\bibitem [{\citenamefont {Walecka}(1974)}]{WALECKA1974491}%
  \BibitemOpen
  \bibfield  {author} {\bibinfo {author} {\bibfnamefont {J.}~\bibnamefont
  {Walecka}},\ }\href {\doibase https://doi.org/10.1016/0003-4916(74)90208-5}
  {\bibfield  {journal} {\bibinfo  {journal} {Annals of Physics}\ }\textbf
  {\bibinfo {volume} {83}},\ \bibinfo {pages} {491} (\bibinfo {year}
  {1974})}\BibitemShut {NoStop}%
\bibitem [{\citenamefont {Ohata}\ and\ \citenamefont
  {Suganuma}(2021)}]{Ohata:2020myj}%
  \BibitemOpen
  \bibfield  {author} {\bibinfo {author} {\bibfnamefont {H.}~\bibnamefont
  {Ohata}}\ and\ \bibinfo {author} {\bibfnamefont {H.}~\bibnamefont
  {Suganuma}},\ }\href {\doibase 10.1103/PhysRevD.103.054505} {\bibfield
  {journal} {\bibinfo  {journal} {Phys. Rev. D}\ }\textbf {\bibinfo {volume}
  {103}},\ \bibinfo {pages} {054505} (\bibinfo {year} {2021})},\ \Eprint
  {http://arxiv.org/abs/2012.03537} {arXiv:2012.03537 [hep-lat]} \BibitemShut
  {NoStop}%
\bibitem [{\citenamefont {Carabba}\ and\ \citenamefont
  {Meggiolaro}(2022)}]{Carabba:2021xmc}%
  \BibitemOpen
  \bibfield  {author} {\bibinfo {author} {\bibfnamefont {N.}~\bibnamefont
  {Carabba}}\ and\ \bibinfo {author} {\bibfnamefont {E.}~\bibnamefont
  {Meggiolaro}},\ }\href {\doibase 10.1103/PhysRevD.105.054034} {\bibfield
  {journal} {\bibinfo  {journal} {Phys. Rev. D}\ }\textbf {\bibinfo {volume}
  {105}},\ \bibinfo {pages} {054034} (\bibinfo {year} {2022})},\ \Eprint
  {http://arxiv.org/abs/2106.10074} {arXiv:2106.10074 [hep-ph]} \BibitemShut
  {NoStop}%
\bibitem [{\citenamefont {Smith}\ \emph {et~al.}(2016)\citenamefont {Smith},
  \citenamefont {Curtis},\ and\ \citenamefont
  {Zeng}}]{https://doi.org/10.48550/arxiv.1608.03355}%
  \BibitemOpen
  \bibfield  {author} {\bibinfo {author} {\bibfnamefont {R.~S.}\ \bibnamefont
  {Smith}}, \bibinfo {author} {\bibfnamefont {M.~J.}\ \bibnamefont {Curtis}}, \
  and\ \bibinfo {author} {\bibfnamefont {W.~J.}\ \bibnamefont {Zeng}},\ }\href
  {\doibase 10.48550/ARXIV.1608.03355} {\enquote {\bibinfo {title} {A practical
  quantum instruction set architecture},}\ } (\bibinfo {year}
  {2016})\BibitemShut {NoStop}%
\bibitem [{\citenamefont {Kottmann}\ \emph {et~al.}(2021)\citenamefont
  {Kottmann}, \citenamefont {Alperin-Lea}, \citenamefont {Tamayo-Mendoza},
  \citenamefont {Cervera-Lierta}, \citenamefont {Lavigne}, \citenamefont {Yen},
  \citenamefont {Verteletskyi}, \citenamefont {Schleich}, \citenamefont
  {Anand}, \citenamefont {Degroote}, \citenamefont {Chaney}, \citenamefont
  {Kesibi}, \citenamefont {Curnow}, \citenamefont {Solo}, \citenamefont
  {Tsilimigkounakis}, \citenamefont {Zendejas-Morales}, \citenamefont
  {Izmaylov},\ and\ \citenamefont {Aspuru-Guzik}}]{Kottmann_2021}%
  \BibitemOpen
  \bibfield  {author} {\bibinfo {author} {\bibfnamefont {J.~S.}\ \bibnamefont
  {Kottmann}}, \bibinfo {author} {\bibfnamefont {S.}~\bibnamefont
  {Alperin-Lea}}, \bibinfo {author} {\bibfnamefont {T.}~\bibnamefont
  {Tamayo-Mendoza}}, \bibinfo {author} {\bibfnamefont {A.}~\bibnamefont
  {Cervera-Lierta}}, \bibinfo {author} {\bibfnamefont {C.}~\bibnamefont
  {Lavigne}}, \bibinfo {author} {\bibfnamefont {T.-C.}\ \bibnamefont {Yen}},
  \bibinfo {author} {\bibfnamefont {V.}~\bibnamefont {Verteletskyi}}, \bibinfo
  {author} {\bibfnamefont {P.}~\bibnamefont {Schleich}}, \bibinfo {author}
  {\bibfnamefont {A.}~\bibnamefont {Anand}}, \bibinfo {author} {\bibfnamefont
  {M.}~\bibnamefont {Degroote}}, \bibinfo {author} {\bibfnamefont
  {S.}~\bibnamefont {Chaney}}, \bibinfo {author} {\bibfnamefont
  {M.}~\bibnamefont {Kesibi}}, \bibinfo {author} {\bibfnamefont {N.~G.}\
  \bibnamefont {Curnow}}, \bibinfo {author} {\bibfnamefont {B.}~\bibnamefont
  {Solo}}, \bibinfo {author} {\bibfnamefont {G.}~\bibnamefont
  {Tsilimigkounakis}}, \bibinfo {author} {\bibfnamefont {C.}~\bibnamefont
  {Zendejas-Morales}}, \bibinfo {author} {\bibfnamefont {A.~F.}\ \bibnamefont
  {Izmaylov}}, \ and\ \bibinfo {author} {\bibfnamefont {A.}~\bibnamefont
  {Aspuru-Guzik}},\ }\href {\doibase 10.1088/2058-9565/abe567} {\bibfield
  {journal} {\bibinfo  {journal} {Quantum Science and Technology}\ }\textbf
  {\bibinfo {volume} {6}},\ \bibinfo {pages} {024009} (\bibinfo {year}
  {2021})}\BibitemShut {NoStop}%
\bibitem [{qsh()}]{qsharp}%
  \BibitemOpen
  \href {https://github.com/microsoft/qsharp-language} {\enquote {\bibinfo
  {title} {Q\# language and core libraries design},}\ }\bibinfo {note}
  {Https://github.com/microsoft/qsharp-language}\BibitemShut {NoStop}%
\bibitem [{\citenamefont {Aleksandrowicz}\ \emph {et~al.}(2019)\citenamefont
  {Aleksandrowicz}, \citenamefont {Alexander}, \citenamefont {Barkoutsos},
  \citenamefont {Bello}, \citenamefont {Ben-Haim}, \citenamefont {Bucher},
  \citenamefont {Cabrera-Hernández}, \citenamefont {Carballo-Franquis},
  \citenamefont {Chen}, \citenamefont {Chen}, \citenamefont {Chow},
  \citenamefont {Córcoles-Gonzales}, \citenamefont {Cross}, \citenamefont
  {Cross}, \citenamefont {Cruz-Benito}, \citenamefont {Culver}, \citenamefont
  {González}, \citenamefont {Torre}, \citenamefont {Ding}, \citenamefont
  {Dumitrescu}, \citenamefont {Duran}, \citenamefont {Eendebak}, \citenamefont
  {Everitt}, \citenamefont {Sertage}, \citenamefont {Frisch}, \citenamefont
  {Fuhrer}, \citenamefont {Gambetta}, \citenamefont {Gago}, \citenamefont
  {Gomez-Mosquera}, \citenamefont {Greenberg}, \citenamefont {Hamamura},
  \citenamefont {Havlicek}, \citenamefont {Hellmers}, \citenamefont {Łukasz
  Herok}, \citenamefont {Horii}, \citenamefont {Hu}, \citenamefont {Imamichi},
  \citenamefont {Itoko}, \citenamefont {Javadi-Abhari}, \citenamefont
  {Kanazawa}, \citenamefont {Karazeev}, \citenamefont {Krsulich}, \citenamefont
  {Liu}, \citenamefont {Luh}, \citenamefont {Maeng}, \citenamefont {Marques},
  \citenamefont {Martín-Fernández}, \citenamefont {McClure}, \citenamefont
  {McKay}, \citenamefont {Meesala}, \citenamefont {Mezzacapo}, \citenamefont
  {Moll}, \citenamefont {Rodríguez}, \citenamefont {Nannicini}, \citenamefont
  {Nation}, \citenamefont {Ollitrault}, \citenamefont {O'Riordan},
  \citenamefont {Paik}, \citenamefont {Pérez}, \citenamefont {Phan},
  \citenamefont {Pistoia}, \citenamefont {Prutyanov}, \citenamefont {Reuter},
  \citenamefont {Rice}, \citenamefont {Davila}, \citenamefont {Rudy},
  \citenamefont {Ryu}, \citenamefont {Sathaye}, \citenamefont {Schnabel},
  \citenamefont {Schoute}, \citenamefont {Setia}, \citenamefont {Shi},
  \citenamefont {Silva}, \citenamefont {Siraichi}, \citenamefont {Sivarajah},
  \citenamefont {Smolin}, \citenamefont {Soeken}, \citenamefont {Takahashi},
  \citenamefont {Tavernelli}, \citenamefont {Taylor}, \citenamefont {Taylour},
  \citenamefont {Trabing}, \citenamefont {Treinish}, \citenamefont {Turner},
  \citenamefont {Vogt-Lee}, \citenamefont {Vuillot}, \citenamefont {Wildstrom},
  \citenamefont {Wilson}, \citenamefont {Winston}, \citenamefont {Wood},
  \citenamefont {Wood}, \citenamefont {Wörner}, \citenamefont {Akhalwaya},\
  and\ \citenamefont {Zoufal}}]{gadi_aleksandrowicz_2019_2562111}%
  \BibitemOpen
  \bibfield  {author} {\bibinfo {author} {\bibfnamefont {G.}~\bibnamefont
  {Aleksandrowicz}}, \bibinfo {author} {\bibfnamefont {T.}~\bibnamefont
  {Alexander}}, \bibinfo {author} {\bibfnamefont {P.}~\bibnamefont
  {Barkoutsos}}, \bibinfo {author} {\bibfnamefont {L.}~\bibnamefont {Bello}},
  \bibinfo {author} {\bibfnamefont {Y.}~\bibnamefont {Ben-Haim}}, \bibinfo
  {author} {\bibfnamefont {D.}~\bibnamefont {Bucher}}, \bibinfo {author}
  {\bibfnamefont {F.~J.}\ \bibnamefont {Cabrera-Hernández}}, \bibinfo {author}
  {\bibfnamefont {J.}~\bibnamefont {Carballo-Franquis}}, \bibinfo {author}
  {\bibfnamefont {A.}~\bibnamefont {Chen}}, \bibinfo {author} {\bibfnamefont
  {C.-F.}\ \bibnamefont {Chen}}, \bibinfo {author} {\bibfnamefont {J.~M.}\
  \bibnamefont {Chow}}, \bibinfo {author} {\bibfnamefont {A.~D.}\ \bibnamefont
  {Córcoles-Gonzales}}, \bibinfo {author} {\bibfnamefont {A.~J.}\ \bibnamefont
  {Cross}}, \bibinfo {author} {\bibfnamefont {A.}~\bibnamefont {Cross}},
  \bibinfo {author} {\bibfnamefont {J.}~\bibnamefont {Cruz-Benito}}, \bibinfo
  {author} {\bibfnamefont {C.}~\bibnamefont {Culver}}, \bibinfo {author}
  {\bibfnamefont {S.~D. L.~P.}\ \bibnamefont {González}}, \bibinfo {author}
  {\bibfnamefont {E.~D.~L.}\ \bibnamefont {Torre}}, \bibinfo {author}
  {\bibfnamefont {D.}~\bibnamefont {Ding}}, \bibinfo {author} {\bibfnamefont
  {E.}~\bibnamefont {Dumitrescu}}, \bibinfo {author} {\bibfnamefont
  {I.}~\bibnamefont {Duran}}, \bibinfo {author} {\bibfnamefont
  {P.}~\bibnamefont {Eendebak}}, \bibinfo {author} {\bibfnamefont
  {M.}~\bibnamefont {Everitt}}, \bibinfo {author} {\bibfnamefont {I.~F.}\
  \bibnamefont {Sertage}}, \bibinfo {author} {\bibfnamefont {A.}~\bibnamefont
  {Frisch}}, \bibinfo {author} {\bibfnamefont {A.}~\bibnamefont {Fuhrer}},
  \bibinfo {author} {\bibfnamefont {J.}~\bibnamefont {Gambetta}}, \bibinfo
  {author} {\bibfnamefont {B.~G.}\ \bibnamefont {Gago}}, \bibinfo {author}
  {\bibfnamefont {J.}~\bibnamefont {Gomez-Mosquera}}, \bibinfo {author}
  {\bibfnamefont {D.}~\bibnamefont {Greenberg}}, \bibinfo {author}
  {\bibfnamefont {I.}~\bibnamefont {Hamamura}}, \bibinfo {author}
  {\bibfnamefont {V.}~\bibnamefont {Havlicek}}, \bibinfo {author}
  {\bibfnamefont {J.}~\bibnamefont {Hellmers}}, \bibinfo {author} {\bibnamefont
  {Łukasz Herok}}, \bibinfo {author} {\bibfnamefont {H.}~\bibnamefont
  {Horii}}, \bibinfo {author} {\bibfnamefont {S.}~\bibnamefont {Hu}}, \bibinfo
  {author} {\bibfnamefont {T.}~\bibnamefont {Imamichi}}, \bibinfo {author}
  {\bibfnamefont {T.}~\bibnamefont {Itoko}}, \bibinfo {author} {\bibfnamefont
  {A.}~\bibnamefont {Javadi-Abhari}}, \bibinfo {author} {\bibfnamefont
  {N.}~\bibnamefont {Kanazawa}}, \bibinfo {author} {\bibfnamefont
  {A.}~\bibnamefont {Karazeev}}, \bibinfo {author} {\bibfnamefont
  {K.}~\bibnamefont {Krsulich}}, \bibinfo {author} {\bibfnamefont
  {P.}~\bibnamefont {Liu}}, \bibinfo {author} {\bibfnamefont {Y.}~\bibnamefont
  {Luh}}, \bibinfo {author} {\bibfnamefont {Y.}~\bibnamefont {Maeng}}, \bibinfo
  {author} {\bibfnamefont {M.}~\bibnamefont {Marques}}, \bibinfo {author}
  {\bibfnamefont {F.~J.}\ \bibnamefont {Martín-Fernández}}, \bibinfo {author}
  {\bibfnamefont {D.~T.}\ \bibnamefont {McClure}}, \bibinfo {author}
  {\bibfnamefont {D.}~\bibnamefont {McKay}}, \bibinfo {author} {\bibfnamefont
  {S.}~\bibnamefont {Meesala}}, \bibinfo {author} {\bibfnamefont
  {A.}~\bibnamefont {Mezzacapo}}, \bibinfo {author} {\bibfnamefont
  {N.}~\bibnamefont {Moll}}, \bibinfo {author} {\bibfnamefont {D.~M.}\
  \bibnamefont {Rodríguez}}, \bibinfo {author} {\bibfnamefont
  {G.}~\bibnamefont {Nannicini}}, \bibinfo {author} {\bibfnamefont
  {P.}~\bibnamefont {Nation}}, \bibinfo {author} {\bibfnamefont
  {P.}~\bibnamefont {Ollitrault}}, \bibinfo {author} {\bibfnamefont {L.~J.}\
  \bibnamefont {O'Riordan}}, \bibinfo {author} {\bibfnamefont {H.}~\bibnamefont
  {Paik}}, \bibinfo {author} {\bibfnamefont {J.}~\bibnamefont {Pérez}},
  \bibinfo {author} {\bibfnamefont {A.}~\bibnamefont {Phan}}, \bibinfo {author}
  {\bibfnamefont {M.}~\bibnamefont {Pistoia}}, \bibinfo {author} {\bibfnamefont
  {V.}~\bibnamefont {Prutyanov}}, \bibinfo {author} {\bibfnamefont
  {M.}~\bibnamefont {Reuter}}, \bibinfo {author} {\bibfnamefont
  {J.}~\bibnamefont {Rice}}, \bibinfo {author} {\bibfnamefont {A.~R.}\
  \bibnamefont {Davila}}, \bibinfo {author} {\bibfnamefont {R.~H.~P.}\
  \bibnamefont {Rudy}}, \bibinfo {author} {\bibfnamefont {M.}~\bibnamefont
  {Ryu}}, \bibinfo {author} {\bibfnamefont {N.}~\bibnamefont {Sathaye}},
  \bibinfo {author} {\bibfnamefont {C.}~\bibnamefont {Schnabel}}, \bibinfo
  {author} {\bibfnamefont {E.}~\bibnamefont {Schoute}}, \bibinfo {author}
  {\bibfnamefont {K.}~\bibnamefont {Setia}}, \bibinfo {author} {\bibfnamefont
  {Y.}~\bibnamefont {Shi}}, \bibinfo {author} {\bibfnamefont {A.}~\bibnamefont
  {Silva}}, \bibinfo {author} {\bibfnamefont {Y.}~\bibnamefont {Siraichi}},
  \bibinfo {author} {\bibfnamefont {S.}~\bibnamefont {Sivarajah}}, \bibinfo
  {author} {\bibfnamefont {J.~A.}\ \bibnamefont {Smolin}}, \bibinfo {author}
  {\bibfnamefont {M.}~\bibnamefont {Soeken}}, \bibinfo {author} {\bibfnamefont
  {H.}~\bibnamefont {Takahashi}}, \bibinfo {author} {\bibfnamefont
  {I.}~\bibnamefont {Tavernelli}}, \bibinfo {author} {\bibfnamefont
  {C.}~\bibnamefont {Taylor}}, \bibinfo {author} {\bibfnamefont
  {P.}~\bibnamefont {Taylour}}, \bibinfo {author} {\bibfnamefont
  {K.}~\bibnamefont {Trabing}}, \bibinfo {author} {\bibfnamefont
  {M.}~\bibnamefont {Treinish}}, \bibinfo {author} {\bibfnamefont
  {W.}~\bibnamefont {Turner}}, \bibinfo {author} {\bibfnamefont
  {D.}~\bibnamefont {Vogt-Lee}}, \bibinfo {author} {\bibfnamefont
  {C.}~\bibnamefont {Vuillot}}, \bibinfo {author} {\bibfnamefont {J.~A.}\
  \bibnamefont {Wildstrom}}, \bibinfo {author} {\bibfnamefont {J.}~\bibnamefont
  {Wilson}}, \bibinfo {author} {\bibfnamefont {E.}~\bibnamefont {Winston}},
  \bibinfo {author} {\bibfnamefont {C.}~\bibnamefont {Wood}}, \bibinfo {author}
  {\bibfnamefont {S.}~\bibnamefont {Wood}}, \bibinfo {author} {\bibfnamefont
  {S.}~\bibnamefont {Wörner}}, \bibinfo {author} {\bibfnamefont {I.~Y.}\
  \bibnamefont {Akhalwaya}}, \ and\ \bibinfo {author} {\bibfnamefont
  {C.}~\bibnamefont {Zoufal}},\ }\href {\doibase 10.5281/zenodo.2562111} {\
  (\bibinfo {year} {2019}),\ 10.5281/zenodo.2562111}\BibitemShut {NoStop}%
\bibitem [{\citenamefont {McCaskey}\ \emph {et~al.}(2020)\citenamefont
  {McCaskey}, \citenamefont {Lyakh}, \citenamefont {Dumitrescu}, \citenamefont
  {Powers},\ and\ \citenamefont {Humble}}]{McCaskey_2020}%
  \BibitemOpen
  \bibfield  {author} {\bibinfo {author} {\bibfnamefont {A.~J.}\ \bibnamefont
  {McCaskey}}, \bibinfo {author} {\bibfnamefont {D.~I.}\ \bibnamefont {Lyakh}},
  \bibinfo {author} {\bibfnamefont {E.~F.}\ \bibnamefont {Dumitrescu}},
  \bibinfo {author} {\bibfnamefont {S.~S.}\ \bibnamefont {Powers}}, \ and\
  \bibinfo {author} {\bibfnamefont {T.~S.}\ \bibnamefont {Humble}},\ }\href
  {\doibase 10.1088/2058-9565/ab6bf6} {\bibfield  {journal} {\bibinfo
  {journal} {Quantum Science and Technology}\ }\textbf {\bibinfo {volume}
  {5}},\ \bibinfo {pages} {024002} (\bibinfo {year} {2020})}\BibitemShut
  {NoStop}%
\bibitem [{\citenamefont {Rubin}\ \emph {et~al.}(2021)\citenamefont {Rubin},
  \citenamefont {Gunst}, \citenamefont {White}, \citenamefont {Freitag},
  \citenamefont {Throssell}, \citenamefont {Chan}, \citenamefont {Babbush},\
  and\ \citenamefont {Shiozaki}}]{Rubin:2021znj}%
  \BibitemOpen
  \bibfield  {author} {\bibinfo {author} {\bibfnamefont {N.~C.}\ \bibnamefont
  {Rubin}}, \bibinfo {author} {\bibfnamefont {K.}~\bibnamefont {Gunst}},
  \bibinfo {author} {\bibfnamefont {A.}~\bibnamefont {White}}, \bibinfo
  {author} {\bibfnamefont {L.}~\bibnamefont {Freitag}}, \bibinfo {author}
  {\bibfnamefont {K.}~\bibnamefont {Throssell}}, \bibinfo {author}
  {\bibfnamefont {G.~K.-L.}\ \bibnamefont {Chan}}, \bibinfo {author}
  {\bibfnamefont {R.}~\bibnamefont {Babbush}}, \ and\ \bibinfo {author}
  {\bibfnamefont {T.}~\bibnamefont {Shiozaki}},\ }\href {\doibase
  10.22331/q-2021-10-27-568} {\bibfield  {journal} {\bibinfo  {journal}
  {Quantum}\ }\textbf {\bibinfo {volume} {5}},\ \bibinfo {pages} {568}
  (\bibinfo {year} {2021})},\ \Eprint {http://arxiv.org/abs/2104.13944}
  {arXiv:2104.13944 [quant-ph]} \BibitemShut {NoStop}%
\bibitem [{\citenamefont {Developers}(2021)}]{cirq_developers_2021_5182845}%
  \BibitemOpen
  \bibfield  {author} {\bibinfo {author} {\bibfnamefont {C.}~\bibnamefont
  {Developers}},\ }\href {\doibase 10.5281/zenodo.5182845} {\enquote {\bibinfo
  {title} {Cirq},}\ } (\bibinfo {year} {2021}),\ \bibinfo {note} {{See full
  list of authors on Github: https://github
  .com/quantumlib/Cirq/graphs/contributors}}\BibitemShut {NoStop}%
\bibitem [{\citenamefont {Bharti}\ \emph {et~al.}(2022)\citenamefont {Bharti}
  \emph {et~al.}}]{Bharti:2021zez}%
  \BibitemOpen
  \bibfield  {author} {\bibinfo {author} {\bibfnamefont {K.}~\bibnamefont
  {Bharti}} \emph {et~al.},\ }\href {\doibase 10.1103/RevModPhys.94.015004}
  {\bibfield  {journal} {\bibinfo  {journal} {Rev. Mod. Phys.}\ }\textbf
  {\bibinfo {volume} {94}},\ \bibinfo {pages} {015004} (\bibinfo {year}
  {2022})},\ \Eprint {http://arxiv.org/abs/2101.08448} {arXiv:2101.08448
  [quant-ph]} \BibitemShut {NoStop}%
\bibitem [{\citenamefont {Anand}\ \emph {et~al.}(2021)\citenamefont {Anand},
  \citenamefont {Schleich}, \citenamefont {Alperin-Lea}, \citenamefont
  {Jensen}, \citenamefont {Sim}, \citenamefont {D\'\i{}az-Tinoco},
  \citenamefont {Kottmann}, \citenamefont {Degroote}, \citenamefont
  {Izmaylov},\ and\ \citenamefont {Aspuru-Guzik}}]{Anand:2021xbq}%
  \BibitemOpen
  \bibfield  {author} {\bibinfo {author} {\bibfnamefont {A.}~\bibnamefont
  {Anand}}, \bibinfo {author} {\bibfnamefont {P.}~\bibnamefont {Schleich}},
  \bibinfo {author} {\bibfnamefont {S.}~\bibnamefont {Alperin-Lea}}, \bibinfo
  {author} {\bibfnamefont {P.~W.~K.}\ \bibnamefont {Jensen}}, \bibinfo {author}
  {\bibfnamefont {S.}~\bibnamefont {Sim}}, \bibinfo {author} {\bibfnamefont
  {M.}~\bibnamefont {D\'\i{}az-Tinoco}}, \bibinfo {author} {\bibfnamefont
  {J.~S.}\ \bibnamefont {Kottmann}}, \bibinfo {author} {\bibfnamefont
  {M.}~\bibnamefont {Degroote}}, \bibinfo {author} {\bibfnamefont {A.~F.}\
  \bibnamefont {Izmaylov}}, \ and\ \bibinfo {author} {\bibfnamefont
  {A.}~\bibnamefont {Aspuru-Guzik}},\ }\href {\doibase 10.1039/D1CS00932J} {\
  (\bibinfo {year} {2021}),\ 10.1039/D1CS00932J},\ \Eprint
  {http://arxiv.org/abs/2109.15176} {arXiv:2109.15176 [quant-ph]} \BibitemShut
  {NoStop}%
\end{thebibliography}%
\end{document}